\RequirePackage{lineno}
\documentclass[aps,prl,twocolumn,showpacs,superscriptaddress,groupedaddress]{revtex4}  
\usepackage{graphicx}  
\usepackage{dcolumn}   
\usepackage{bm}        
\usepackage{amssymb}   
\usepackage{amsmath}
\usepackage{footnote}
\usepackage{setspace}
\usepackage{enumerate}
\usepackage{lineno}
\usepackage{color}

\def \ttbar {\ensuremath{ t\bar{t}                  }}

\def \ifb    {\ensuremath{ \rm fb^{-1}                          }}

\def \pt     {\ensuremath{ p_T                                  }}

\def \pte    {\ensuremath{ p_{T}^{e}                            }}
\def \ptm    {\ensuremath{ p_{T}^{\mu}                          }}
\def \ptl    {\ensuremath{ p_{T}^{\ell}                        }}
\def \ptone  {\ensuremath{ p_{T}^{\ell1}                        }}
\def \pttwo  {\ensuremath{ p_{T}^{\ell2}                        }}

\def \egev   {\ensuremath{ \, \mathrm{GeV}                      }}

\def \mtmin {\ensuremath{M_T^{\rm min}}}
\def \mt    {\ensuremath{M_T}}
\def \etmisssc {\mbox{\ensuremath{E\kern-0.6em\slash_T^{\rm\kern+0.1em Sc}}}}
\def \etmiss {\mbox{\ensuremath{E\kern-0.6em\slash_T}}}

\def \ee {$ee$}
\def \em {$e\mu$}
\def \mm {$\mu\mu$}

\newcommand{\Eslash}{\mbox{$E \kern-0.6em\slash$                }}

\def \nL0  {N$_{\mathrm{L0}}$}

\begin{document}

\widetext
\hspace{5.2in} \mbox{Fermilab-Pub-12/335-E}

\title{Search for Higgs boson production in oppositely charged dilepton and missing energy 
events in $\boldsymbol{p\bar{p}}$ collisions at $\boldsymbol{\sqrt{s} =}$
1.96~TeV}
\affiliation{LAFEX, Centro Brasileiro de Pesquisas F\'{i}sicas, Rio de Janeiro, Brazil}
\affiliation{Universidade do Estado do Rio de Janeiro, Rio de Janeiro, Brazil}
\affiliation{Universidade Federal do ABC, Santo Andr\'e, Brazil}
\affiliation{University of Science and Technology of China, Hefei, People's Republic of China}
\affiliation{Universidad de los Andes, Bogot\'a, Colombia}
\affiliation{Charles University, Faculty of Mathematics and Physics, Center for Particle Physics, Prague, Czech Republic}
\affiliation{Czech Technical University in Prague, Prague, Czech Republic}
\affiliation{Center for Particle Physics, Institute of Physics, Academy of Sciences of the Czech Republic, Prague, Czech Republic}
\affiliation{Universidad San Francisco de Quito, Quito, Ecuador}
\affiliation{LPC, Universit\'e Blaise Pascal, CNRS/IN2P3, Clermont, France}
\affiliation{LPSC, Universit\'e Joseph Fourier Grenoble 1, CNRS/IN2P3, Institut National Polytechnique de Grenoble, Grenoble, France}
\affiliation{CPPM, Aix-Marseille Universit\'e, CNRS/IN2P3, Marseille, France}
\affiliation{LAL, Universit\'e Paris-Sud, CNRS/IN2P3, Orsay, France}
\affiliation{LPNHE, Universit\'es Paris VI and VII, CNRS/IN2P3, Paris, France}
\affiliation{CEA, Irfu, SPP, Saclay, France}
\affiliation{IPHC, Universit\'e de Strasbourg, CNRS/IN2P3, Strasbourg, France}
\affiliation{IPNL, Universit\'e Lyon 1, CNRS/IN2P3, Villeurbanne, France and Universit\'e de Lyon, Lyon, France}
\affiliation{III. Physikalisches Institut A, RWTH Aachen University, Aachen, Germany}
\affiliation{Physikalisches Institut, Universit\"at Freiburg, Freiburg, Germany}
\affiliation{II. Physikalisches Institut, Georg-August-Universit\"at G\"ottingen, G\"ottingen, Germany}
\affiliation{Institut f\"ur Physik, Universit\"at Mainz, Mainz, Germany}
\affiliation{Ludwig-Maximilians-Universit\"at M\"unchen, M\"unchen, Germany}
\affiliation{Fachbereich Physik, Bergische Universit\"at Wuppertal, Wuppertal, Germany}
\affiliation{Panjab University, Chandigarh, India}
\affiliation{Delhi University, Delhi, India}
\affiliation{Tata Institute of Fundamental Research, Mumbai, India}
\affiliation{University College Dublin, Dublin, Ireland}
\affiliation{Korea Detector Laboratory, Korea University, Seoul, Korea}
\affiliation{CINVESTAV, Mexico City, Mexico}
\affiliation{Nikhef, Science Park, Amsterdam, the Netherlands}
\affiliation{Radboud University Nijmegen, Nijmegen, the Netherlands}
\affiliation{Joint Institute for Nuclear Research, Dubna, Russia}
\affiliation{Institute for Theoretical and Experimental Physics, Moscow, Russia}
\affiliation{Moscow State University, Moscow, Russia}
\affiliation{Institute for High Energy Physics, Protvino, Russia}
\affiliation{Petersburg Nuclear Physics Institute, St. Petersburg, Russia}
\affiliation{Instituci\'{o} Catalana de Recerca i Estudis Avan\c{c}ats (ICREA) and Institut de F\'{i}sica d'Altes Energies (IFAE), Barcelona, Spain}
\affiliation{Uppsala University, Uppsala, Sweden}
\affiliation{Lancaster University, Lancaster LA1 4YB, United Kingdom}
\affiliation{Imperial College London, London SW7 2AZ, United Kingdom}
\affiliation{The University of Manchester, Manchester M13 9PL, United Kingdom}
\affiliation{University of Arizona, Tucson, Arizona 85721, USA}
\affiliation{University of California Riverside, Riverside, California 92521, USA}
\affiliation{Florida State University, Tallahassee, Florida 32306, USA}
\affiliation{Fermi National Accelerator Laboratory, Batavia, Illinois 60510, USA}
\affiliation{University of Illinois at Chicago, Chicago, Illinois 60607, USA}
\affiliation{Northern Illinois University, DeKalb, Illinois 60115, USA}
\affiliation{Northwestern University, Evanston, Illinois 60208, USA}
\affiliation{Indiana University, Bloomington, Indiana 47405, USA}
\affiliation{Purdue University Calumet, Hammond, Indiana 46323, USA}
\affiliation{University of Notre Dame, Notre Dame, Indiana 46556, USA}
\affiliation{Iowa State University, Ames, Iowa 50011, USA}
\affiliation{University of Kansas, Lawrence, Kansas 66045, USA}
\affiliation{Kansas State University, Manhattan, Kansas 66506, USA}
\affiliation{Louisiana Tech University, Ruston, Louisiana 71272, USA}
\affiliation{Boston University, Boston, Massachusetts 02215, USA}
\affiliation{Northeastern University, Boston, Massachusetts 02115, USA}
\affiliation{University of Michigan, Ann Arbor, Michigan 48109, USA}
\affiliation{Michigan State University, East Lansing, Michigan 48824, USA}
\affiliation{University of Mississippi, University, Mississippi 38677, USA}
\affiliation{University of Nebraska, Lincoln, Nebraska 68588, USA}
\affiliation{Rutgers University, Piscataway, New Jersey 08855, USA}
\affiliation{Princeton University, Princeton, New Jersey 08544, USA}
\affiliation{State University of New York, Buffalo, New York 14260, USA}
\affiliation{University of Rochester, Rochester, New York 14627, USA}
\affiliation{State University of New York, Stony Brook, New York 11794, USA}
\affiliation{Brookhaven National Laboratory, Upton, New York 11973, USA}
\affiliation{Langston University, Langston, Oklahoma 73050, USA}
\affiliation{University of Oklahoma, Norman, Oklahoma 73019, USA}
\affiliation{Oklahoma State University, Stillwater, Oklahoma 74078, USA}
\affiliation{Brown University, Providence, Rhode Island 02912, USA}
\affiliation{University of Texas, Arlington, Texas 76019, USA}
\affiliation{Southern Methodist University, Dallas, Texas 75275, USA}
\affiliation{Rice University, Houston, Texas 77005, USA}
\affiliation{University of Virginia, Charlottesville, Virginia 22901, USA}
\affiliation{University of Washington, Seattle, Washington 98195, USA}
\author{V.M.~Abazov} \affiliation{Joint Institute for Nuclear Research, Dubna, Russia}
\author{B.~Abbott} \affiliation{University of Oklahoma, Norman, Oklahoma 73019, USA}
\author{B.S.~Acharya} \affiliation{Tata Institute of Fundamental Research, Mumbai, India}
\author{M.~Adams} \affiliation{University of Illinois at Chicago, Chicago, Illinois 60607, USA}
\author{T.~Adams} \affiliation{Florida State University, Tallahassee, Florida 32306, USA}
\author{G.D.~Alexeev} \affiliation{Joint Institute for Nuclear Research, Dubna, Russia}
\author{G.~Alkhazov} \affiliation{Petersburg Nuclear Physics Institute, St. Petersburg, Russia}
\author{A.~Alton$^{a}$} \affiliation{University of Michigan, Ann Arbor, Michigan 48109, USA}
\author{G.~Alverson} \affiliation{Northeastern University, Boston, Massachusetts 02115, USA}
\author{A.~Askew} \affiliation{Florida State University, Tallahassee, Florida 32306, USA}
% ...
%
%
\collaboration{The D0 Collaboration\footnote{with visitors from
$^{a}$Augustana College, Sioux Falls, SD, USA,
$^{b}$The University of Liverpool, Liverpool, UK,
$^{c}$UPIITA-IPN, Mexico City, Mexico,
$^{d}$DESY, Hamburg, Germany,
,
$^{e}$SLAC, Menlo Park, CA, USA,
$^{f}$University College London, London, UK,
$^{g}$Centro de Investigacion en Computacion - IPN, Mexico City, Mexico,
$^{h}$ECFM, Universidad Autonoma de Sinaloa, Culiac\'an, Mexico
and
$^{i}$Universidade Estadual Paulista, S\~ao Paulo, Brazil.
}} \noaffiliation
\vskip 0.25cm
   
\date{July 04, 2012}

\begin{abstract}
We present a search for the standard model Higgs boson using events
with two oppositely charged leptons and large missing transverse
energy as expected in $H\rightarrow WW$ decays. The events are
selected from data corresponding to 8.6~\ifb\ of integrated luminosity
in $p
\overline{p}$ collisions at $\sqrt{s}=1.96$ TeV collected with the D0
detector at the Fermilab Tevatron Collider. No significant excess
above the standard model background expectation in the Higgs boson
mass range this search is sensitive to is observed, and upper limits
on the Higgs boson production cross section are derived.
\end{abstract}

\maketitle

\section{\label{sec:intro}INTRODUCTION}

In the standard model (SM), the Higgs boson appears during the
spontaneous breaking of the electroweak symmetry $SU(2) \times U(1)$
that is responsible for the generation of the masses of the $W$ and
$Z$ bosons.  Although the SM requires the existence of this neutral
scalar particle, its mass ($M_{H}$) is a free parameter.  Direct
searches at the CERN $e^+e^-$ collider (LEP) yield a lower limit of
$M_H > 114.4~\egev$~\cite{bib:lephiggs} at the 95\%~C.L.
~Precision electroweak data yield, including the latest $W$ boson mass
requirements from CDF~\cite{bib:cdf_Wmass} and D0~\cite{bib:d0_Wmass},
constrain the mass of a SM Higgs boson to $M_H <
152~\egev$~\cite{bib:lepewwg_summer11} at 95\% C.L.

In this Article, we present a search for the SM Higgs boson in final
states containing two oppositely charged leptons ($\ell\ell'$=$e\mu$,
$ee$, or $\mu\mu$, where small contributions from leptonic
$\tau$~decays are also included) and missing transverse energy
($\etmiss$), using $8.6~\ifb$ of $p\bar{p}$ collisions collected with
the D0 detector~\cite{bib:d0det} at the Fermilab Tevatron
Collider. These three leptonic final states are combined to produce a
result which supersedes our previously published search for Higgs
boson production in the oppositely charged dilepton and missing
transverse energy final state based on data corresponding to an
integrated luminosity of 5.4~\ifb~\cite{bib:hww}.  A similar search
was published by the CDF Collaboration at the Tevatron using 4.8\,
\ifb\ of integrated luminosity~\cite{bib:cdf-hww} and by the ATLAS and
CMS Collaborations at the CERN Large Hadron Collider (LHC) using 4.7\,
\ifb\ and 4.6\, \ifb\ of data, respectively~\cite{bib:atlas-hww,
bib:cms-hww}. Using up to 5.4 fb$^{-1}$ of integrated luminosity, the
combination of the results from the Tevatron led to the first
exclusion using the $H \rightarrow WW$ decays, excluding the Higgs
boson beyond the LEP limits, in the mass range from 162 to 166 GeV at
the 95\% C.L.~\cite{bib:tevcomb-hww}. Recently, both ATLAS and CMS
Collaborations have individually combined all their searches, and the
results from ATLAS have excluded a Higgs boson in the mass range from
111.4 to 116.6, 119.4 to 122.1, 129.2 to 541 GeV, while results from
CMS excluded a Higgs boson in the range 127 to 600 GeV at the 95\%
C.L.~\cite{bib:atlas_comb, bib:cms_comb}.


The primary signal for opposite charge dilepton signatures with
considerable missing energy arises from production of Higgs bosons by
gluon fusion $gg\rightarrow H$ with subsequent decay $H \rightarrow WW
\rightarrow \ell\nu\ell'\nu'$. Additional contributions to this
signature come from vector boson fusion (VBF), $qq'\rightarrow qq'H$,
where the initial state partons radiate weak gauge bosons that then
fuse to form a Higgs boson, and from production in association with a
vector boson $qq'\rightarrow VH=(W/Z)H$.  The dominant background
contribution is from diboson production, in particular, contributions
from non-resonant $p\bar{p} \rightarrow WW
\rightarrow~\ell\nu\ell'\nu'$ processes. Additionally, two types of
instrumental backgrounds exist: 1) events with mismeasured $\etmiss$
in the Drell-Yan process $p\bar{p} \rightarrow Z/\gamma^*\rightarrow
\ell^+\ell^-$, which contribute particularly to the \ee\ and \mm\
final states, and 2) events with jets misidentified as leptons and
photons converting to electrons in $W$ boson or multijet production.
Although such false identification is rare, the resulting backgrounds
are sizeable as the rates of $W+$jets and multijet production are
significantly higher than that of Higgs boson
production. Contributions in the $\mu\mu$ channel from falsely
identified muons in $W+$jets events are relatively smaller.

The following Article first discusses the simulation methods used to
predict the yields from signal and SM background processes. This is
then followed by a brief description of the D0 detector and of the
algorithms used to reconstruct and identify the objects used in the
analysis. The event selection and the multivariate techniques used to
separate the signal from the background are then discussed. The
different sources of systematics uncertainties are then presented,
followed by the results of the search for the Higgs boson.

\section{\label{sec:simulation}EVENT SIMULATION}

Higgs boson signal samples are simulated using the {\sc
pythia}~\cite{bib:pythia} Monte Carlo (MC) event generator with the
CTEQ6L1 parton distribution functions (PDFs)~\cite{bib:cteq} for $115
\le M_{H} \le 200$~GeV in increments of $5$~GeV.  The normalization of
these MC samples is obtained using the highest-order cross section
calculation available for the corresponding production process.  The
cross section for the gluon fusion process is calculated at
next-to-next-to-leading order (NNLO) in quantum chromodynamics with
soft gluon resummation to next-to-next-to-leading-log (NNLL)
accuracy~\cite{bib:gluon_fusion_xsec}. For $WH$, $ZH$ and vector boson
fusion processes, cross section calculations at NNLO are used
~\cite{bib:vh-xs,bib:vbf_xsec}.  All signal cross sections are
computed using the MSTW2008 PDF set~\cite{bib:mstw08}.  The PDF
uncertainties are assessed according to the recommendations given in
Refs.~\cite{bib:pdf_uncertainties, bib:ggH01jetUncert}.  The Higgs
boson branching ratio predictions are from {\sc
hdecay}~\cite{bib:hdecay}. The distribution of the transverse momentum
($p_{T}$) of the Higgs boson in the {\sc pythia}-generated gluon
fusion sample is reweighted to match the $p_{T}$ as calculated by {\sc
hqt}, at NNLL and NNLO accuracy~\cite{bib:Higgs_pT}.

The dominant background processes for the search are
$Z/\gamma^{*}+$jets, $W$+jets, diboson, $t\bar{t}$, and multijet
production where jets can be misidentified as leptons.  Electroweak
single top quark production is not considered since its contribution
is negligible. The $W+$jets and $Z+$jets backgrounds are modeled using
{\sc alpgen}~\cite{bib:alpgen}, with showering and hadronization
provided by {\sc pythia}.  Diboson production processes ($WW$, $WZ$,
and $ZZ$) are simulated using {\sc pythia}.  The $Z+$jets and $W+$jets
processes are normalized using the NNLO cross section calculations of
Ref.~\cite{bib:v-xs} which uses the NLO CTEQ6.1 PDFs.  The $Z$ boson
$p_{T}$ distribution is weighted to match the distribution observed in
data~\cite{bib:zbosonpT},~taking into account its dependence on the
number of reconstructed jets.  The $W$ boson $p_{T}$ distribution is
corrected to match the measured $Z$ boson $p_{T}$
spectrum~\cite{bib:zbosonpT} multiplied by the ratio of the $W$ boson
$p_{T}$ to $Z$ boson $p_{T}$ distributions as predicted in NLO
QCD~\cite{bib:w_z_bosonpT_ratio}.  For the search in the $ee$ and
$e\mu$ channels, the $W+$jets sample includes contributions from
events in which a jet or a photon is misidentified as an electron. For
$t\bar{t}$ production, approximate NNLO cross
sections~\cite{bib:tt-xs} are used, while the NLO production cross
section values are used for $WW$, $WZ$, and $ZZ$
processes~\cite{bib:dibo-xs}.  For the irreducible background source,
$WW$ production, the $p_{T}$ of the diboson system is modeled using
the {\sc mc@nlo} simulation~\cite{bib:mcatnlo}. All MC samples are
processed through a {\sc geant} simulation of the
detector~\cite{bib:geant}.  Recorded detector signals from randomly
selected beam crossings with the same luminosity profile as data are
added to the simulated detector signals of MC events in order to model
effects of detector noise and additional $p\bar{p}$ interactions. The
simulated background samples are subsequently normalized to the
integrated luminosity.

\section{DETECTOR AND OBJECT RECONSTRUCTION}\label{sec:detector}

The innermost part of the D0 detector~\cite{bib:d0det}~is composed of
a central tracking system with a silicon microstrip tracker (SMT) and
a central fiber tracker (CFT) embedded within a 2\,T solenoidal
magnet.  The tracking system is surrounded by a central preshower
detector (CPS) and a liquid-argon and uranium calorimeter with
electromagnetic (EM), fine and coarse hadronic sections.  A muon
spectrometer resides beyond the calorimetry and is made of drift
tubes, scintillation counters and toroidal magnets. The D0 detector
was upgraded in Spring 2006 to include modifications to the trigger
system~\cite{trigger} as well as an additional inner layer of silicon
microstrip tracking installed near the beam pipe and referred to as
Layer 0~\cite{bib:layer0}. The data used for this analysis include
1.1\, \ifb\ collected before these upgrades (Run IIa) and 7.5\, \ifb\
collected afterwards (Run IIb).

Electrons are identified as clusters in the EM calorimeter and are
required to spatially match a track reconstructed in the central
tracking detector. The electron energy is measured from the
calorimeter energy deposits within a cone of a radius $\mathcal{R} =
\sqrt{(\Delta\eta)^2 + (\Delta\phi)^2}$ = 0.2, in the four layers of
the EM calorimeter and the first layer of the hadronic calorimeter,
where $\eta$ and $\phi$ are the pseudorapidity~\cite{bib:eta} and the
azimuthal angle, respectively. Due to the different probabilities for
jets and photons to be misidentified as electrons in the central (CC,
$|\eta| <$ 1.1) and the forward (EC, $|\eta| >$ 1.5) calorimeter
regions, different criteria are required for the electron
clusters. The most important of these are: (1) calorimeter isolation
$f_{\mathrm{iso}}$ less than 0.15 (CC) or less than 0.1 (EC), with
$f_{\mathrm{iso}}=[E_{\mathrm{tot}}({\cal R}<0.4)-
E_{\mathrm{EM}}({\cal R}<0.2)]/E_{\mathrm{EM}}({\cal R}<0.2)$, where
$E_{\mathrm{tot}}$ is the total energy in the isolation cone of radius
$ {\cal R}=0.4$ and $E_{\mathrm{EM}}$ is the EM energy in a cone of
radius ${\cal R}=0.2$; (2) an EM fraction $f_\mathrm{EM}$ larger than
0.9, where $f_\mathrm{EM}$ is the fraction of total energy deposited
in the EM calorimeter, where the energies are measured within a cone
of radius ${\cal R}=0.2$; (3) a track isolation $h_{\mathrm{iso}}$
less than 3.5 GeV (CC) or less than $(-2.5\times|\eta|+7.0)$\,GeV
(EC), where $h_{\mathrm{iso}}$ is the scalar sum of the $p_{T}$ of all
tracks originating from the primary $p\bar{p}$ interaction vertex in
an annulus $0.05<{\cal R}<0.4$ around the cluster; (4) a cluster shape
consistent with that of an EM shower; (5) an eight-variable electron
likelihood ${\cal L}_8$ that is required to be greater than 0.05,
where ${\cal L}_8$ is constructed using the variables:
$f_{\mathrm{iso}}$, $f_{\mathrm{EM}}$, $h_{\mathrm{iso}}$, the ratio
of the electron cluster energy to track momentum ${(E/p)}$, the number
of tracks within a cone of radius ${\cal R} =0.4$, the track-cluster
match probability computed from the spatial separation and the
expected resolution, the track distance to the $p\bar{p}$ interaction
vertex at closest approach (dca), and covariance matrices that contain
variables that relate the energy depositions between various layers of
the calorimeter as well as the longitudinal and lateral shower
development; and (6) an artificial neural network trained using
information from the tracker, calorimeter and CPS detector to further
reject backgrounds from jets misidentified as electrons.

Muons are identified by the presence of at least one track segment,
reconstructed in the muon spectrometer, that is spatially consistent
with a track in the central detector. The momentum and charge are
measured by the curvature of the central track.  The muon candidate
must pass quality requirements aimed at reducing false matching and
background from cosmic rays.  Muons are required to be isolated.  The
isolation variables are defined to be the scalar sum of the transverse
energy in the calorimeter ($\Sigma E_T^{\mathrm{calo}}$) within an
annular cone $0.1<{\cal R}<0.4$ and the scalar sum of the transverse
momenta ($\Sigma p_{T}^{\mathrm{trk}}$) of tracks within a cone ${\cal
R} < 0.5$ around the muon candidate.  In the \mm\ channel, the
isolation variables for each muon must satisfy $\Sigma
p_{T}^{\mathrm{trk}}<0.25\times\ptm$ and $\Sigma
E_{T}^{\mathrm{calo}}<0.4\times\ptm$, \ptm\ being the momentum of the
muon.  Similarly in the \em\ channel, the isolation variables must
satisfy $\Sigma p_{T}^{\mathrm{trk}}<0.15\times\ptm$ and $\Sigma
E_{T}^{\mathrm{calo}}<0.15\times\ptm$. For \em\ and \mm\ channels,
the momentum of the muon track, the momentum of the electron track,
and the electron energy deposit in the calorimeter are not considered
when calculating isolation variables for the other lepton.  This
prevents the presence of one lepton to spoil the isolation of the
other lepton in events where the separation of leptons in $(\eta,
\phi)$ space is smaller than ${\cal R}=0.5$.

Jets are reconstructed from energy deposits in the calorimeter using
Ban iterative midpoint cone algorithm with a cone radius of 0.5
\cite{bib:jet}. All jets are required to have at least two associated
tracks matched to the $p\bar{p}$ interaction vertex.
The efficiency of this requirement is adjusted in the simulation to
match that measured in data. Jets can be identified as likely
containing $b$ quarks ($b$-tagged) if they pass a selection cut on the
output of a multivariate (MVA) based b-tagging discriminant, trained
to separate $b$ jets from light jets~\cite{bib:bid}. The jet energies
are calibrated using transverse momentum balance in $\gamma+$jet
events \cite{bib:jetCalib}. To account for differences in the
quark/gluon jet composition between the $\gamma+$jet events and the
$W/Z+$jet events, the jet energies are further corrected in simulated
events to match those measured in $Z+$jets data. Comparison of {\sc
alpgen} with other generators~\cite{bib:other_gen} and with the data
~\cite{bib:marco_ref} shows discrepancies in jet $\eta$ and dijet
angular separation. Therefore a data based correction allows for a
better modeling of these quantities in the {\sc
alpgen}~$Z/\gamma^*\mathrm{+jets}$~samples.


The $\etmiss$ is obtained from the vector sum of the transverse
components of energy depositions in the electromagnetic and fine
hadronic sections of the calorimeter and is corrected for any
identified muons. All energy corrections to leptons and to jets are
propagated to the $\etmiss$. Data based corrections are applied to MC
samples which allow for a better modeling of the calorimeter response
to unclustered objects.

In order to increase acceptance, all events satisfying any trigger
requirement from the complete suite of triggers used for data taking
are considered.  While most of the candidate events in the analysis
are selected by single-lepton and dilepton triggers, a gain in
efficiency of up to 20\%, depending on the channel is achieved by
including events which pass lepton+jets and lepton$+\etmiss$ triggers.

\section{EVENT SELECTION}\label{sec:selection}

Candidate events are selected by requiring at least two high-$p_T$
oppositely charged leptons $(e~\rm or~\mu)$, to originate from the
same $p\bar{p}$ interaction vertex along the beam line ({\it i.e.,}
within $\Delta$$z$=2\,cm, where $\Delta z$ is the distance between
lepton tracks along the beam axis, measured at the distance of closest
approach to this axis). Additional selections are carried out in two
steps, ``pre-selection'' and the ``final-selection''. The following
section describes the selections imposed in each step and also
outlines the additional kinematic and quality requirements for the
search.

\subsection{Preselection}\label{sec:preselection}

In the \mm~and \ee~channels, the two highest-$p_{T}$ leptons are
required to satisfy $\ptone>15$\,GeV and $\pttwo>10$\,GeV
respectively, whereas in the \em\ channel,~$\pte>15$\,GeV and
$\ptm>10$\,GeV are required. Electrons are required to be within the
acceptances of the EM calorimeter and tracking system ($|\eta|<1.1$ or
$1.5<|\eta|<2.5$) and muons are restricted to the fiducial coverage of
the muon system $|\eta|<2.0$. In the \ee\ channel, events are rejected
when both electrons are found in the EC calorimeter as this eliminates
only a small contribution to the signal which has poor signal to
background ratio. Additionally, in the \ee\ and \mm\ final states, the
dilepton invariant mass $M_{\ell_{1}\ell_{2}}$ is required to be
greater than 15\,GeV. These criteria define the ``preselection'' stage
of the analysis.

To correct for any possible mismodeling of the lepton reconstruction
and trigger efficiencies, and to reduce the impact of the luminosity
uncertainty, scale factors are applied to the MC samples at the
preselection stage to match the data.  The normalization factors are
determined from Drell-Yan dominated samples within a dilepton mass
window of $M_{\ell_{1}\ell_{2}}\in[80,100]$\,GeV for $ee$, $\mu\mu$
and $M_{\ell_{1}\ell_{2}}\in[57,75]$\,GeV for $e\mu$, and their
differences from unity are smaller than the luminosity
uncertainty. Figures \ref{fig:presel_emmu}--\ref{fig:presel_mumu} show
a comparison between data and the background prediction for the
distributions of the kinematic quantities for each of the dilepton
final states after preselection requirements. In the dilepton mass
distributions shown in Figs.~\ref{fig:presel_emem}a and
\ref{fig:presel_mumu}a, the peak in the signal expectation at $M_{Z}$
originates from $ZH$ associated production where the two observed
leptons are from the $Z$ boson decay. The differences in the widths of
the resonance in the \ee\ and \mm\ channels is due to detector
resolution. The transverse mass is defined as
\begin{displaymath} \mt(\ell,\etmiss) =
\sqrt{2\cdot\ptl\cdot\etmiss\cdot[1 - \cos\Delta\phi(\ell,\etmiss )]},
\end{displaymath} 
and consequently the minimal transverse mass, ${\mtmin}$, shown in
figs.~\ref{fig:presel_emem}d and ~\ref{fig:presel_mumu}d, is the
minimum of the two $\mt(\ell,\etmiss)$ defined for each lepton.

\begin{figure*}[!] 
  \begin{center}
    \begin{tabular}{cc}
      \includegraphics[width=1.0\columnwidth]{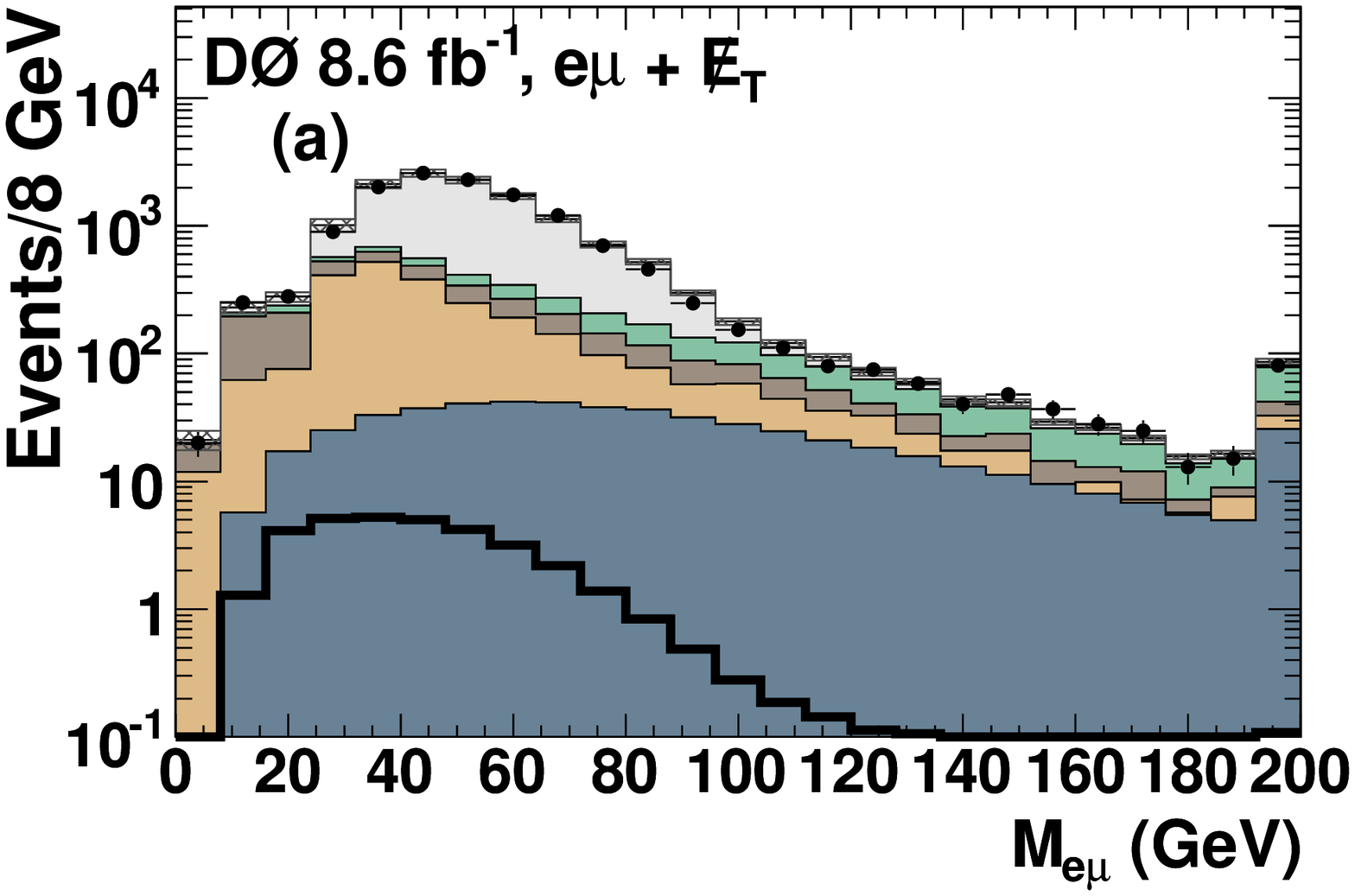} &
      \includegraphics[width=1.0\columnwidth]{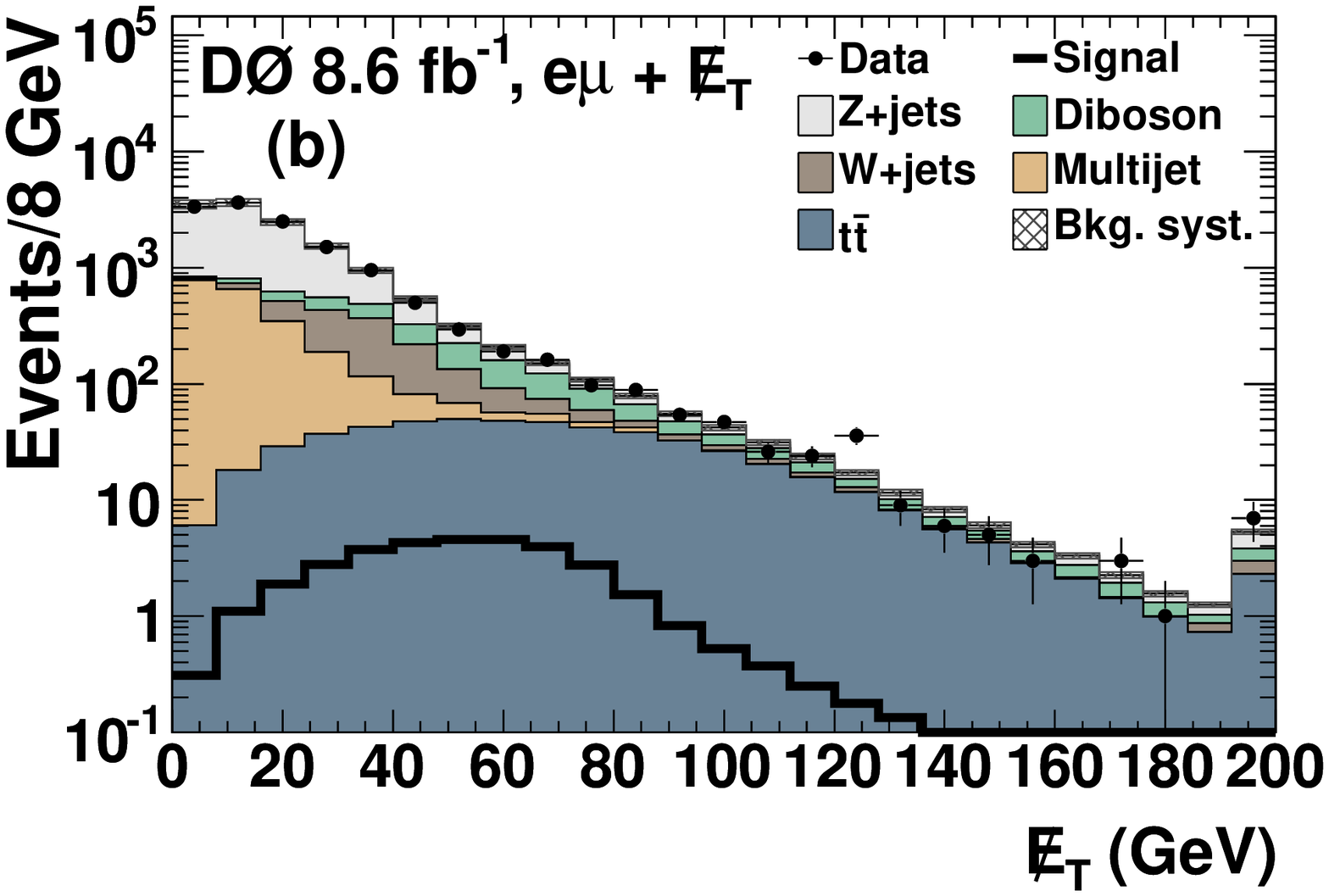} \\
      \includegraphics[width=1.0\columnwidth]{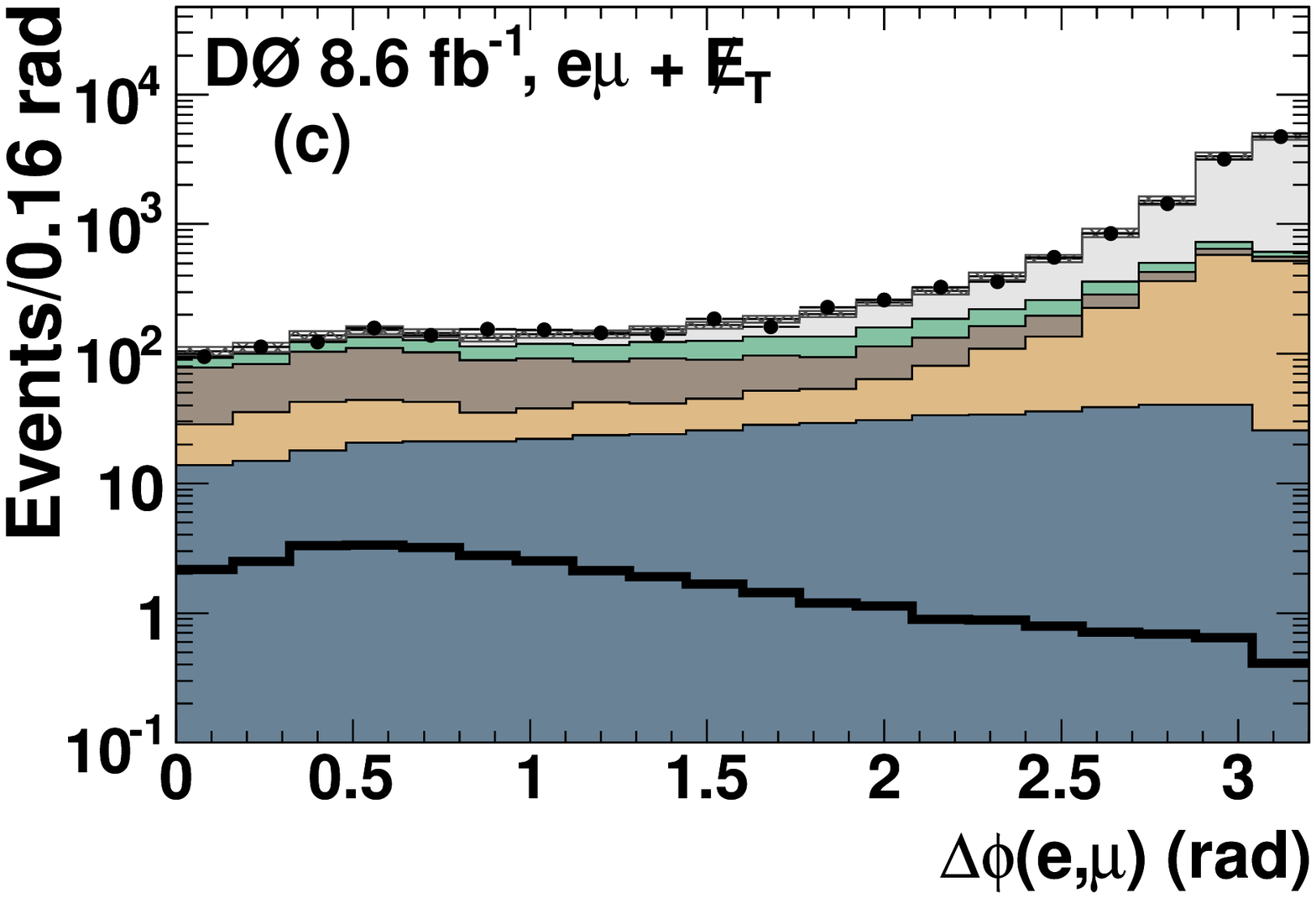} &
      \includegraphics[width=1.0\columnwidth]{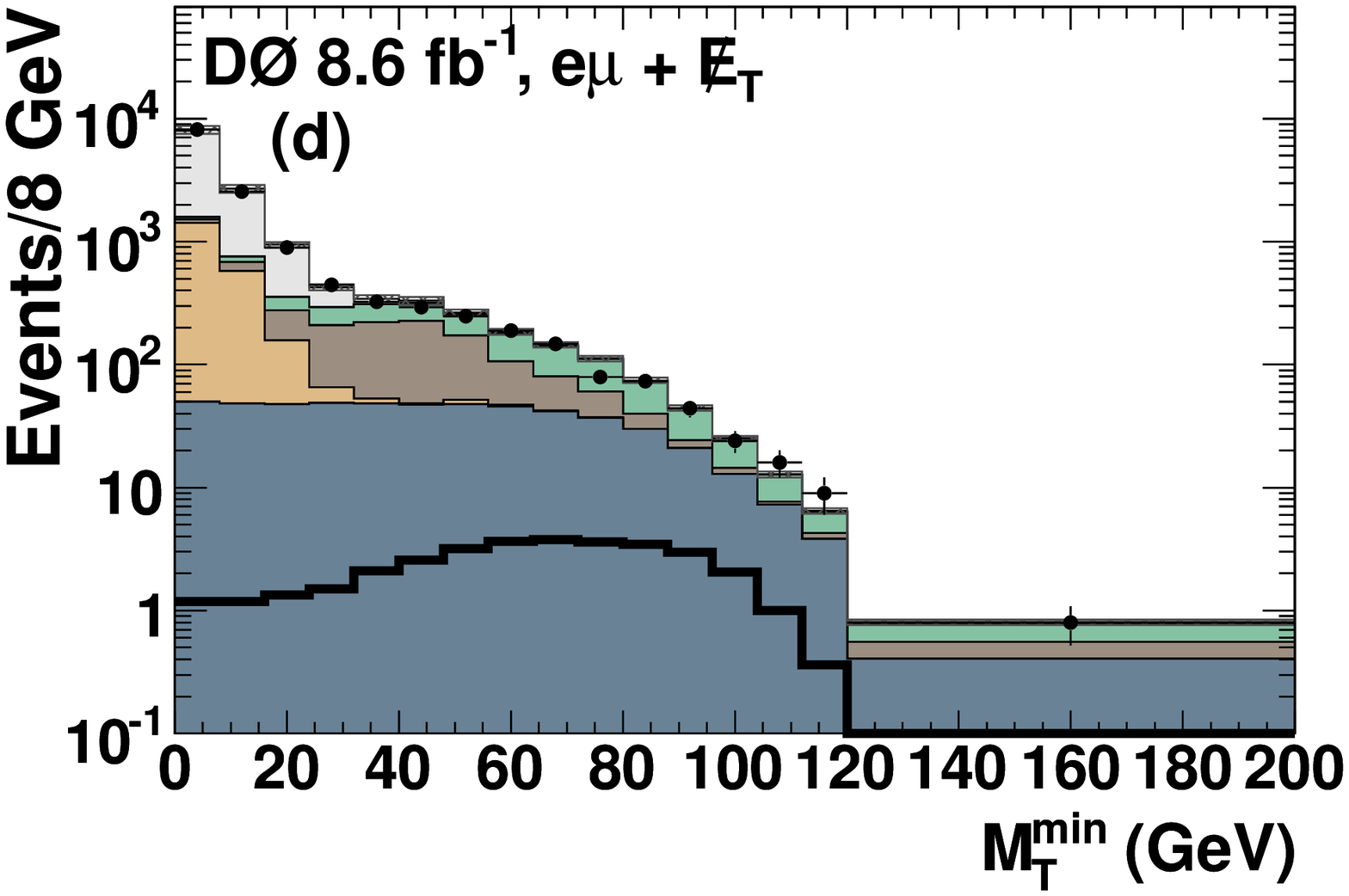} \\
    \end{tabular}
  \end{center}
  \caption{ [color online] The (a) dilepton invariant mass, (b)
    $\etmiss$, (c) $\Delta \phi$ between the leptons, and (d) minimum
    transverse mass for the \em\ channel at the preselection stage.
    The last bin also includes all events above the upper range of the
    histogram (a,b,d). The signal distribution shown corresponds to a
    Higgs boson mass of 165\,GeV. The hatched bands show the total
    systematic uncertainty on the background prediction.}
\label{fig:presel_emmu}
\end{figure*}

\begin{figure*}[!] 
  \begin{center}
    \begin{tabular}{cc}
      \includegraphics[width=1.0\columnwidth]{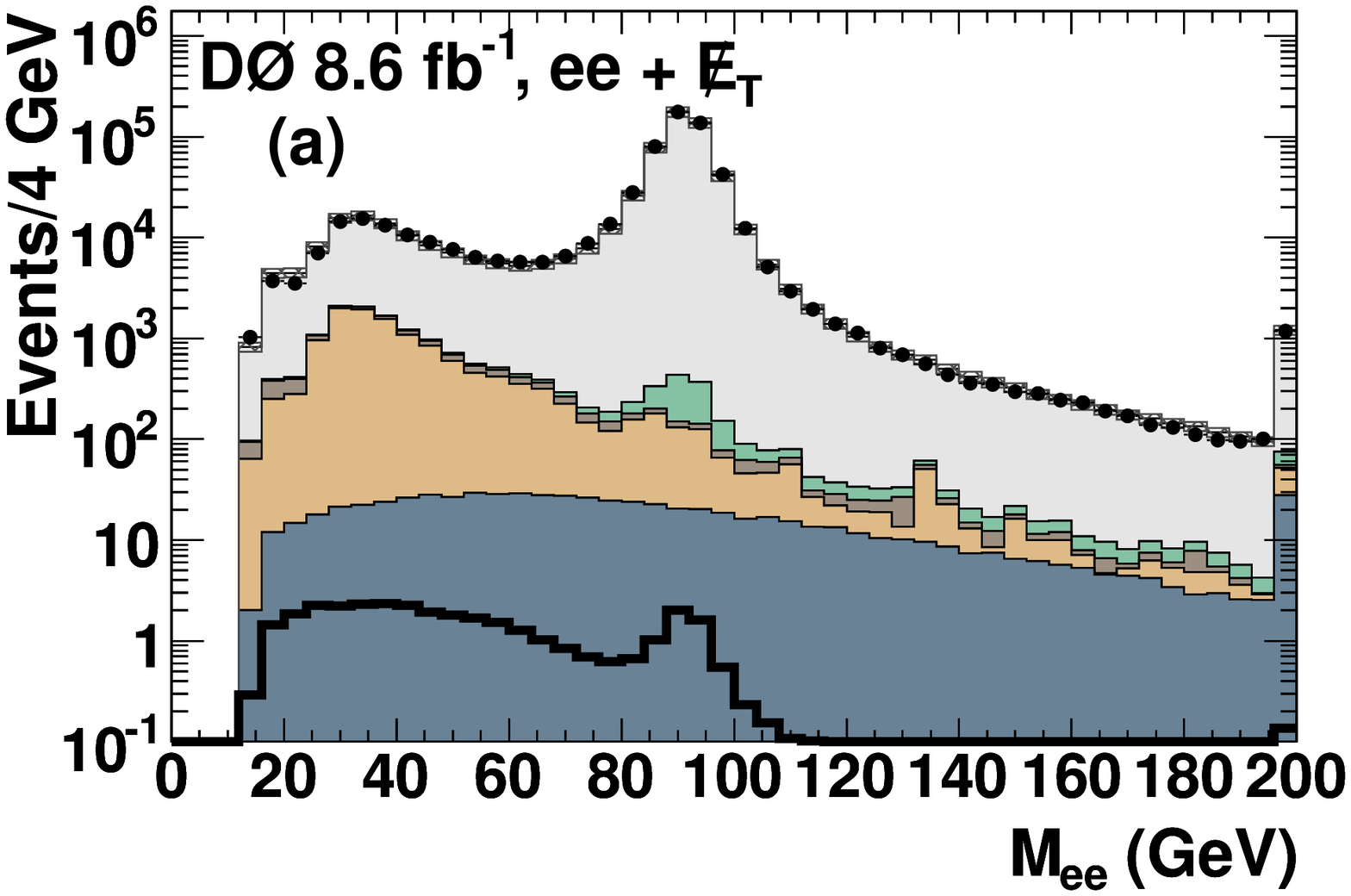} &
      \includegraphics[width=1.0\columnwidth]{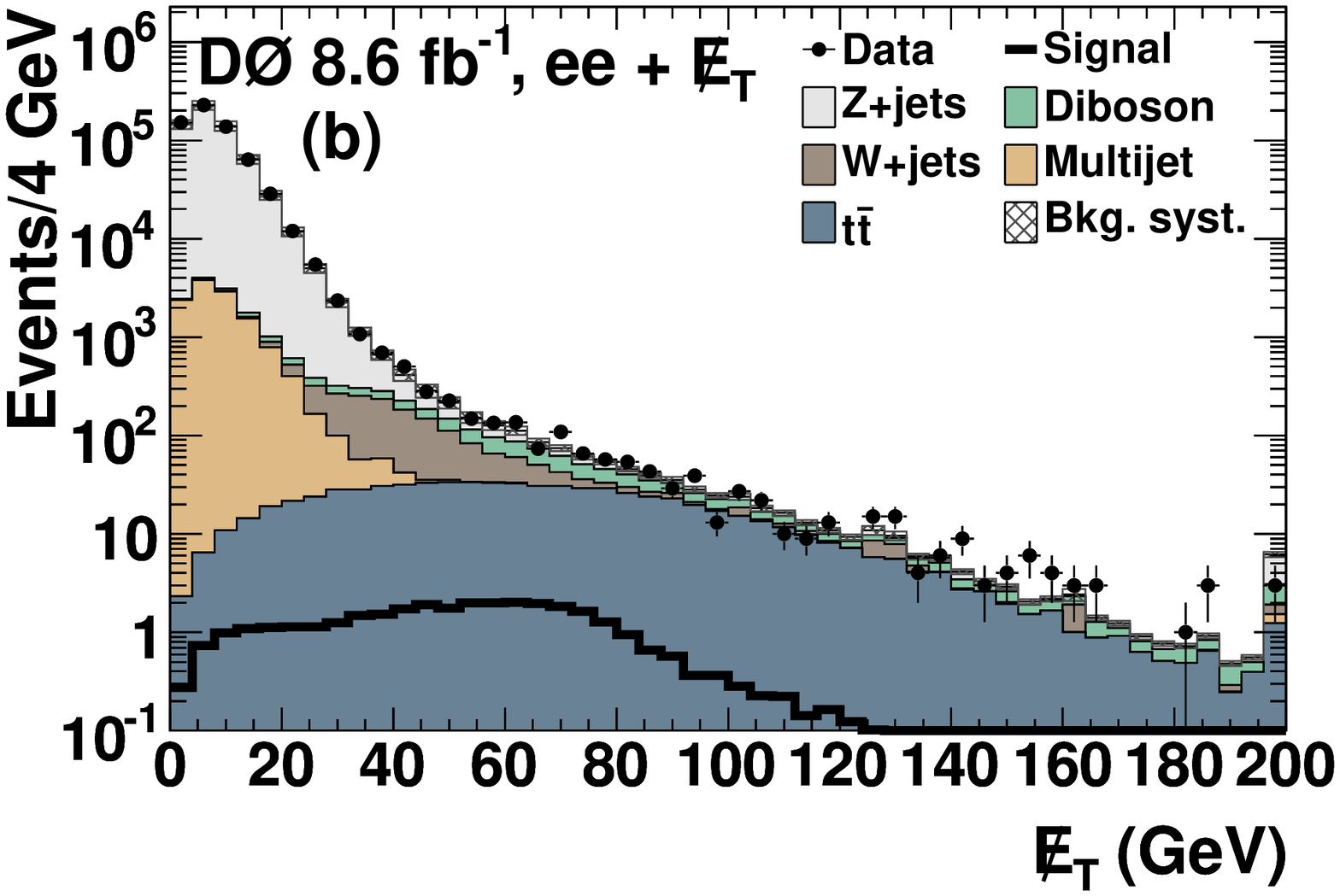} \\
      \includegraphics[width=1.0\columnwidth]{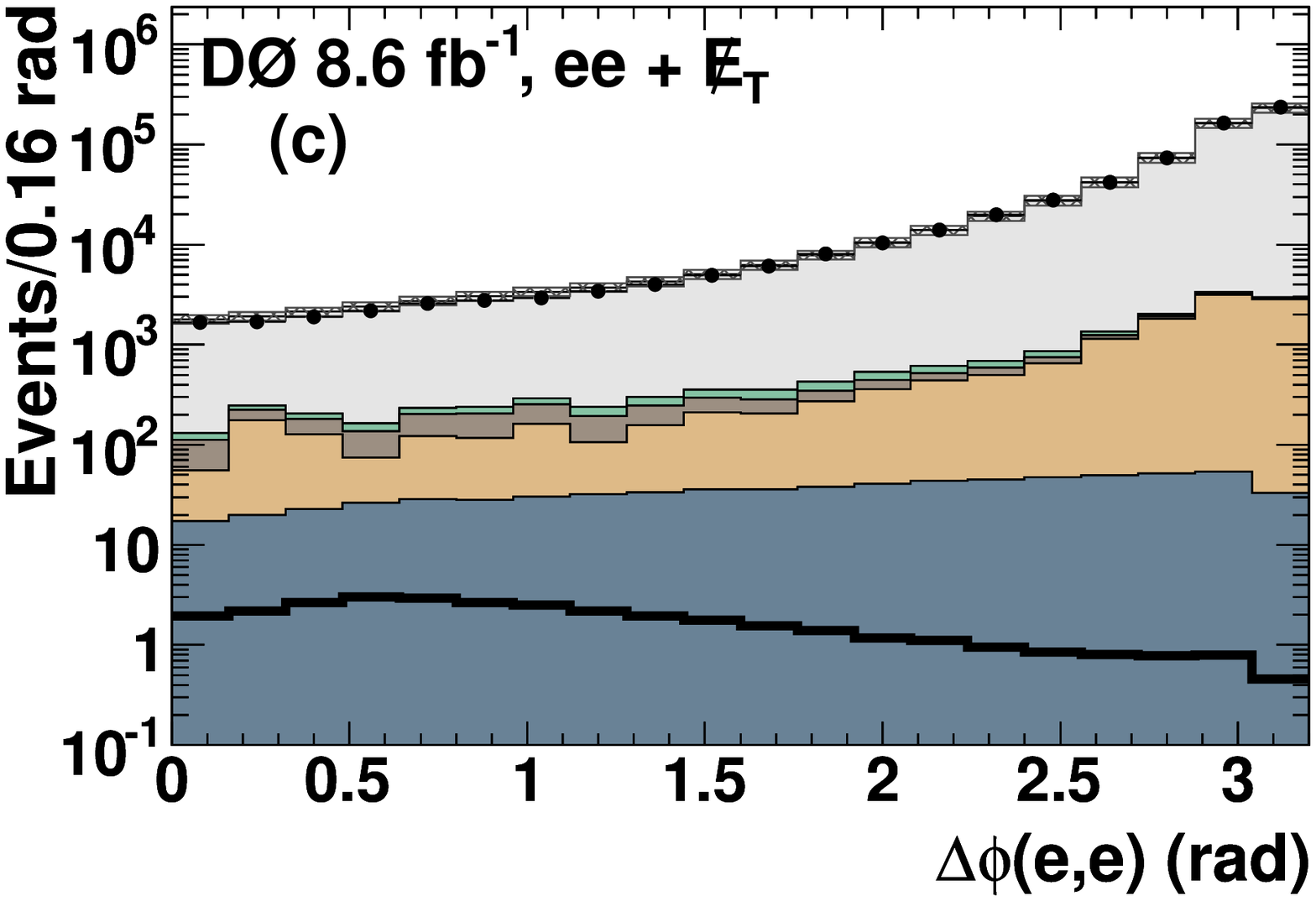} &
      \includegraphics[width=1.0\columnwidth]{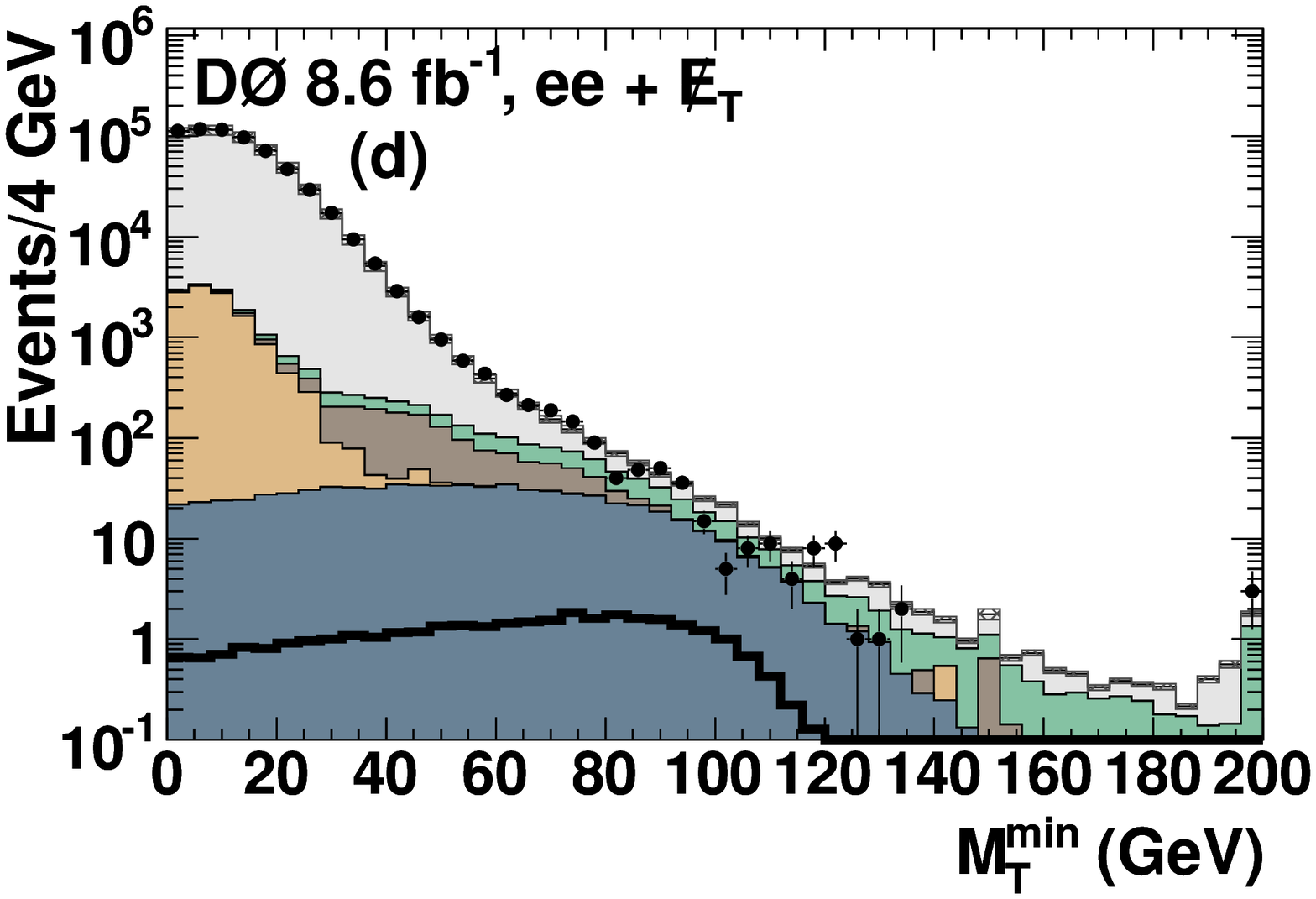} \\
    \end{tabular}
  \end{center}
  \caption{ [color online] The (a) dilepton mass, (b) $\etmiss$, (c)
    $\Delta \phi$ between the leptons, and (d) minimum transverse mass
    for the $ee$ channel at the preselection stage. The last bin also
    includes all events above the upper range of the histogram
    (a,b,d). The signal distribution shown corresponds to a Higgs
    boson mass of 165\,GeV. The hatched bands show the total
    systematic uncertainty on the background prediction.}
  \label{fig:presel_emem}
\end{figure*}

\begin{figure*}[!] 
  \begin{center}
    \begin{tabular}{cc}
      \includegraphics[width=1.0\columnwidth]{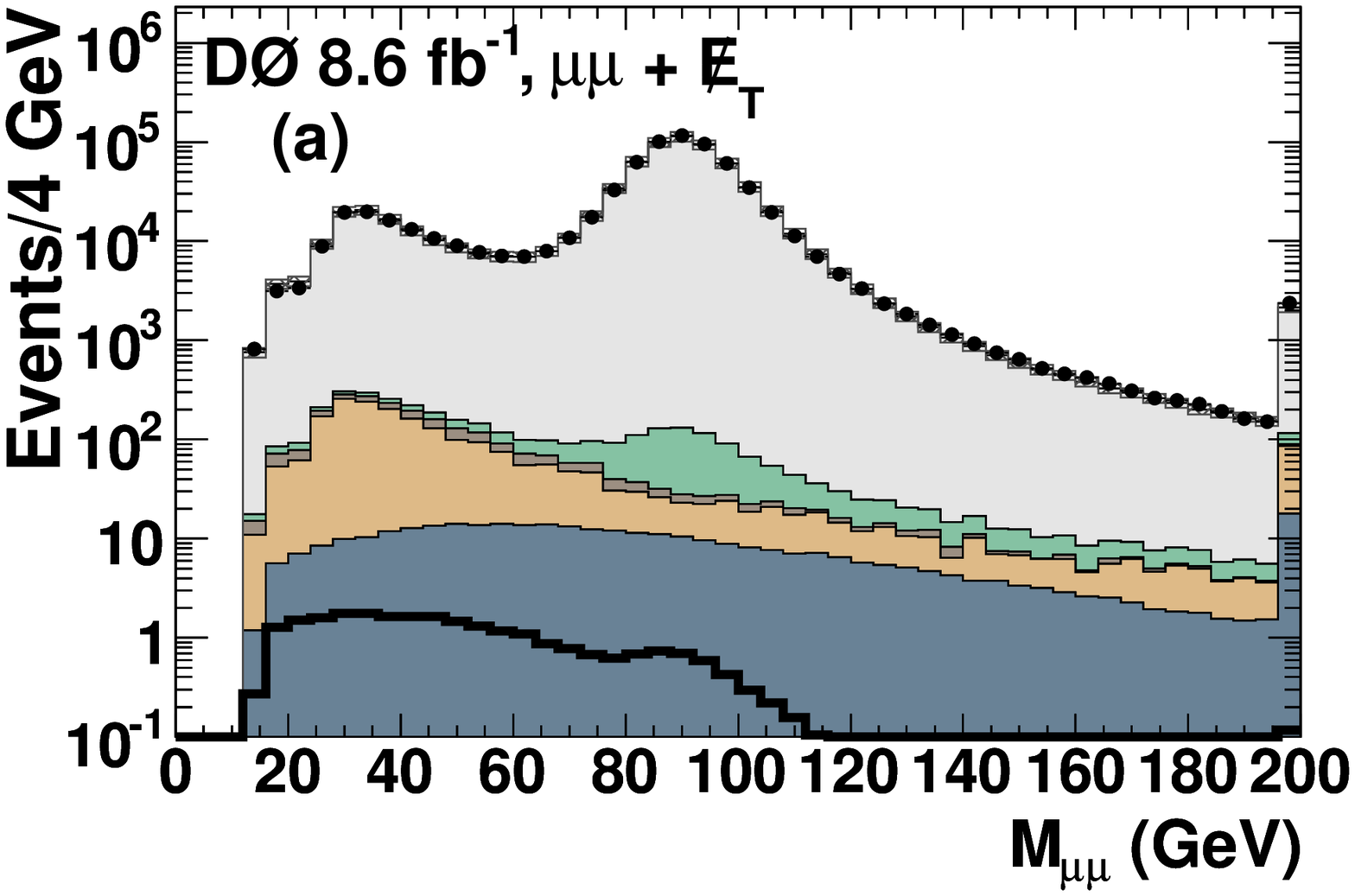} &
      \includegraphics[width=1.0\columnwidth]{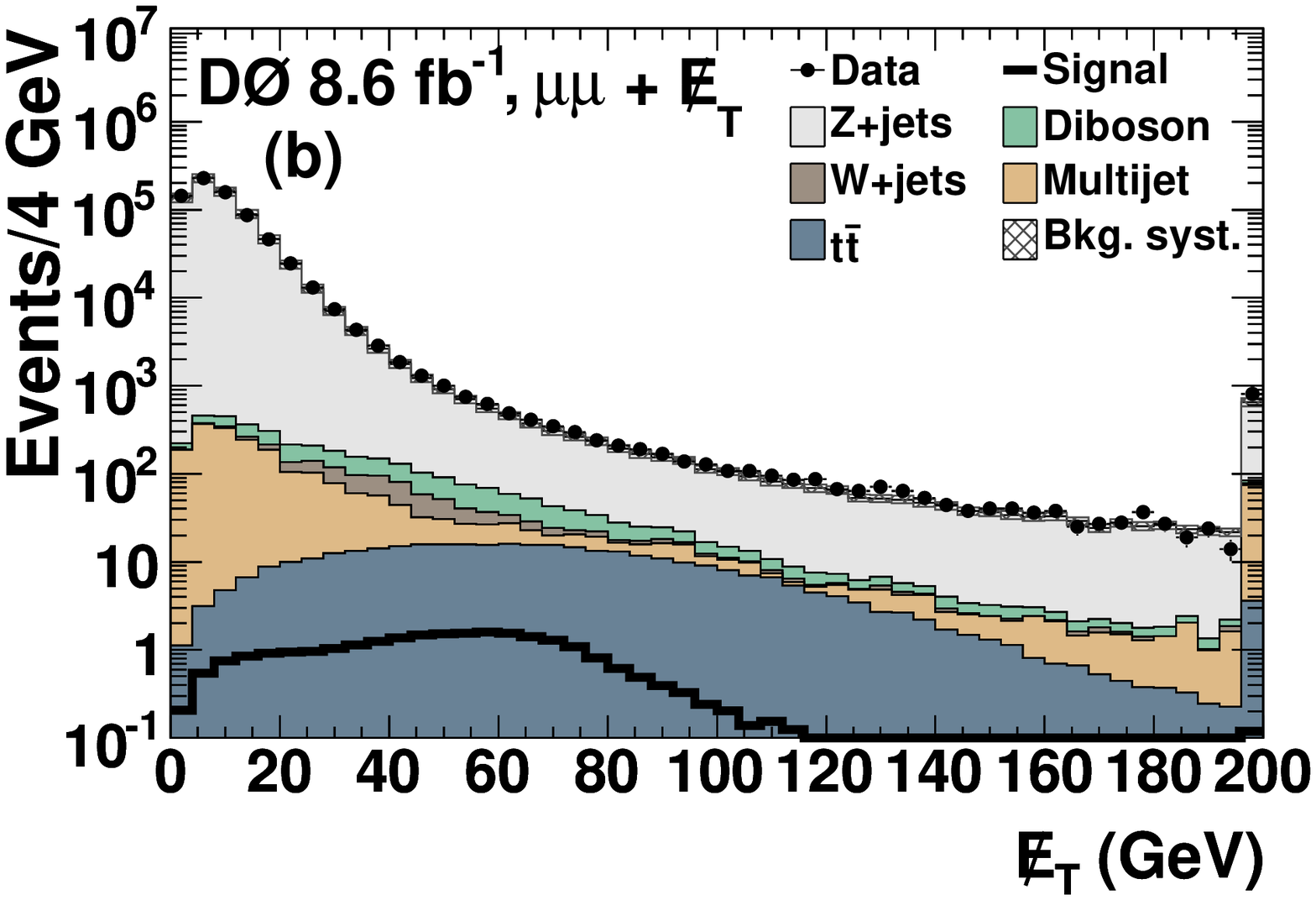} \\
      \includegraphics[width=1.0\columnwidth]{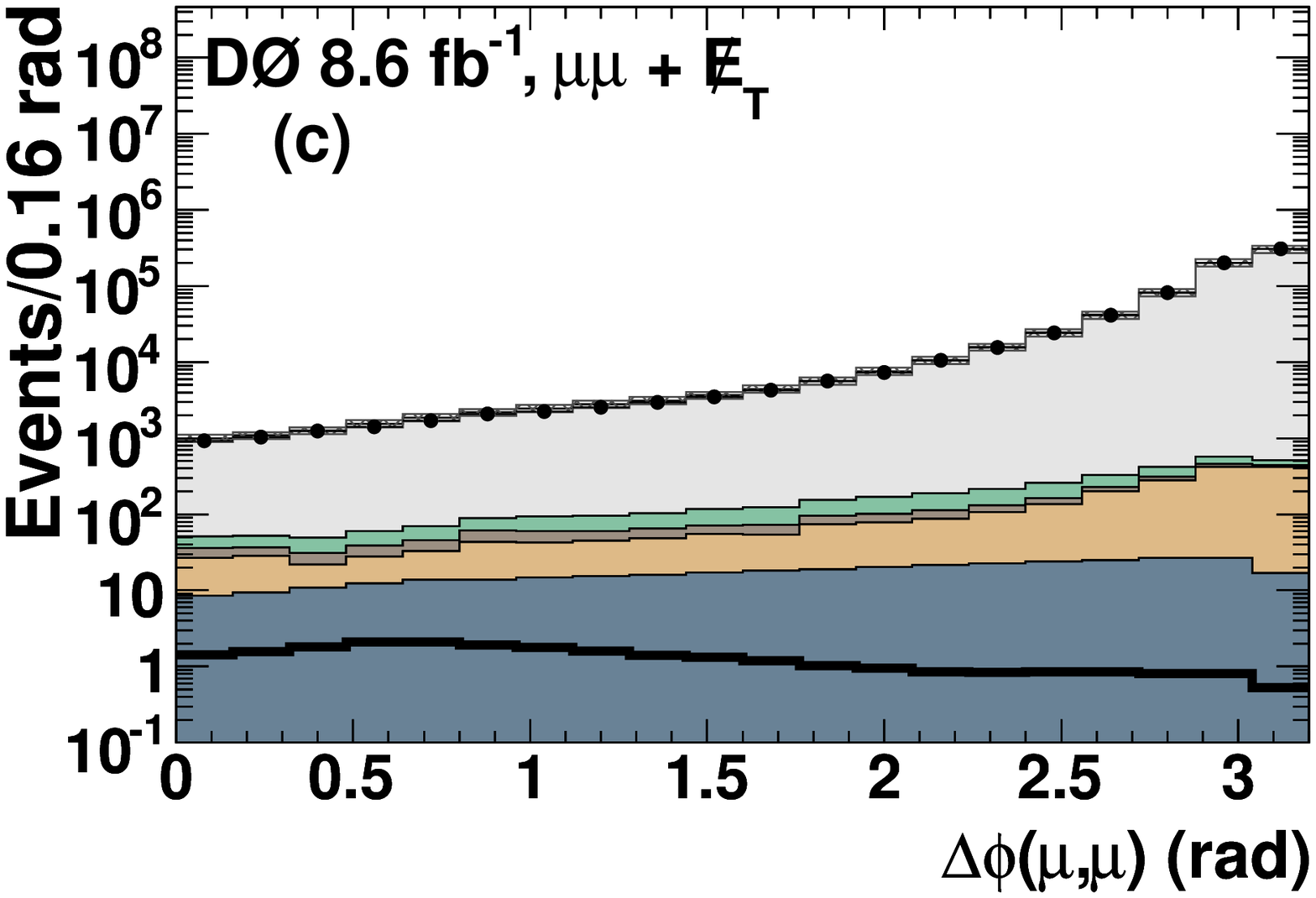} &
      \includegraphics[width=1.0\columnwidth]{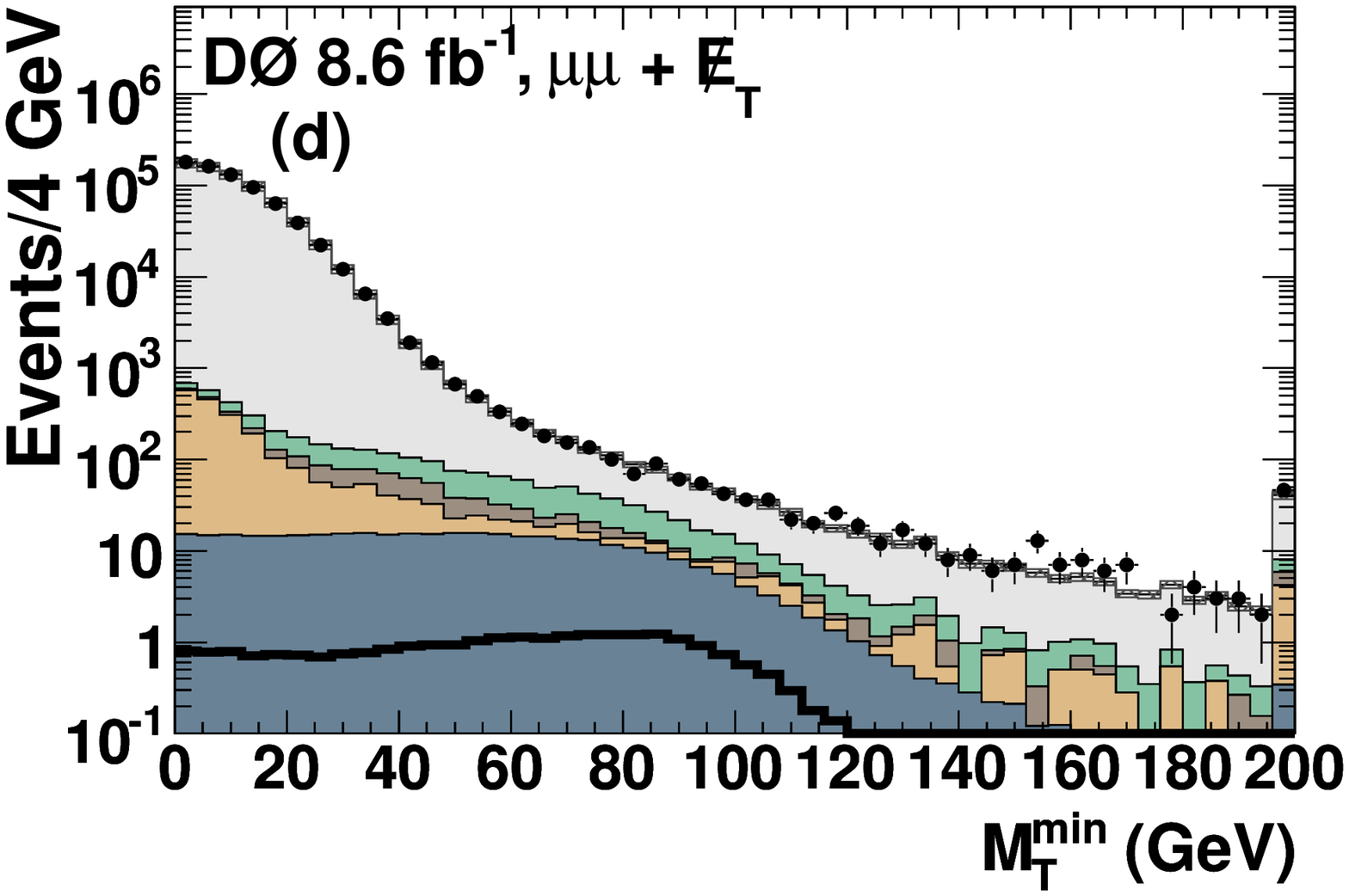} \\
    \end{tabular}
  \end{center}
  \caption{ [color online] The (a) dilepton mass, (b) $\etmiss$, (c)
    $\Delta \phi$ between the leptons, and (d) minimum transverse mass
    for the $\mu\mu$ channel at the preselection stage. The last bin
    also includes all events above the upper range of the histogram
    (a,b,d). The signal distribution shown corresponds to a Higgs
    boson mass of 165\,GeV. The hatched bands show the total
    systematic uncertainty on the background prediction.}
  \label{fig:presel_mumu}
\end{figure*}

Jets are considered in this analysis only if they have $p_{T} >$ 20
GeV and $|\eta| <$ 2.4.  The preselected samples are further
subdivided by the number of jets present in the event.  Dividing the
analysis into different jet multiplicity bins significantly increases
the sensitivity of this search as the signal and background
composition change between each sample.  In particular, $gg
\rightarrow H \rightarrow WW$ signal processes populate primarily the
0 and 1 jet multiplicity bins whereas contributions to higher
multiplicity bins arise mainly from vector boson fusion production and
associated $VH$ processes which contain additional jets in the event.
For the background, $WW$ diboson production tends to dominate lower
jet multiplicity bins while $t\bar{t}$ events generally contain two
jets that are often $b$-tagged.  Subsequent analysis steps are carried
out separately for events with zero jets, one jet, and two or more
jets in order to optimally separate signal from backgrounds, resulting
in a total of nine analysis channels ({\it i.e.}, three dilepton final
states with three jet multiplicity bins each). The jet multiplicity
spectrum of the simulated $Z/\gamma^*$ sample is corrected to match
that of the data for each channel considered. These corrections are
derived within the mass windows as described above and have the
primary effect of improving the \textsc{alpgen} modeling of
$Z/\gamma^*+$jets.

The number of events for each jet multiplicity bin at preselection can
be found in Table \ref{tab:presel_cutflow}.  In general, good
agreement between data and the expected background contribution is
observed. At this stage, the $Z/\gamma^*$ contribution is the dominant
background source.

\begin{table*}[!]
  \caption{Expected and observed numbers of events at preselection in
   the \em, \ee, and \mm\ final states.  The signal is for a Higgs
   boson mass of 165\,GeV.}
 \label{tab:presel_cutflow}
 \begin{ruledtabular}
    \begin{tabular}{ccccccccc}
      & Data & Total background & Signal & $Z/\gamma^*$ & $\ttbar$ & $W+\gamma/$jets & Dibosons & Multijet \\
      \hline 
      \hline
      $e\mu$: & 13468& 13754 & 35 & 9275 & 541 & 1066 & 842 & 2031 \\
      0 jets  &10942 & 11171 & 20 & 8023 & 16 & 861 & 677 & 1594 \\   
      1 jet  & 1849 & 1902 & 10 &  1088  & 157 &  154 & 142 &  362  \\   
      $\ge 2$ jets & 677 & 681 & 5 & 164 & 368 &  51 &  23 &  75 \\
      \hline
      $ee$:       & 525942 & 524204 & 18 & 513365 & 244 & 1091 & 730   & 8776\\
      0 jets      & 473311 & 472195 &  9 & 463751 & 9   & 840    & 425   & 7171\\
      1 jet       &  42480 &  41795 &  5 &  40234 & 64  & 175  & 151   & 1172 \\
      $\ge 2$ jets &  10151 &  10214 &  4 &   9380 & 171 &  76  & 154   &  433   \\
      \hline      
      $\mu\mu$:   & 724131  & 727456  & 26 & 723726  & 353 & 397 & 1107 & 1872 \\
      0 jets      & 624062  & 626473  & 13 & 624116  &   10 & 316 &  594 & 1437 \\
      1 jet       &  85349  &  85856  & 7  &  85069  &  90 &  68 &  280 &  348 \\
      $\ge 2$ jets &  14720  &  15127  & 6  &  14541  & 253 &  13 &  233 &   87 \\
    \end{tabular}
    \end{ruledtabular}
\end{table*}

\subsection{Final Selection}\label{sec:finalselection}

In the \ee\ and \mm\ channels, a multivariate discriminant is used to
remove the dominant $Z/\gamma^*$ background present in the preselected
data sample. The complete details are discussed later in this Article.

As the $Z/\gamma^*$ contribution is smaller in the $e\mu$ channel,
kinematic selections are instead applied to suppress backgrounds after
preselection. For the signal, the \etmiss~is not aligned with any of
the leptons in the final state, while for the $Z/\gamma^*$ background
processes, the \etmiss~is mostly caused by inaccurate measurements of
the energies of the leptons and tends to point in the direction of one
of the two leptons. Observables that take into account both the
absolute value and the direction of the $\overrightarrow\etmiss$ are
$\mtmin$ and $M_{T2}$, where $M_{T2}$ is an extension of the
transverse mass for final states with two visible and two invisible
particles~\cite{bib:mt2}. It is obtained as the minimum of the
$\mtmin$ between either lepton and neutrino pair using a minimization
procedure, where the sum of the momenta of the neutrinos is varied
under the constraint that the sum of the momenta of the lepton pair is
the missing transverse energy in the event. The distributions of these
two observables in the \em\ channel after the preselection are shown
in Fig.\ \ref{fig:minMt_mt2_emmu} for each jet multiplicity bin.  The
requirements $\mtmin>20$\,GeV and $M_{T2}>15$\,GeV define the final
selection for this channel.  The number of events at this selection
stage for the \em\ state can be found in Table
\ref{tab:final_cutflow}.

\begin{figure*}[!]
  \begin{center}
    \begin{tabular}{cc}
      \includegraphics[width=1.0\columnwidth]{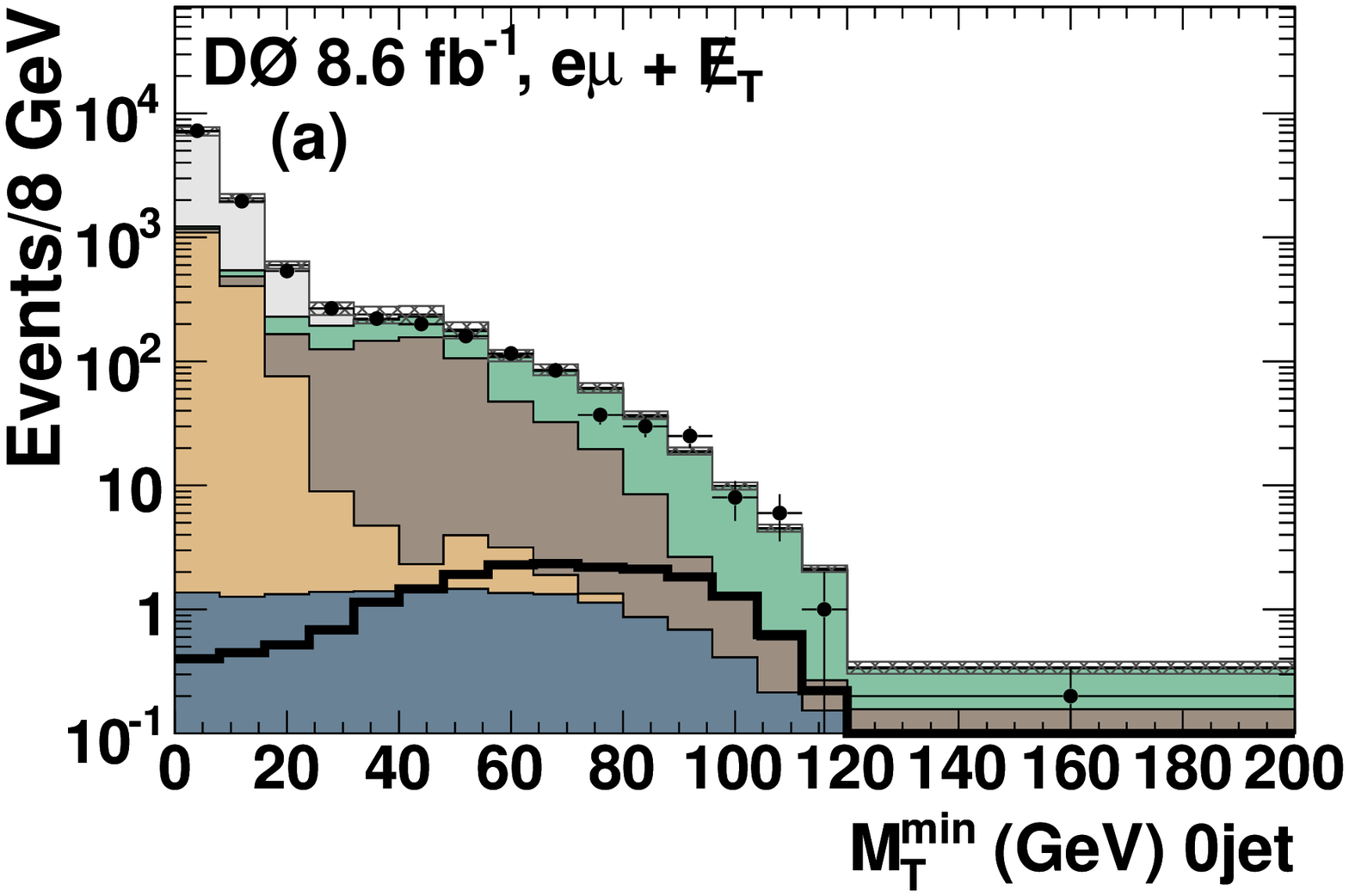} &
      \includegraphics[width=1.0\columnwidth]{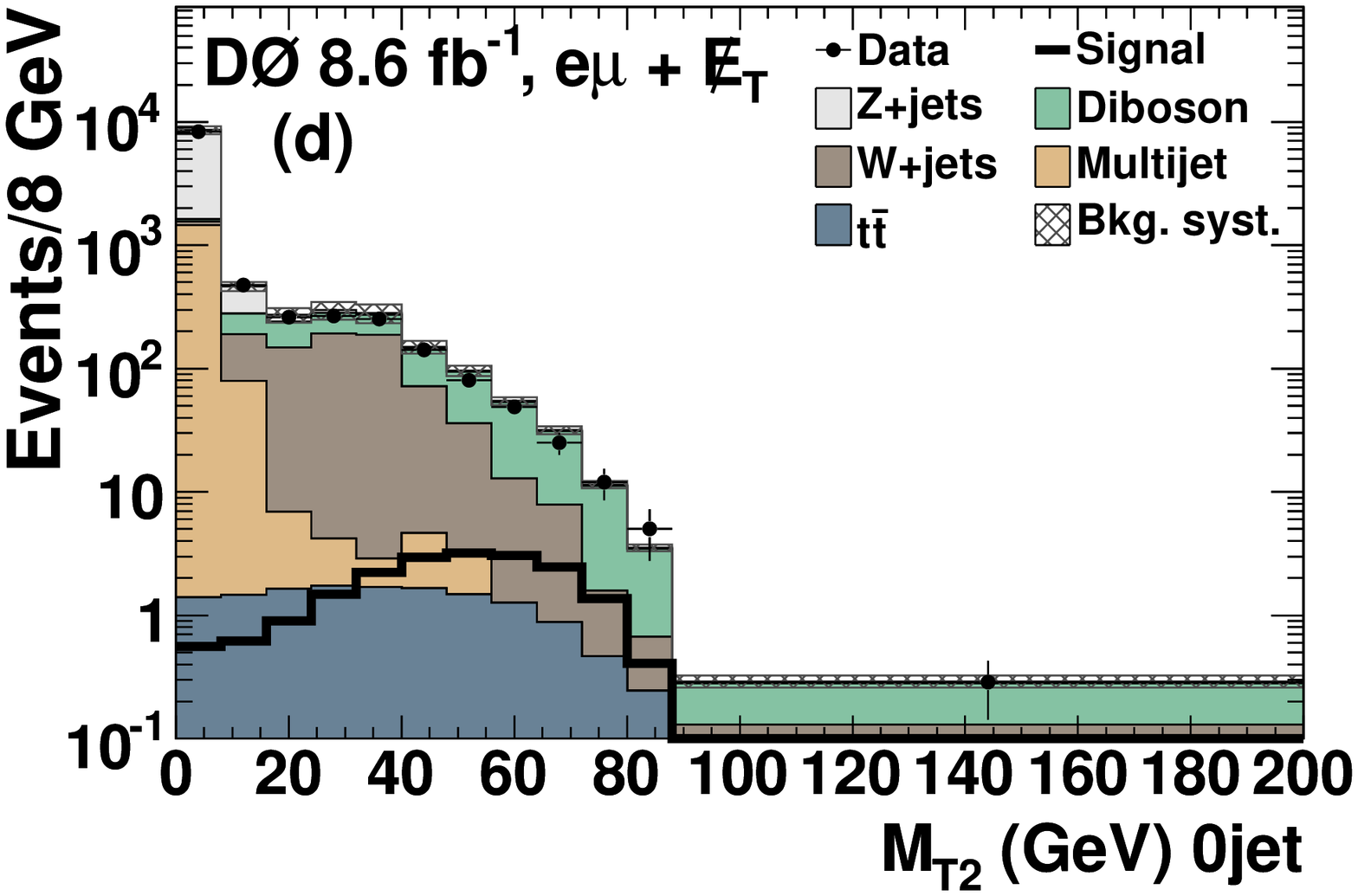} \\
      \includegraphics[width=1.0\columnwidth]{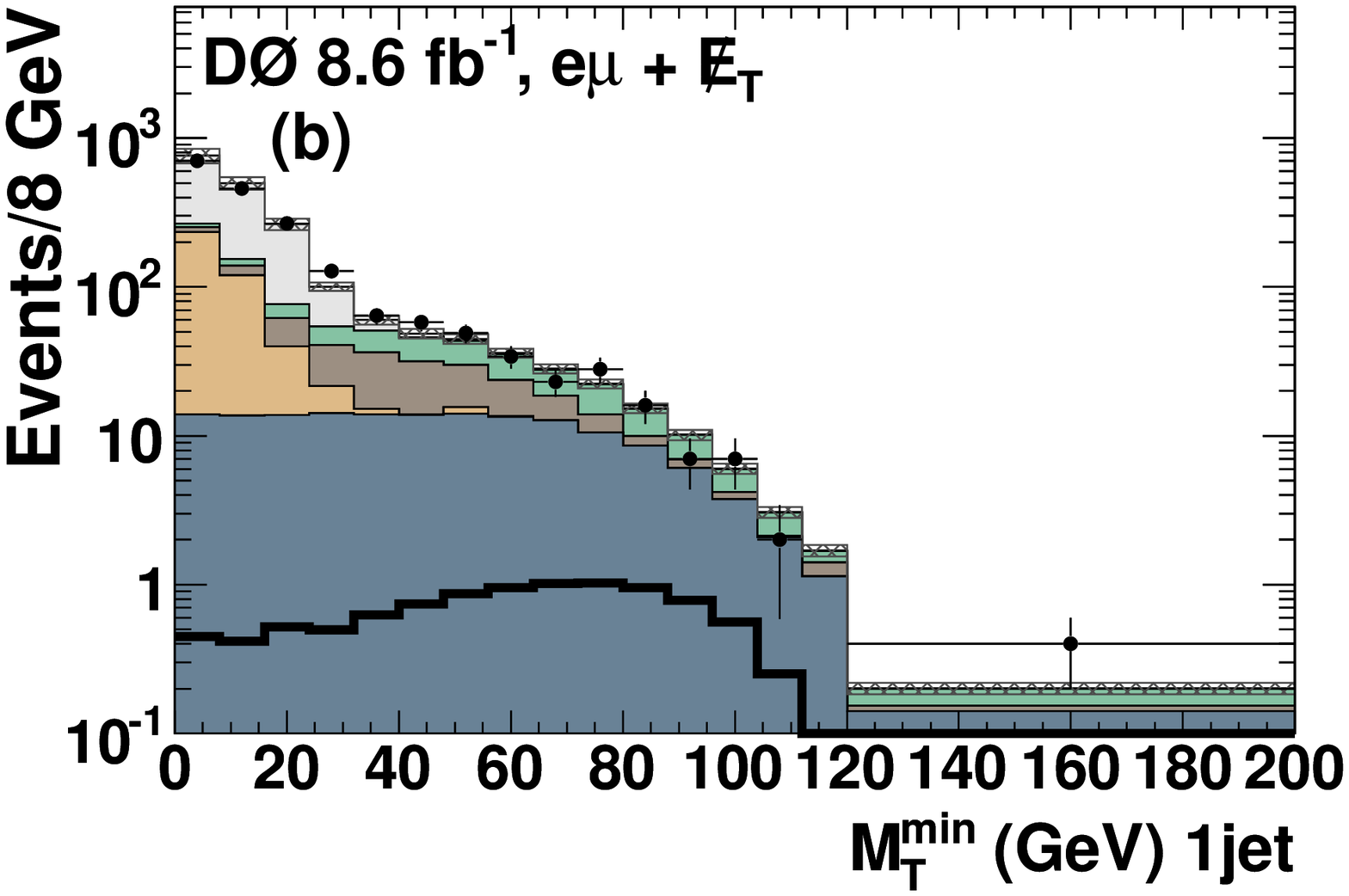} &
      \includegraphics[width=1.0\columnwidth]{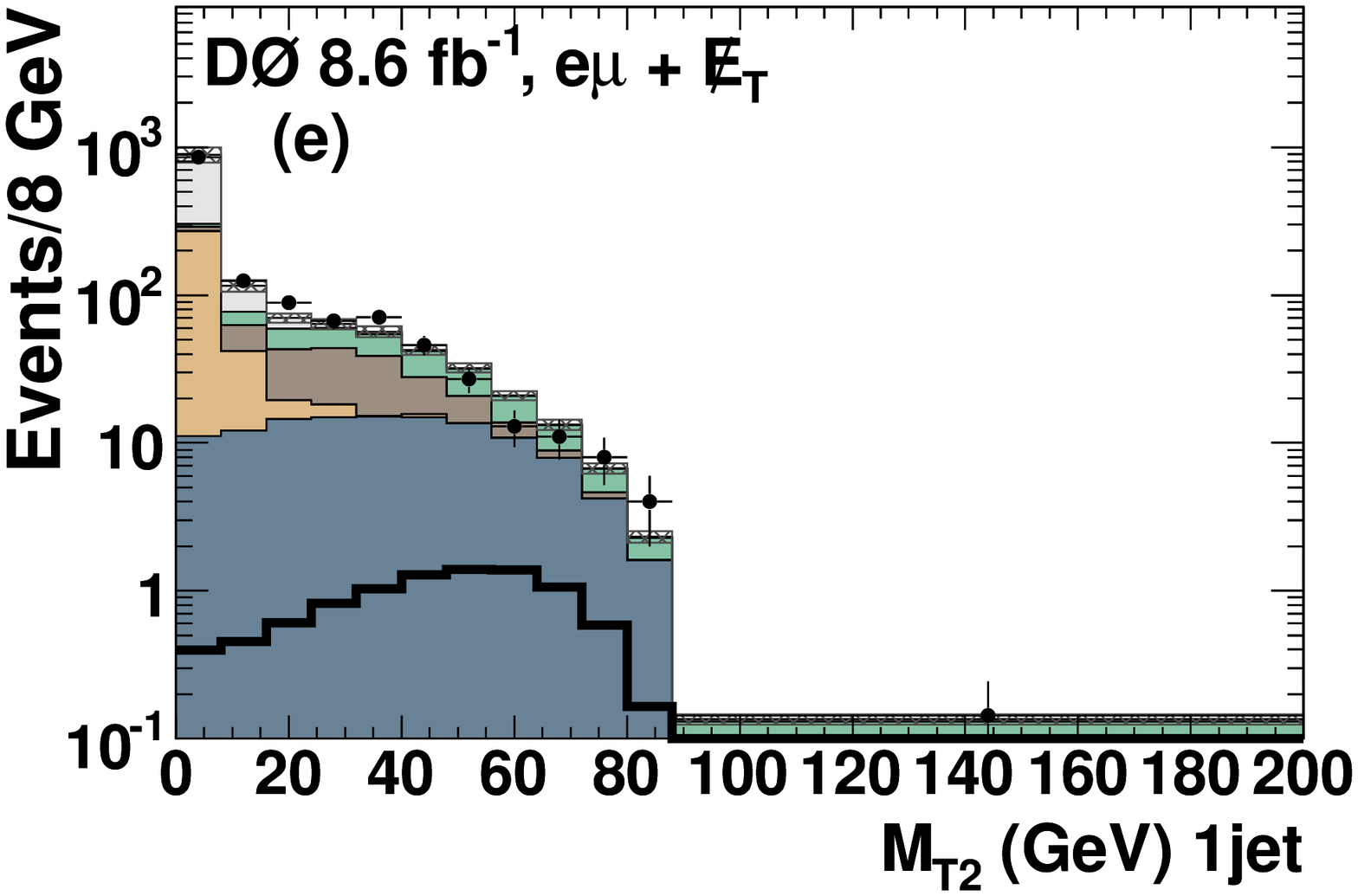} \\
      \includegraphics[width=1.0\columnwidth]{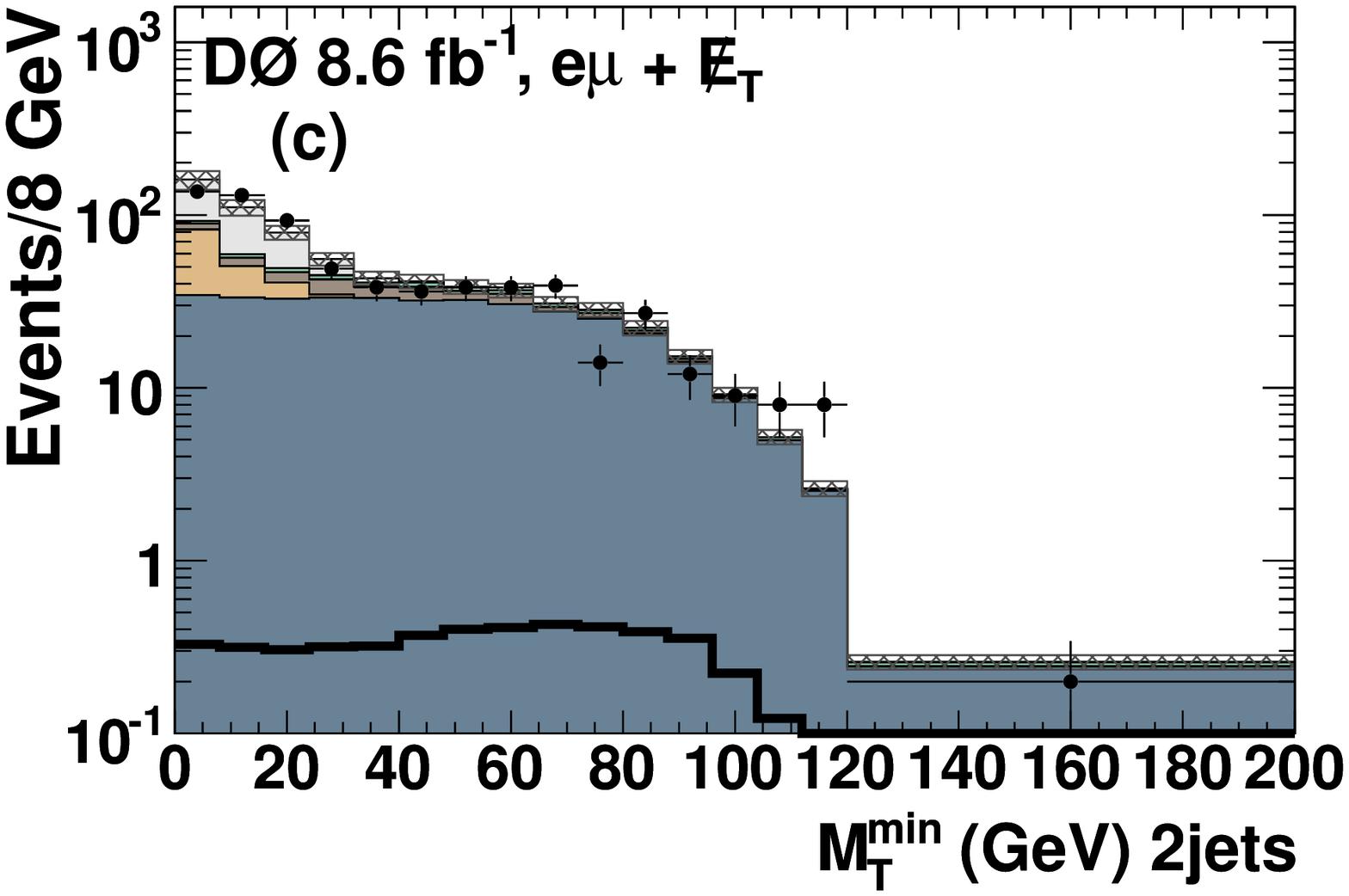} &
      \includegraphics[width=1.0\columnwidth]{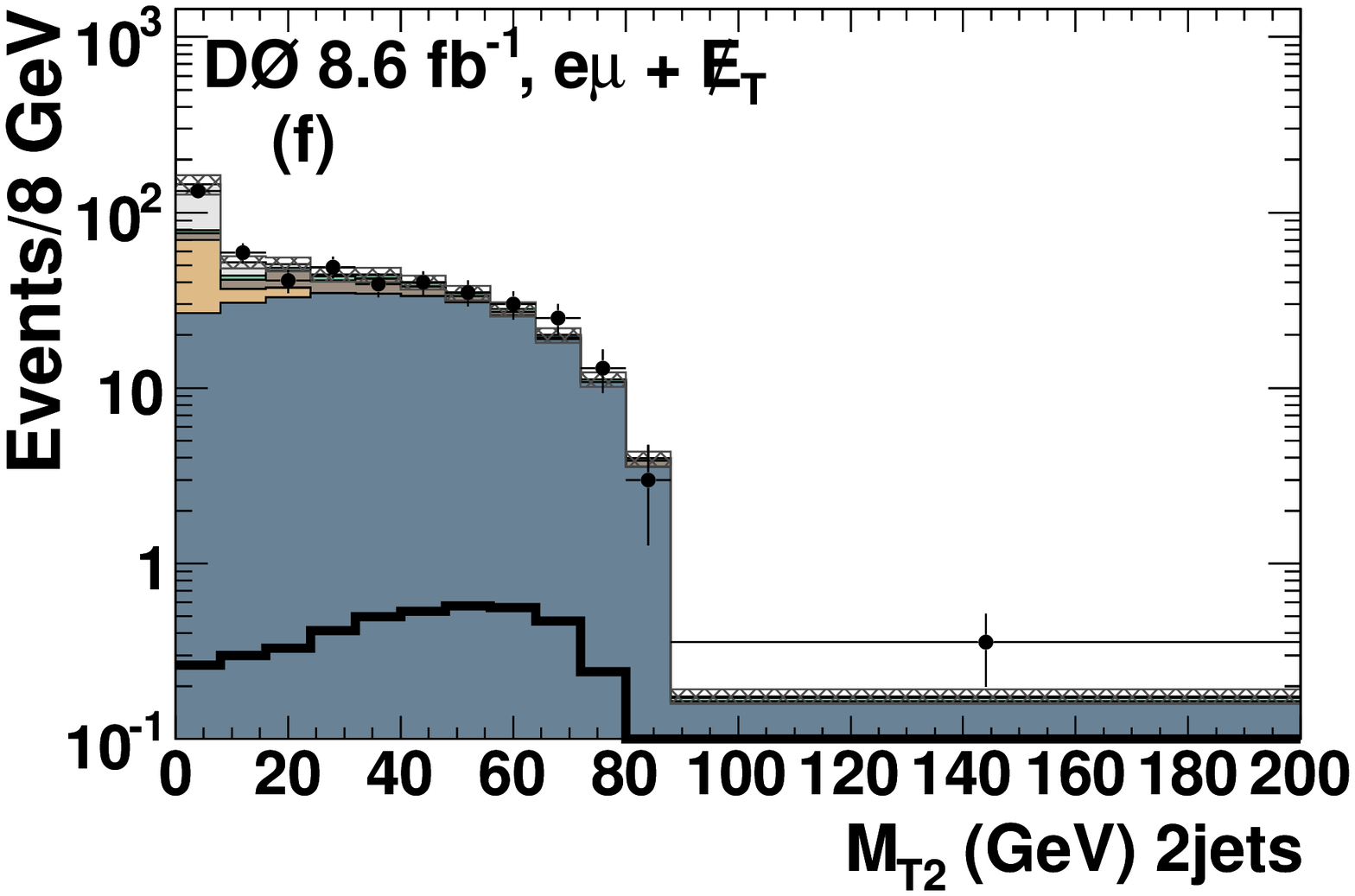} \\
    \end{tabular}
  \end{center}
  \caption{[color online] \mtmin\ distribution for the $e\mu$ channel
    in the (a) 0-jet bin, (b) 1-jet bin, and (c) $\ge 2$-jet
    bin. $M_{T2}$ distribution for the $e\mu$ channel in the (d) 0-jet
    bin, (e) 1-jet bin, and (f)$\ge 2$-jet bin.  The last bin also
    includes all events above the upper range of the histogram. The
    signal distribution shown corresponds to a Higgs boson mass of
    165\,GeV. The hatched bands show the total systematic uncertainty
    on the background prediction\label{fig:minMt_mt2_emmu} }
\end{figure*}

\section{\label{sec:backgrounds}INSTRUMENTAL BACKGROUNDS}

The main instrumental background processes for this analysis are due
to (1) the mismeasurement of $\etmiss$ in $Z/\gamma^*\mathrm{+jets}$
events, (2) the misidentification of associated jets or photons in
$W+\gamma/$jets production as leptons, and (3) the misidentification
of jets in multijet production as leptons.

\subsection{\boldmath{$Z$} and \boldmath{$W$} Boson Production }\label{sec:vjets}

Background contributions from $Z$ bosons are estimated using MC
simulations. The mismeasurement of $\etmiss$ in
$Z/\gamma^*\mathrm{+jets}$ events adds a significant source of
background particularly for the \ee\ and \mm\ selections, as shown in
Figs.~\ref{fig:presel_emem}b and \ref{fig:presel_mumu}b.



A $W$ boson decaying leptonically and associated with one or more jets
or a photon may contribute to the background if a jet is misidentified
as a lepton or a photon overlaps an isolated track or converts into an
electron-positron pair. The contribution from these backgrounds is
estimated using MC simulations, and corrections to the contributions
of jets and photons misidentified as electrons are derived using data,
as explained below.


An enriched sample of $W+\gamma/$jets not overlapping with the signal
is selected from events passing all the selection criteria except that
the charges of the two leptons are required to be identical. This
requirement assumes that the probability of misidentifying a lepton as
a jet is independent of the lepton charge, and therefore, the
like-charge dilepton sample can be used to estimate background
corrections from misidentified leptons in the opposite-charge dilepton
sample.  Corrections are obtained separately for initial state
radiation jets and photons (ISR$_{\gamma/j}$) and for final state
radiation photons (FSR$_\gamma$) by splitting this control sample into
high dilepton invariant mass ($M_{\ell_{1}\ell_{2}}>40$\,GeV) and low
dilepton invariant mass ($M_{\ell_{1}\ell_{2}}<20$\,GeV) samples where
the contributions of ISR$_{\gamma/j}$ and FSR$_\gamma$ are,
respectively, dominant. These corrections are applied in the \ee\ and
\em\ final states, whereas they are not required in the \mm\ final
state due to the smaller $W+\gamma/$jets contribution.



\subsection{ Multijet Production}

A high statistics sample of predominantly multijet events, where jets
are misidentified as leptons, is obtained from data by inverting
certain lepton selection criteria.  All other preselection criteria
are applied in order to model the kinematic distributions of the
multijet background in the signal region.  In the \mm\ channel, the
opposite-charge requirement for muons is reversed and a correction for
the presence of non-multijet events in the like-charge sample,
estimated from simulation, is applied.  For the \em\ and \ee~channels,
the eight-variable electron likelihood selection is reversed, and to
normalize the multijet sample to the actual contribution in the signal
region, the multijet sample is compared to events which pass all the
signal selections except that a like-charge requirement is
imposed. This method accounts for any kinematic bias introduced from
reversing the electron likelihood requirement. Since the probability
of a jet being misidentified as a lepton $(P_{lj})$ is independent of
charge, assuming that there is no correlation between the charges of
the two misidentified leptons in multijet events, the like-charge
sample has exactly the same normalization and kinematics as the actual
multijet contribution. $P_{lj}$ depends on the jet multiplicity, and
therefore the multijet background is estimated separately for each jet
multiplicity bin.  The analysis further assumes contributions of
non-multijet processes are negligible in the reversed lepton quality
sample.




\section{\label{sec:tmva}MULTIVARIATE ANALYSIS}
A multivariate technique is used to characterize events as originating
from a Higgs boson signal or from background processes and to achieve
maximum separation between them.  A random forest of boosted decision
trees (BDTs)~\cite{bib:bdt} is used to construct a discriminant from
kinematic variables, taking into account their correlations.  The
decision trees are trained separately in each of the nine analysis
channels and for each Higgs boson mass hypothesis.  To increase the
statistics of the available simulated signal events, signal samples
for neighboring mass hypotheses are used for the training of the
multivariate discriminant. For example, the training of the
discriminant for the 165~GeV mass hypothesis uses signal samples
corresponding to a Higgs boson mass of 160, 165, and 170~GeV.

\subsection{Multivariate Discriminant against \boldmath {$Z/\gamma^*$}}
A BDT discriminant is used in the \ee\ and \mm\ final states to reject
the large $Z/\gamma^*$ background while retaining a high signal
efficiency. This random forest of BDTs will be referred to as
DY-BDT. The DY-BDT is trained for each Higgs boson mass hypothesis and
jet multiplicity bin, separately for the \ee\ and \mm\ final states,
to differentiate between the $Z/\gamma^*$ background and all
considered SM Higgs boson signal events.

The following input variables are used for the DY-BDT:
\begin{enumerate}[(i)]
\item lepton $\pt$
\item invariant mass of the leptons, $M_{\ell_{1}\ell_{2}}$
\item azimuthal opening angle between the two leptons, $\Delta \phi(\ell_{1},\ell_{2})$
\item separation in $\eta$, $\phi$ space between the two leptons,
  $\Delta R(\ell_{1},\ell_{2}) = \sqrt {{(\eta_{\ell_{1}}-\eta_{\ell_{2}})^2}+{(\phi_{\ell_{1}}-\phi_{\ell_{2}})^2}}$
\item minimal transverse mass, $\mtmin$
\item extended transverse mass, $M_{T2}$
\item missing transverse energy, \etmiss 
\item smallest and largest of the azimuthal angles, $\Delta\phi$ between the $\etmiss$ and either lepton
\item transverse mass of the $\etmiss$ and the dilepton pair, $M_T(\ell_{1}\ell_{2},\etmiss)$
\item special missing transverse energy, $\Eslash_T^{\text{special}}$,
  defined for object $\zeta$, which corresponds to either the nearest
  lepton or jet in the event relative to the direction of the $\overrightarrow\etmiss$:
  \[ \Eslash_T^{\text{special}} = \left\{ \begin{array}{ll}
    \mbox{$\etmiss$} ,& \mbox{if $\Delta\phi(\etmiss, \zeta) > \pi/2$} \\
    \mbox{$\etmiss\times\sin[\Delta\phi(\etmiss, \zeta)]$}, &  \mbox{otherwise} \\
  \end{array}
  \right. \]
\item jet $\pt$
\item scaled missing transverse energy defined as
  \begin{displaymath}
    \Eslash_T^{\text{scaled}}
    = \frac{\etmiss}{\sqrt{\sum_{\rm jets}\left[\Delta E^{\rm jet}\cdot\sin\theta^{\rm jet}\cdot\cos\Delta\phi\left({\rm jet},
          \etmiss\right)\right]^{2}}}, 
  \end{displaymath}
  
  where $\Delta E^{\rm jet}$ is a measure of jet energy resolution and
  is proportional to $\sqrt{E^{\rm jet}}$; the fluctuation in the
  measurement of jet energy in the transverse plane can be approximated
  by the quantity $\Delta E^{\rm jet}\cdot \sin\theta^{\rm jet}$~\cite{bib:hww}
\item azimuthal angle between the $\etmiss$ and the jets,
  $\Delta\phi(\etmiss, \text{\rm jet})$
\item absolute value of the pseudorapidity difference between the
  jets, $|\Delta\eta(j_1, j_2)|$, where $j_1$ and $j_2$ are the two
  highest-$p_{T}$ jets in the event
\item invariant mass of the two jets, $M(j_1, j_2)$.
\end{enumerate}

Variables (i) and (ii) exploit the di-lepton kinematics of the
event. Variables\,(iii) and\,(iv) are related to the opening angle
between the two leptons and provide discrimination against SM
backgrounds which tend to exhibit back-to-back topologies. This is not
the case for Higgs boson decays because of the spin correlation in the
scalar decay where leptons tend to be aligned in the same direction.

The $\etmiss$-related variables\,(v)--(ix) help distinguish genuine
\etmiss\ in the Higgs boson signal from mismeasured \etmiss\ in
$Z/\gamma^*$ events. Variable\,(x) helps to further suppress
$Z/\gamma^*$ events, which populate lower values of
$\Eslash_T^{\text{special}}$ where a mismeasured lepton or jet tends
to align with the $\overrightarrow\etmiss$
direction~\cite{bib:cdf-hww}. Variables\,(xi)--(xv) are used in the
1-jet and 2-jet bins, as appropriate. Since the events are categorized
in terms of jet multiplicities, variables (xii)-(xv) exploit the jet
kinematics in the event.

To reject most of the $Z/\gamma^*$ background after the preselection,
events are required to appear in the signal-like region of the DY-BDT
discriminant. This defines the final selection of the $ee$ and
$\mu\mu$ final states. The threshold varies for each Higgs boson mass
hypothesis in each jet multiplicity bin and yields a $Z/\gamma^*$
rejection factor of $\mathcal{O}(10^{-5})$, $\mathcal{O}(10^{-3})$,
$\mathcal{O}(10^{-2})$ for the 0-jet, 1-jet, 2-jet bins, respectively
for all dilepton channels and Higgs boson masses. The thresholds are
chosen to obtain similar rejection factors of background events as the
cut-based analysis employed in the previous
publication~\cite{bib:hww}. The DY-BDT discriminants for a Higgs boson
mass of 165~GeV are shown in Fig.\ \ref{fig:dy_output}.  This figure
demonstrates that a good separation is achieved between the
$Z/\gamma^*$ background and the majority of signal. However it can be
noticed that some signal events cannot be distinguished from the
background and have a very low DY-BDT discriminant value. This is
primarily due to some of the Higgs decay modes which have a signature
similar to $Z/\gamma^*$ background. The numbers of events at the final
selection stage for the \ee\ and \mm\ final states are shown in
Table~\ref{tab:final_cutflow}.

\begin{figure*}[!]
  \begin{center}
    \begin{tabular}{cc}
      \includegraphics[width=1.0\columnwidth]{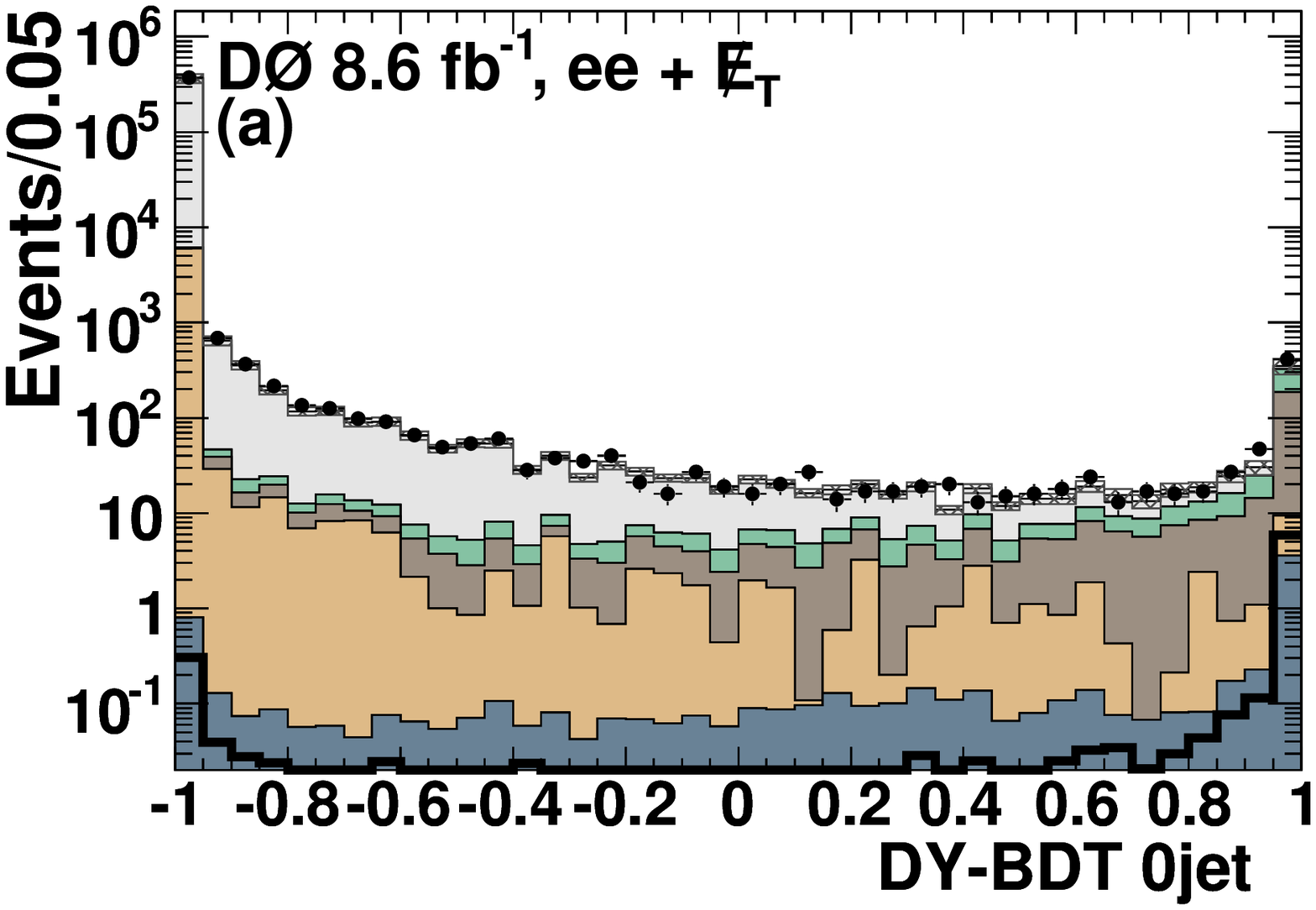} &
      \includegraphics[width=1.0\columnwidth]{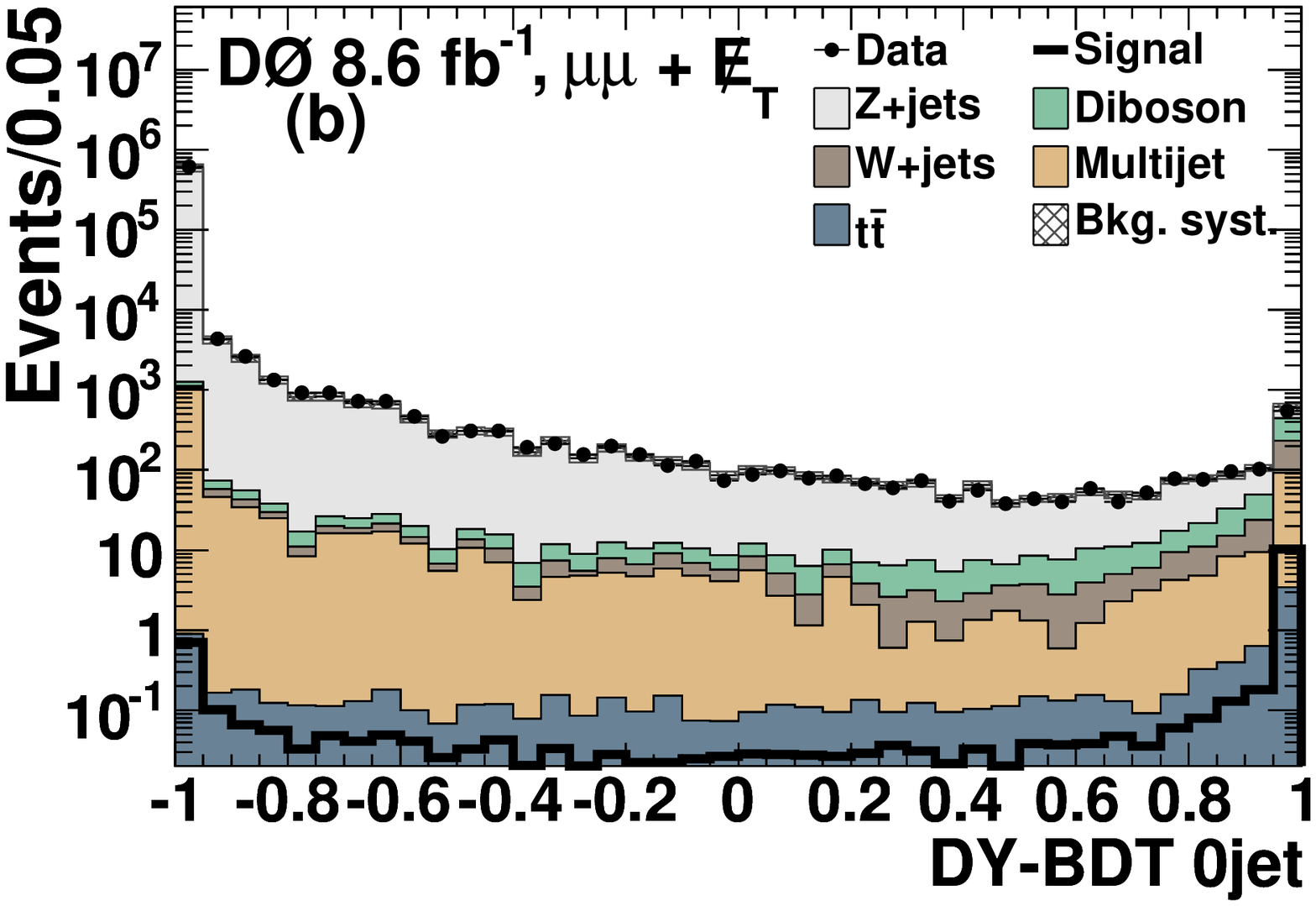} \\
      \includegraphics[width=1.0\columnwidth]{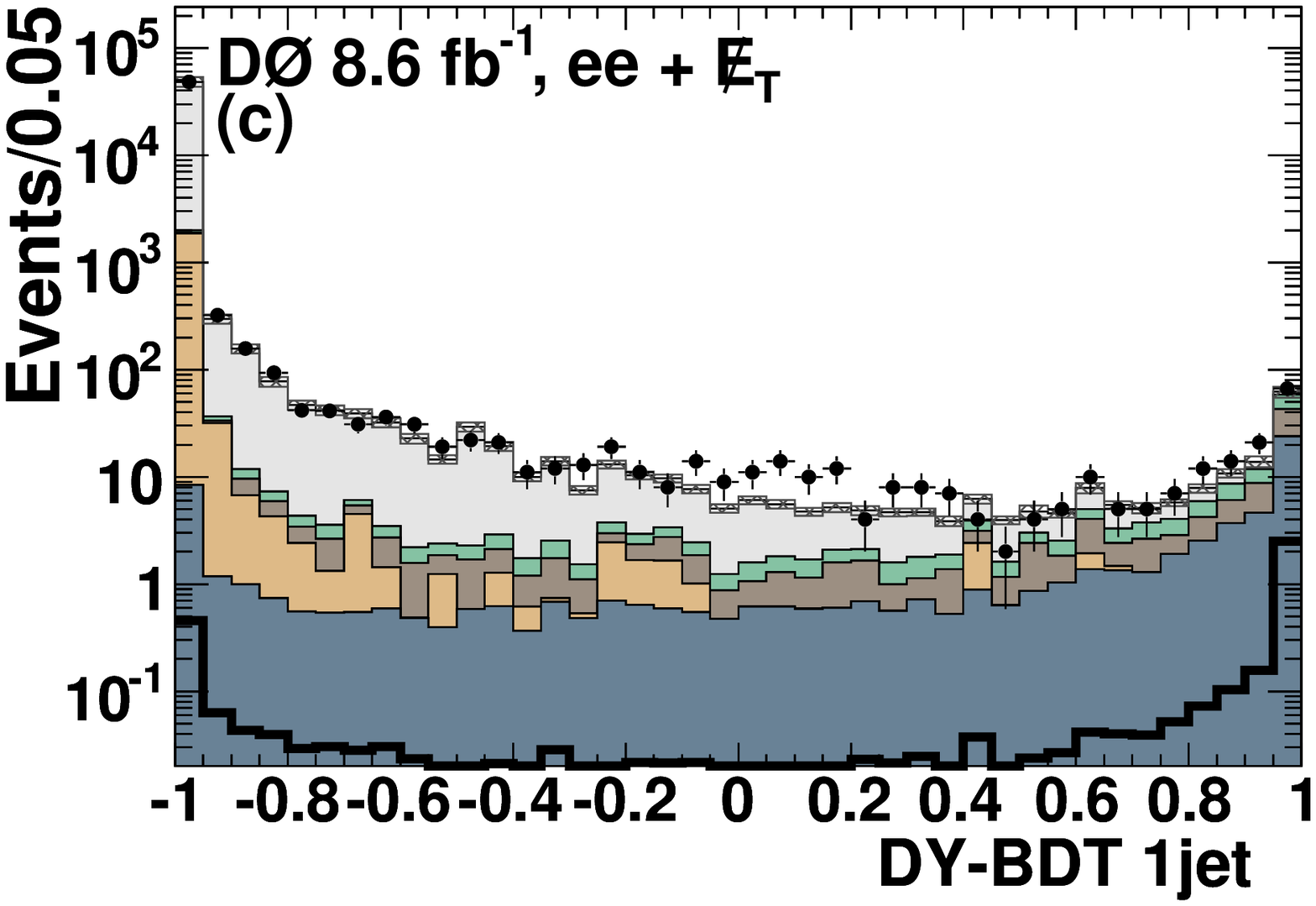} &
      \includegraphics[width=1.0\columnwidth]{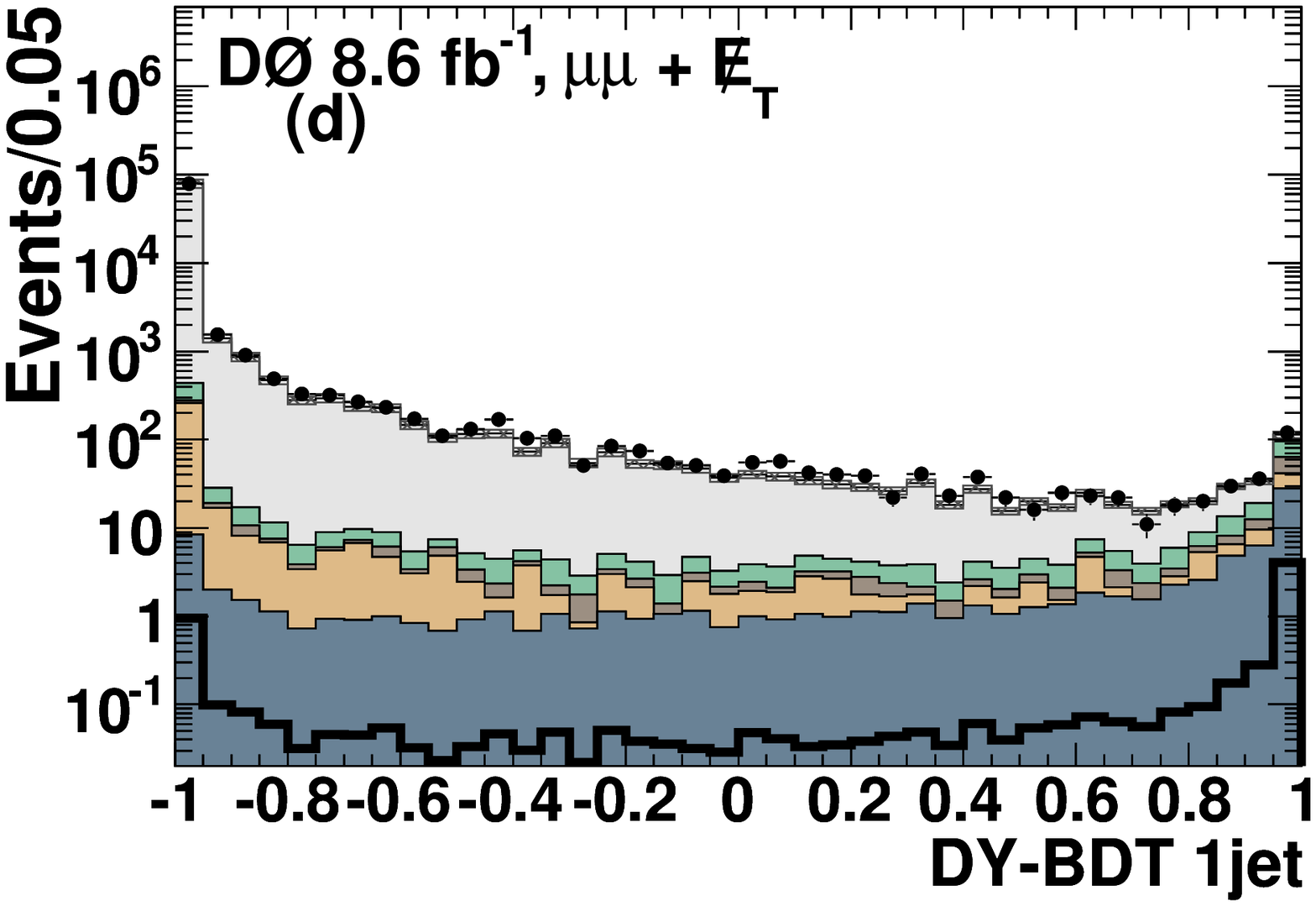} \\
      \includegraphics[width=1.0\columnwidth]{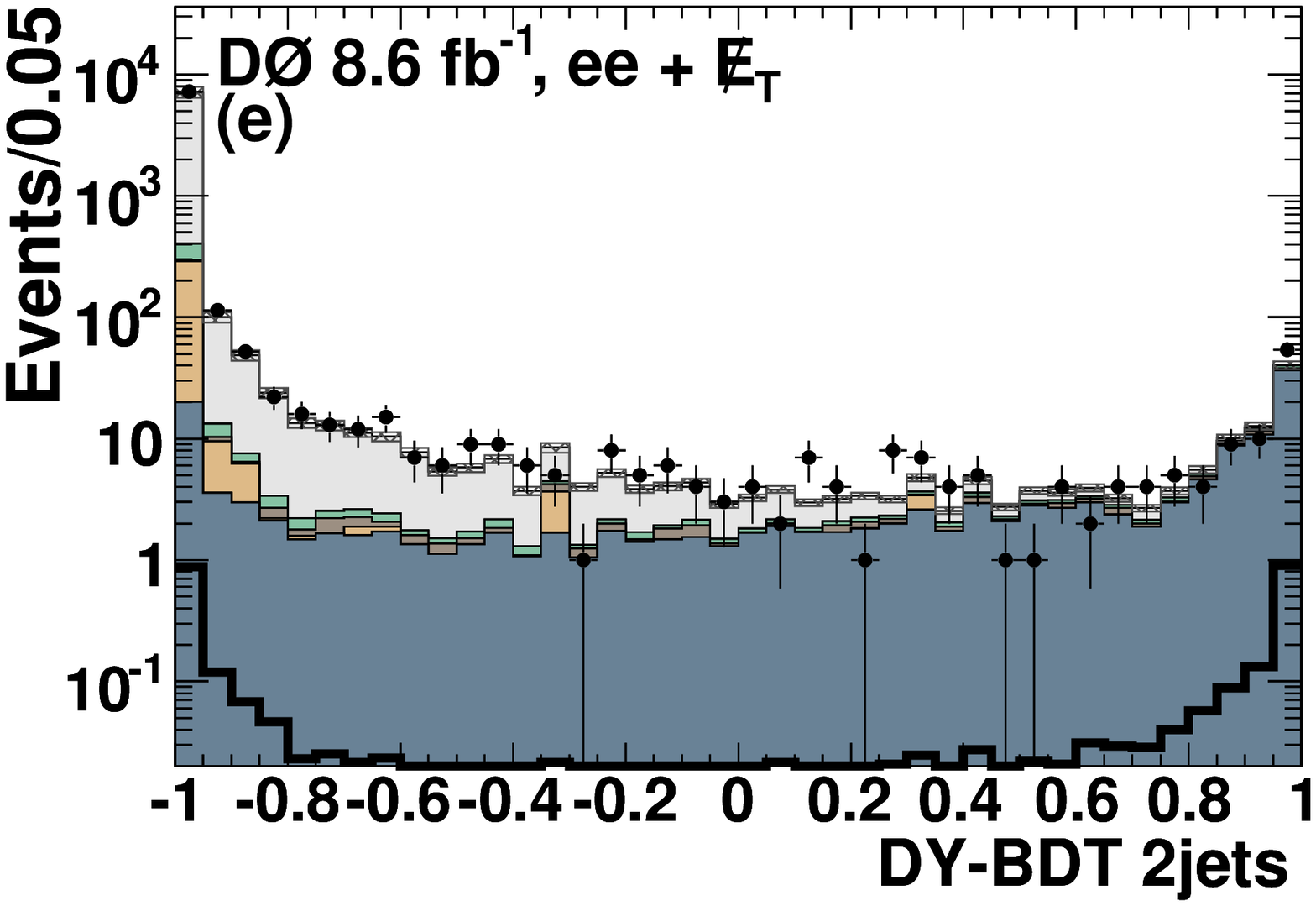} &
      \includegraphics[width=1.0\columnwidth]{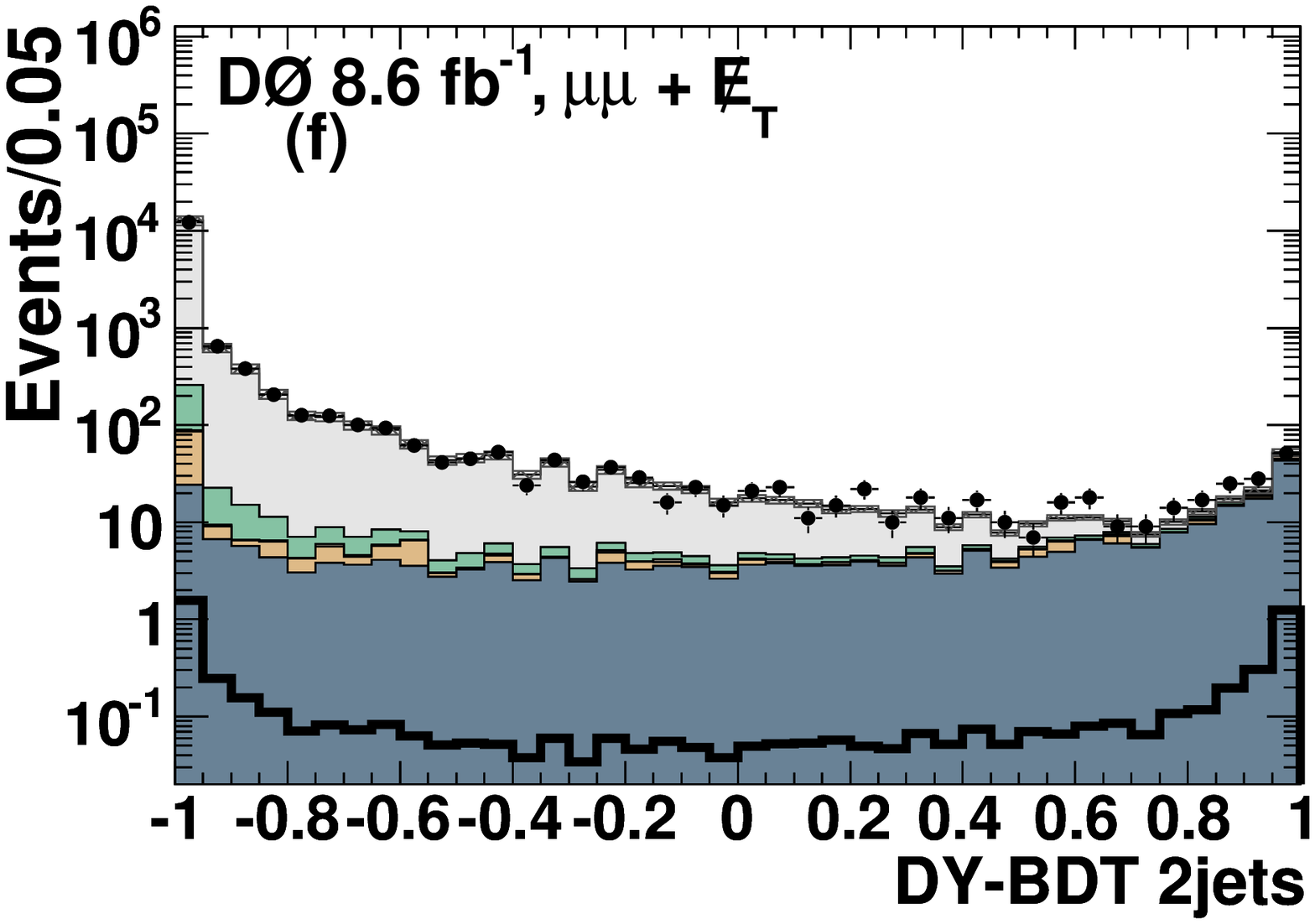} \\
    \end{tabular}
  \end{center}
  \caption{DY-BDT discriminant for the 0-jet (top row), 1-jet (middle
    row) and $\ge 2$-jet (bottom row) for the \ee\ [left (a,c,e)] and
    \mm\ [right (b,d,f)] final states. The discriminant shown is
    trained for a Higgs boson mass of $165$\,GeV. A final selection
    requirement is applied in the above distributions of 0.35, -0.6,
    and -0.85 for the \ee\ final state and 0.9, 0., and -0.7 for the
    \mm\ final state, in the 0-jet, 1-jet, and $\ge 2$-jet bins,
    respectively. The hatched bands show the total systematic
    uncertainty on the background prediction.\label{fig:dy_output} }
\end{figure*}

\begin{figure*}[!]
  \begin{center}
    \begin{tabular}{ccc}
       \includegraphics[width=1.0\columnwidth]{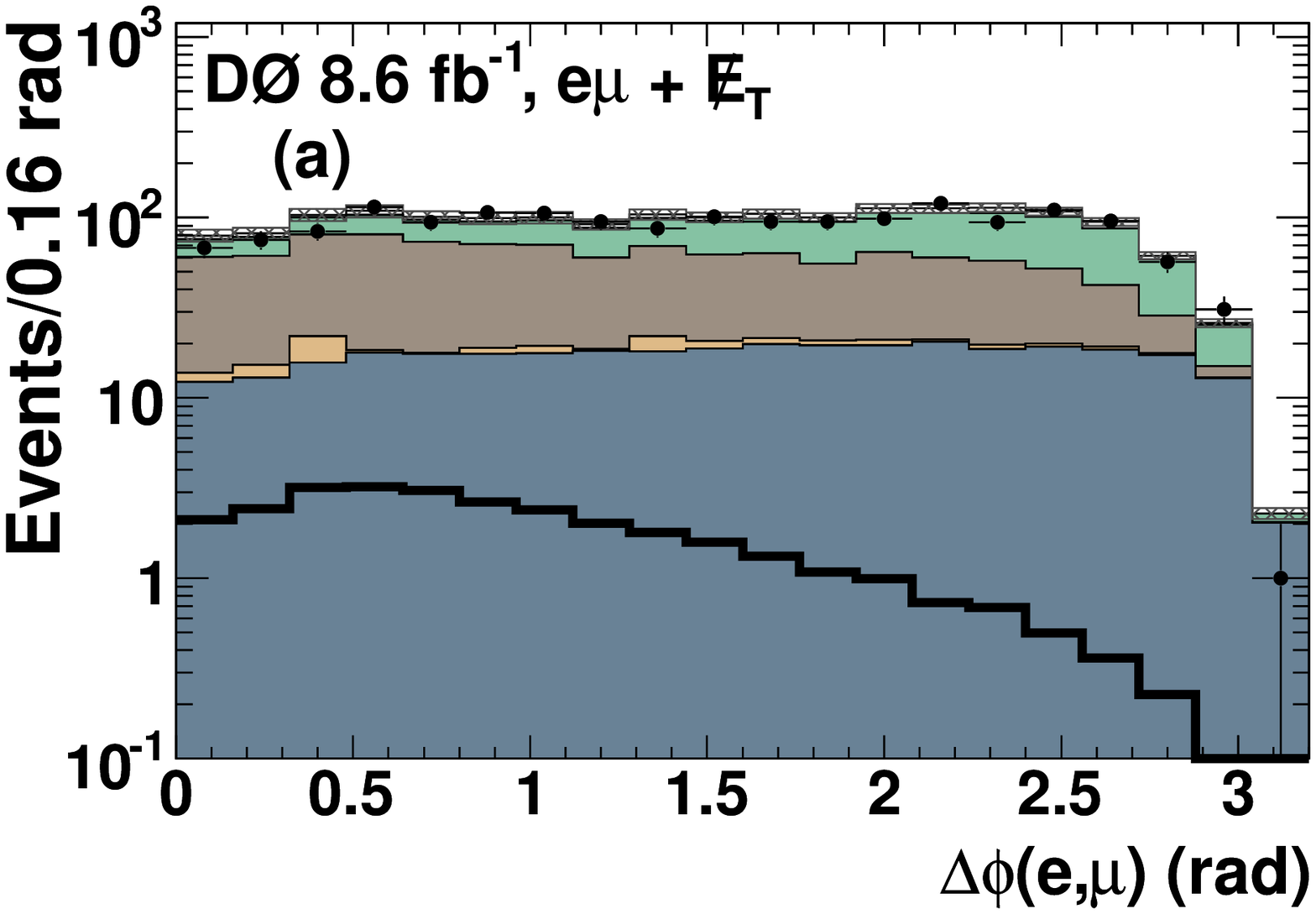} &
       \includegraphics[width=1.0\columnwidth]{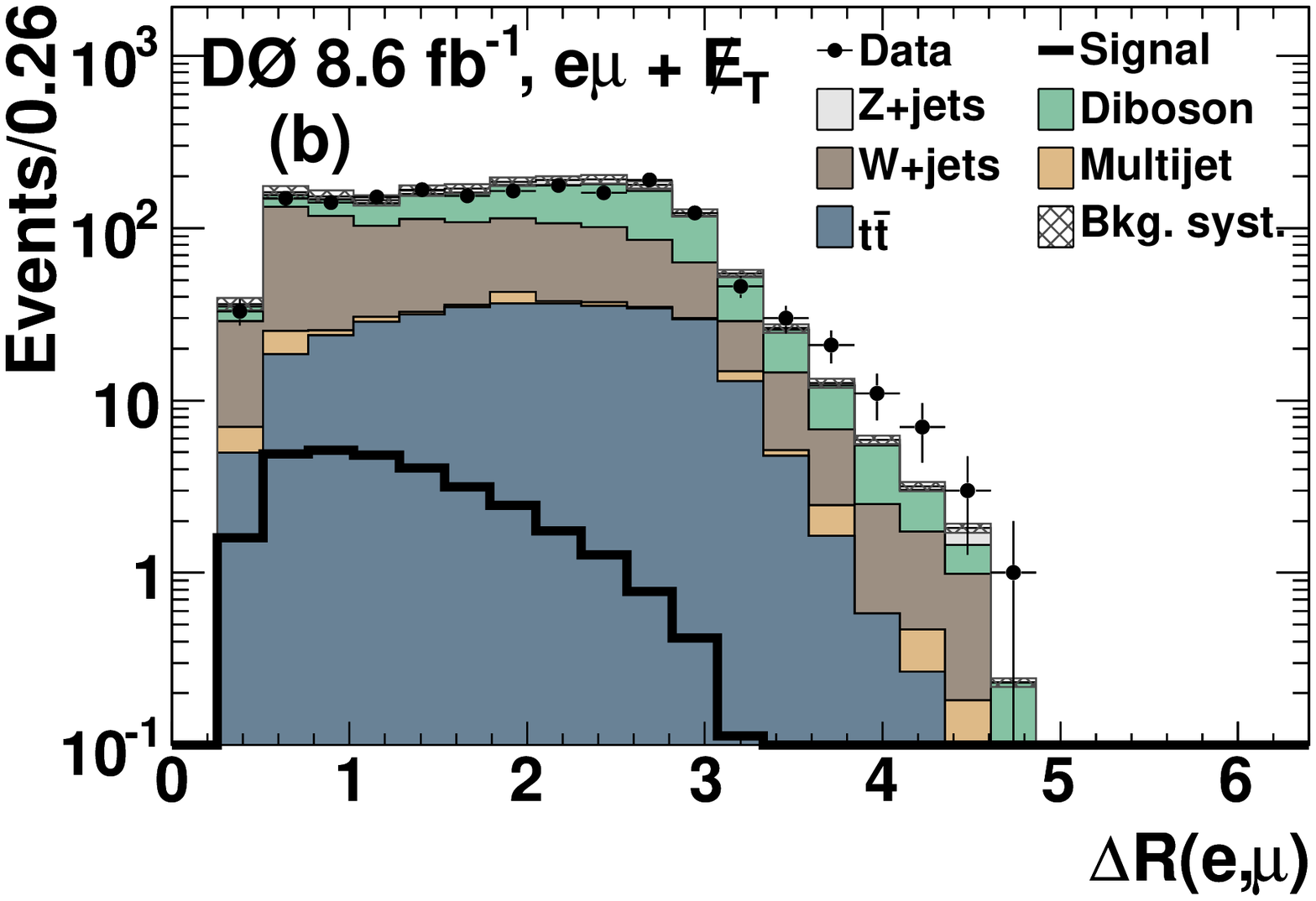}\\
	\includegraphics[width=1.0\columnwidth]{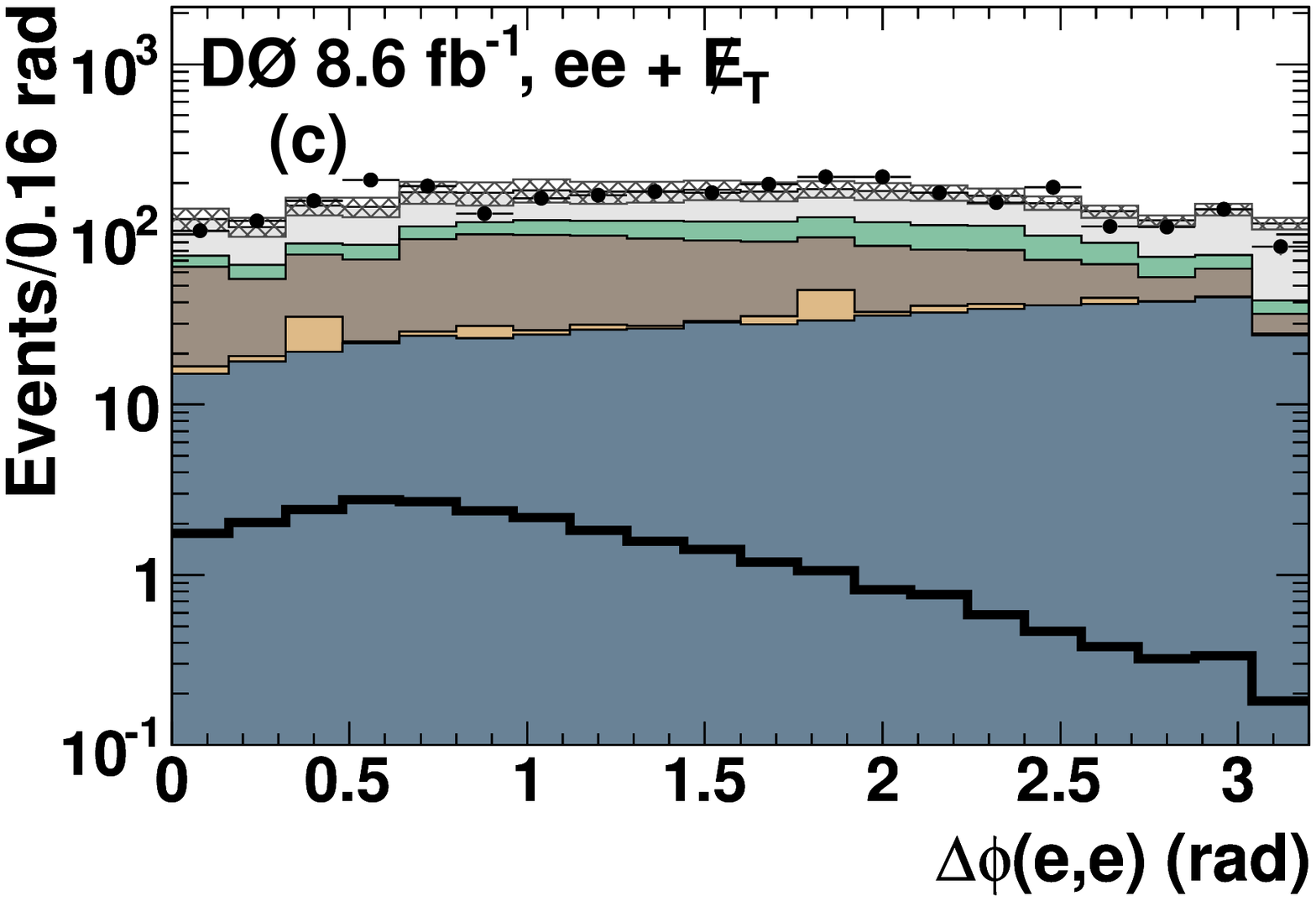}&
	\includegraphics[width=1.0\columnwidth]{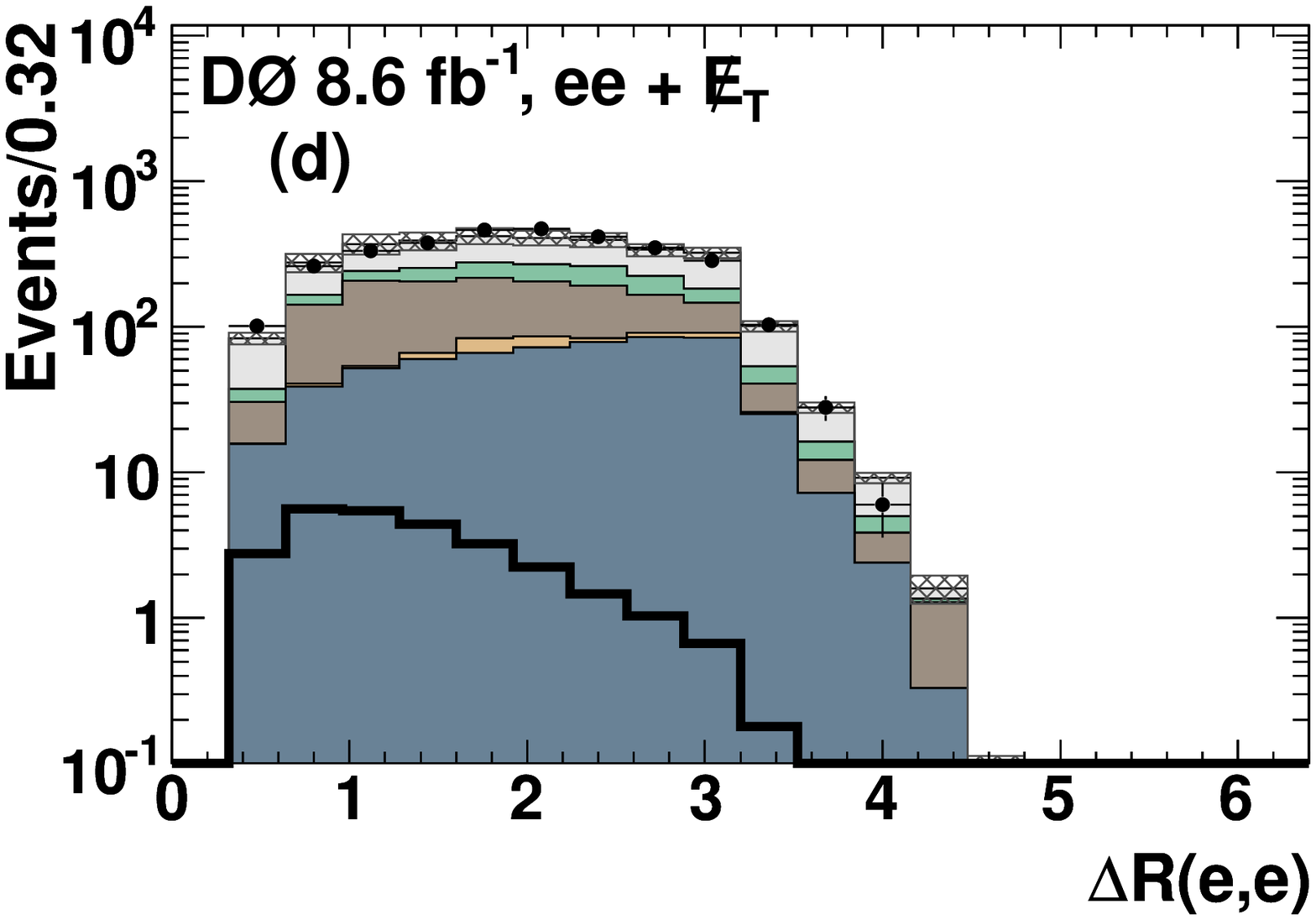}\\
	\includegraphics[width=1.0\columnwidth]{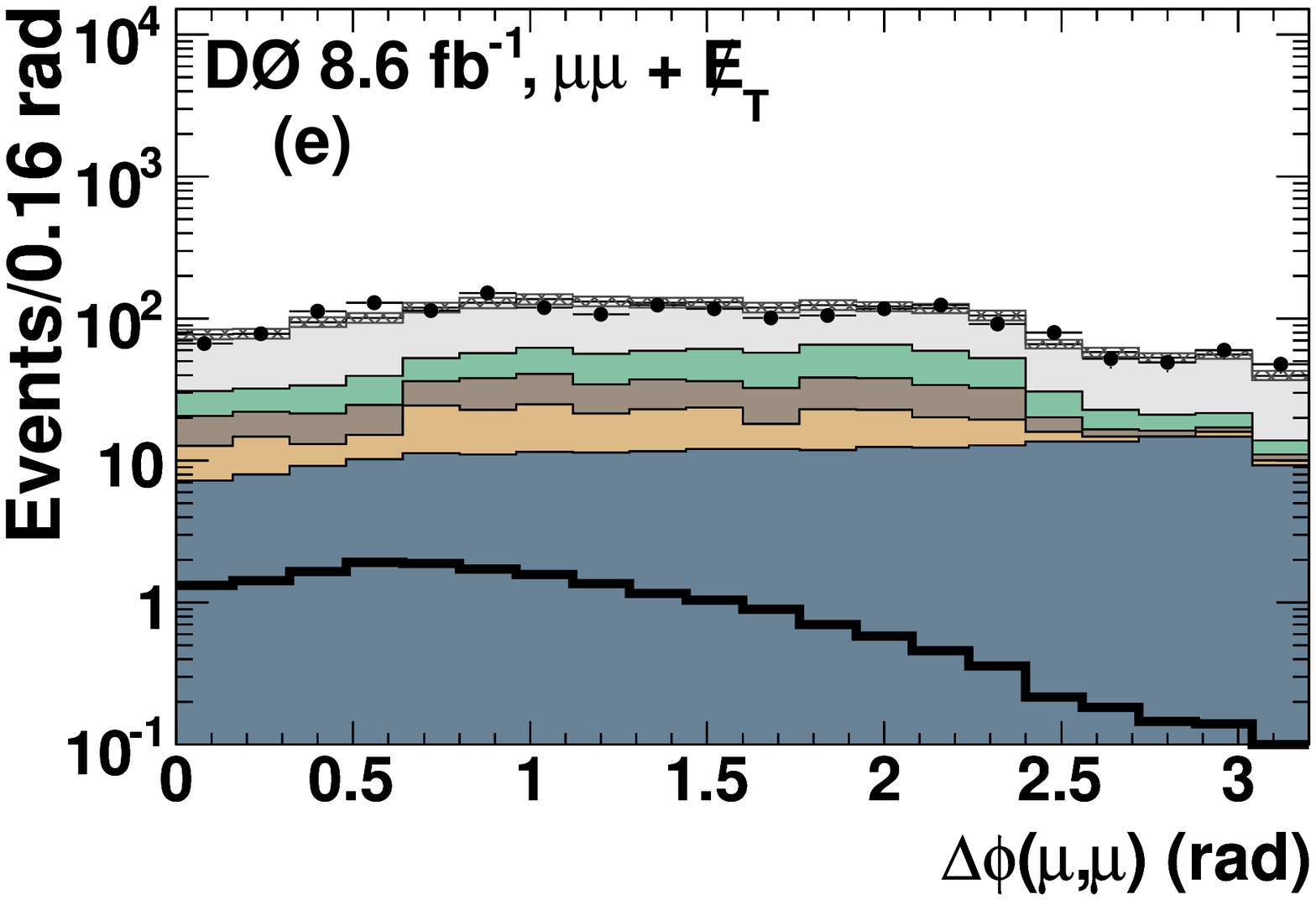}&
	\includegraphics[width=1.0\columnwidth]{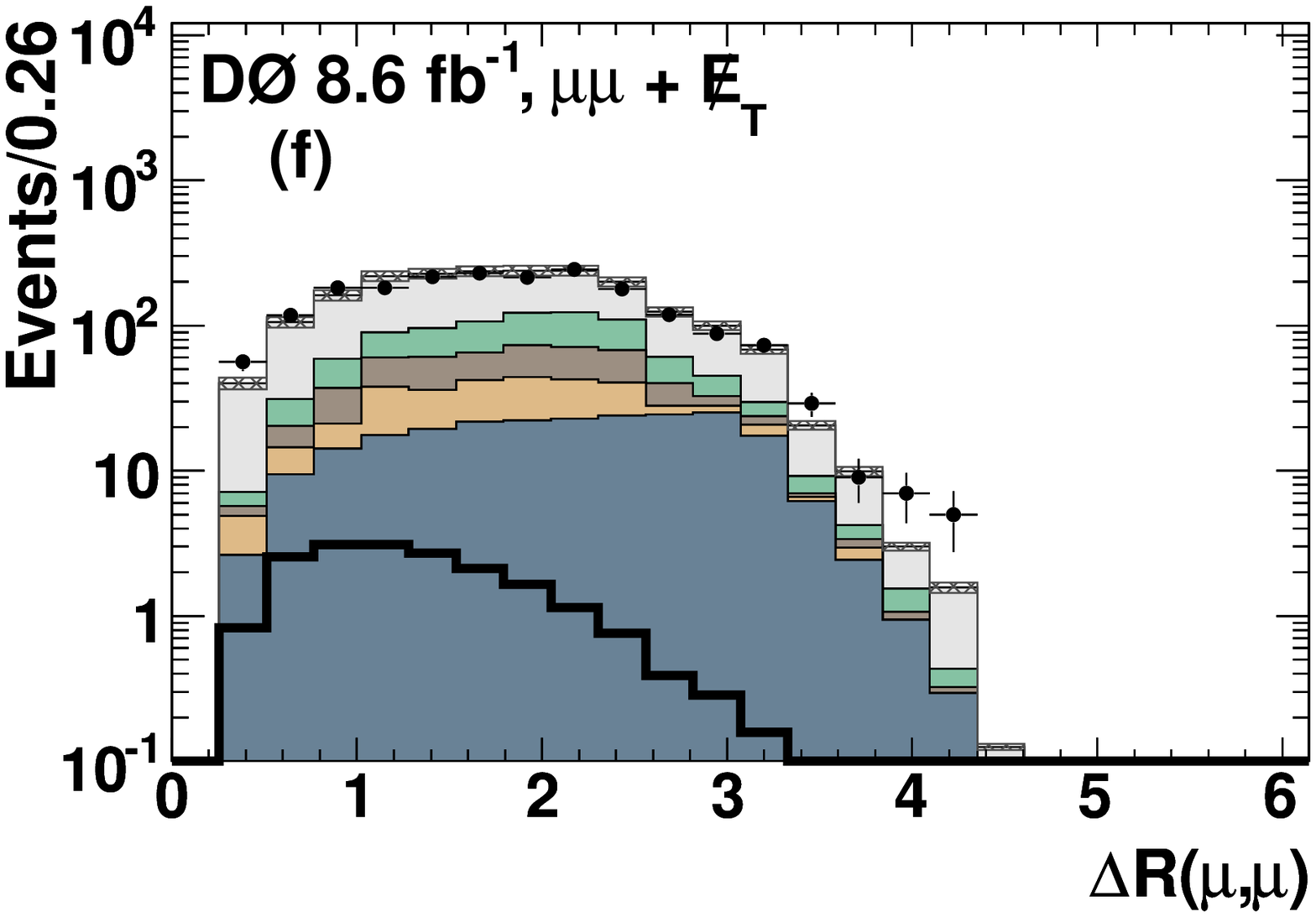}\\
    \end{tabular}
  \end{center}
  \caption{[color online] The (left column) $\Delta
     \phi(\ell_{1},\ell_{2})$ and (right column) $\Delta
     R(\ell_{1},\ell_{2})$ for the \em\ (a-b), \ee\ (c-d) and \mm\
     (e-f) channel at the final selection stage. The signal
     distribution shown corresponds to a Higgs boson mass of
     165\,GeV. The hatched bands show the total systematic uncertainty
     on the background prediction.}
    \label{fig:finalsel_plots_1}
\end{figure*}

\begin{figure*}[!]
  \begin{center}
    \begin{tabular}{cc}
        \includegraphics[width=1.0\columnwidth]{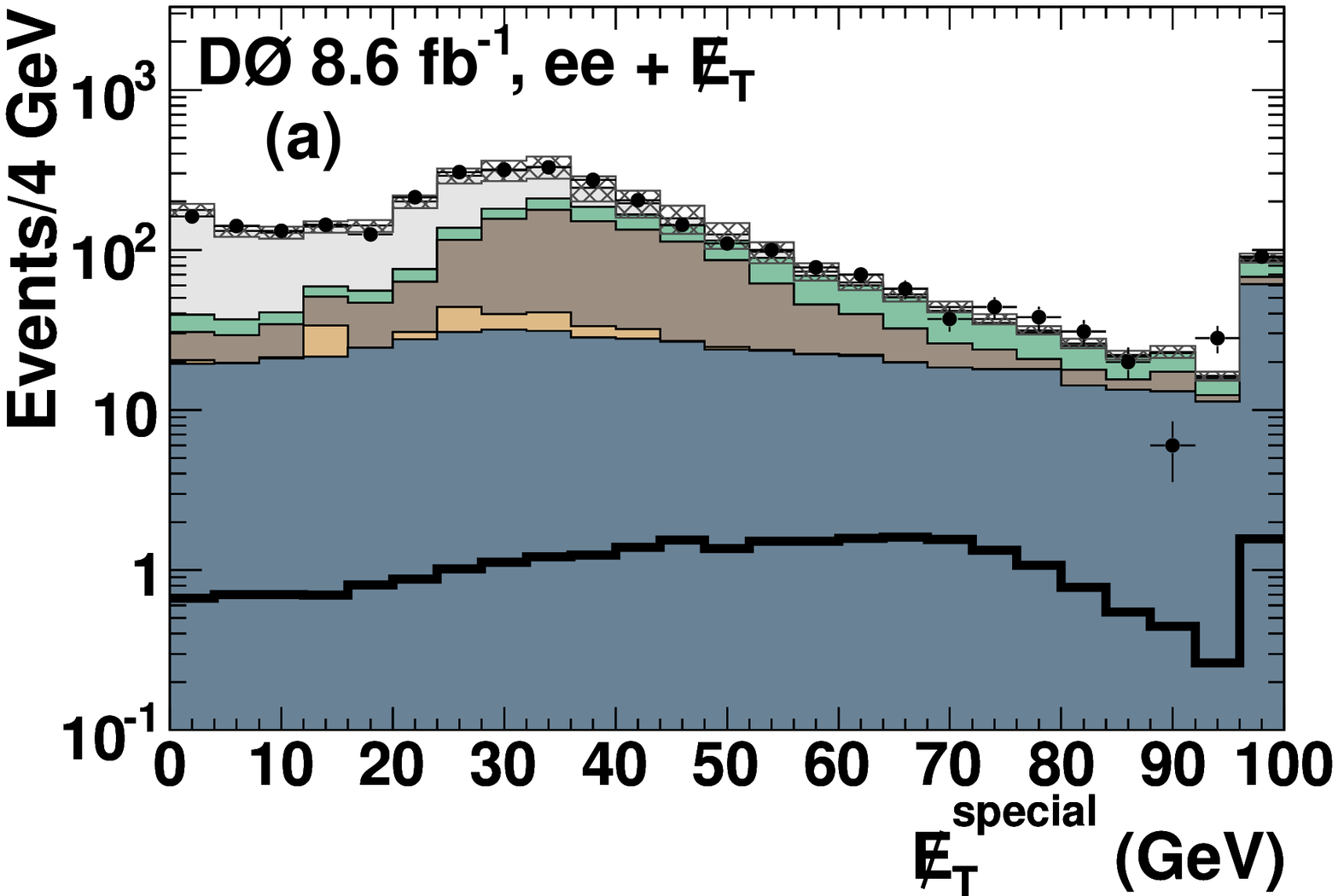} &
        \includegraphics[width=1.0\columnwidth]{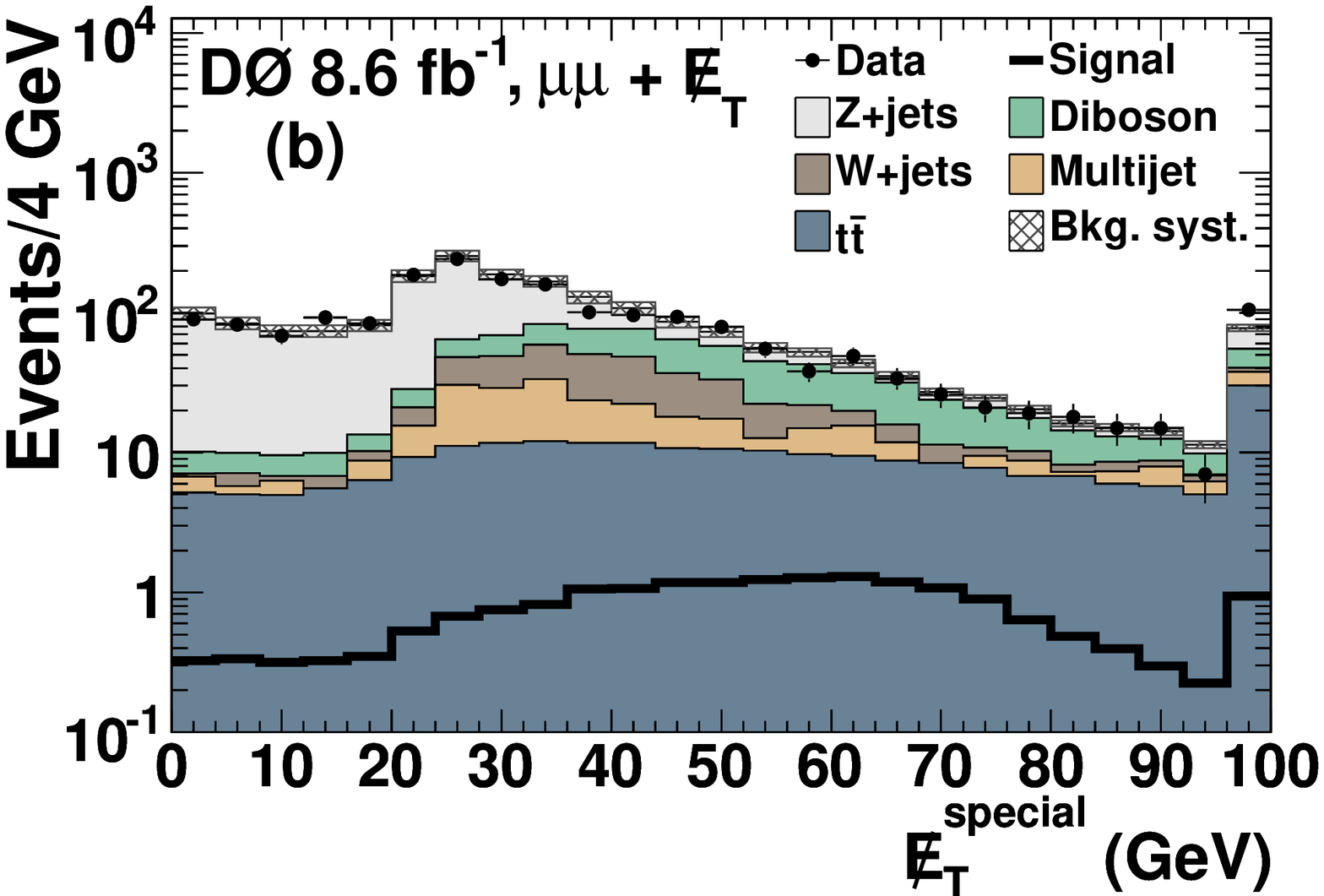}\\
    \end{tabular}
  \end{center}
  \caption{[color online] The $\Eslash_T^{\text{special}}$ for the
     \ee\ (a) and \mm\ (b) channel at the final selection stage. The
     last bin includes all events above the upper range of the
     histogram. The signal distribution shown corresponds to a Higgs
     boson mass of 165\,GeV. The last bin also includes all events
     above the upper range of the histogram. The hatched bands show
     the total systematic uncertainty on the background prediction.}
    \label{fig:finalsel_plots_2}
\end{figure*}

\begin{figure*}[!]
  \begin{center}
    \begin{tabular}{ccc}
       \includegraphics[width=1.0\columnwidth]{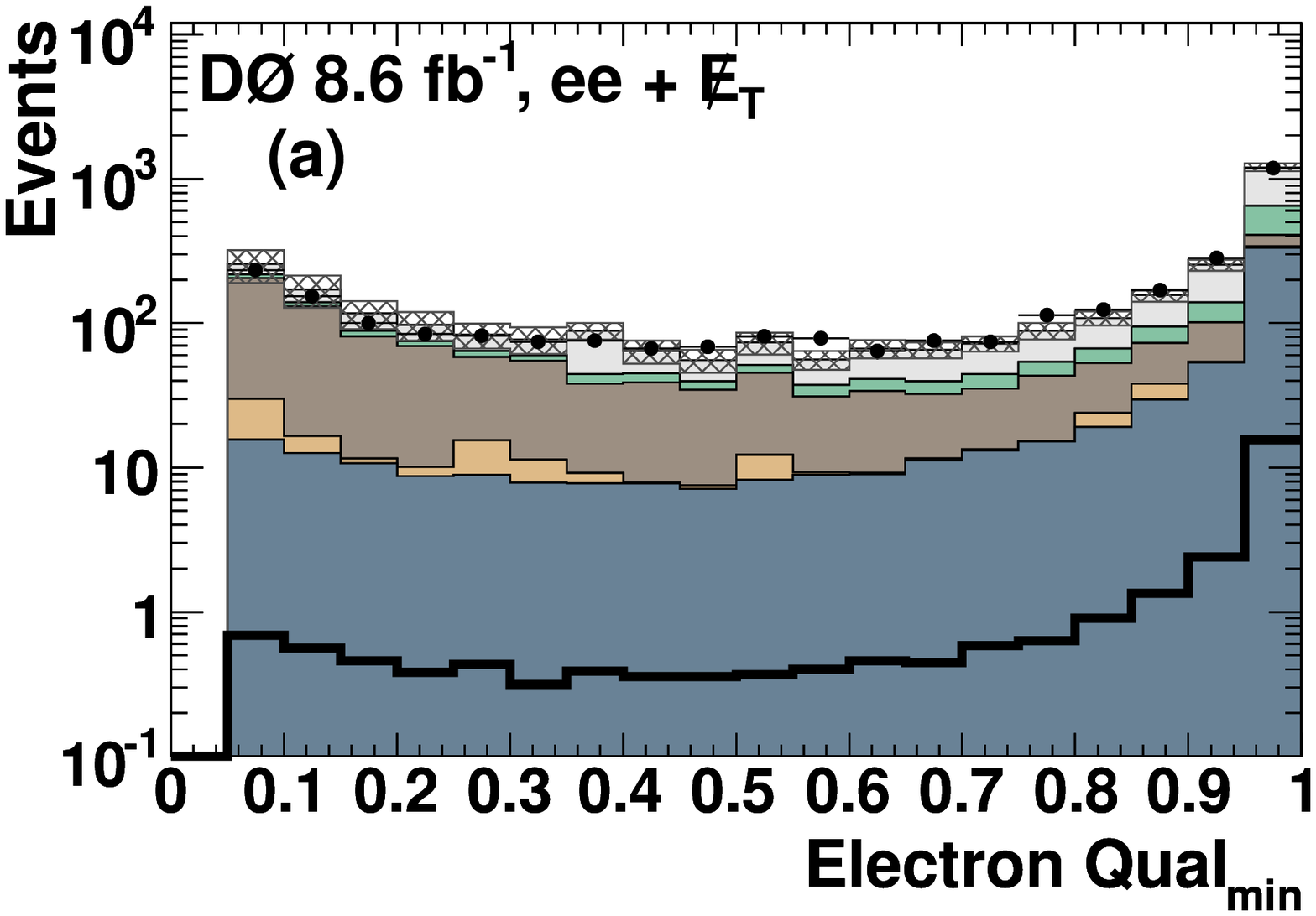}&
       \includegraphics[width=1.0\columnwidth]{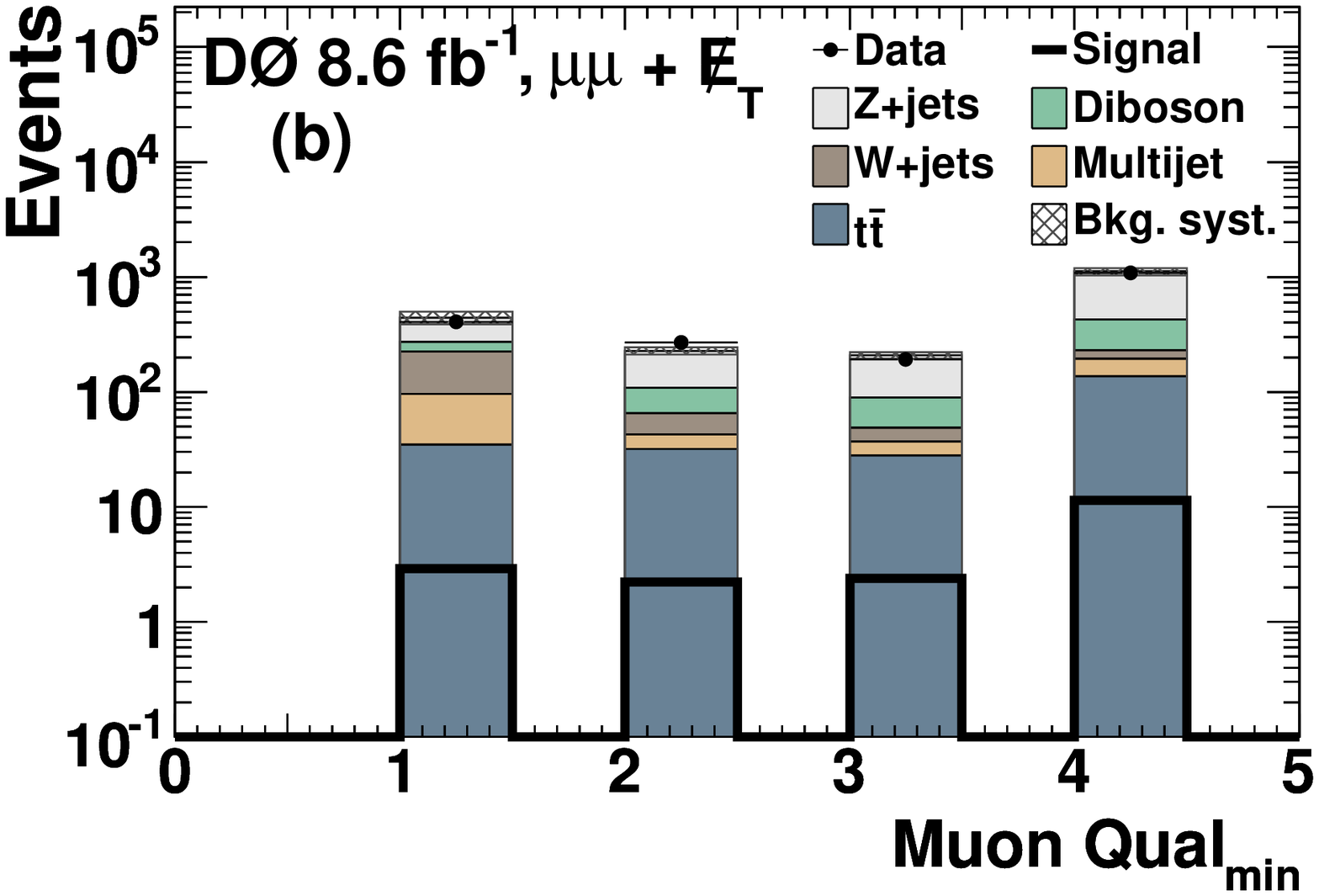} \\
    \end{tabular}
  \end{center}
  \caption{[color online] The quality variable for (a) the \ee\
    channel and (b) the \mm\ channel at the final selection stage. The
    signal distribution shown corresponds to a Higgs boson mass of
    165\,GeV. The hatched bands show the total systematic uncertainty
    on the background prediction.}
  \label{fig:finalsel_minQual_plots}
\end{figure*}

\begin{figure*}[!]
  \begin{center}
    \begin{tabular}{ccc}
       \includegraphics[width=1.0\columnwidth]{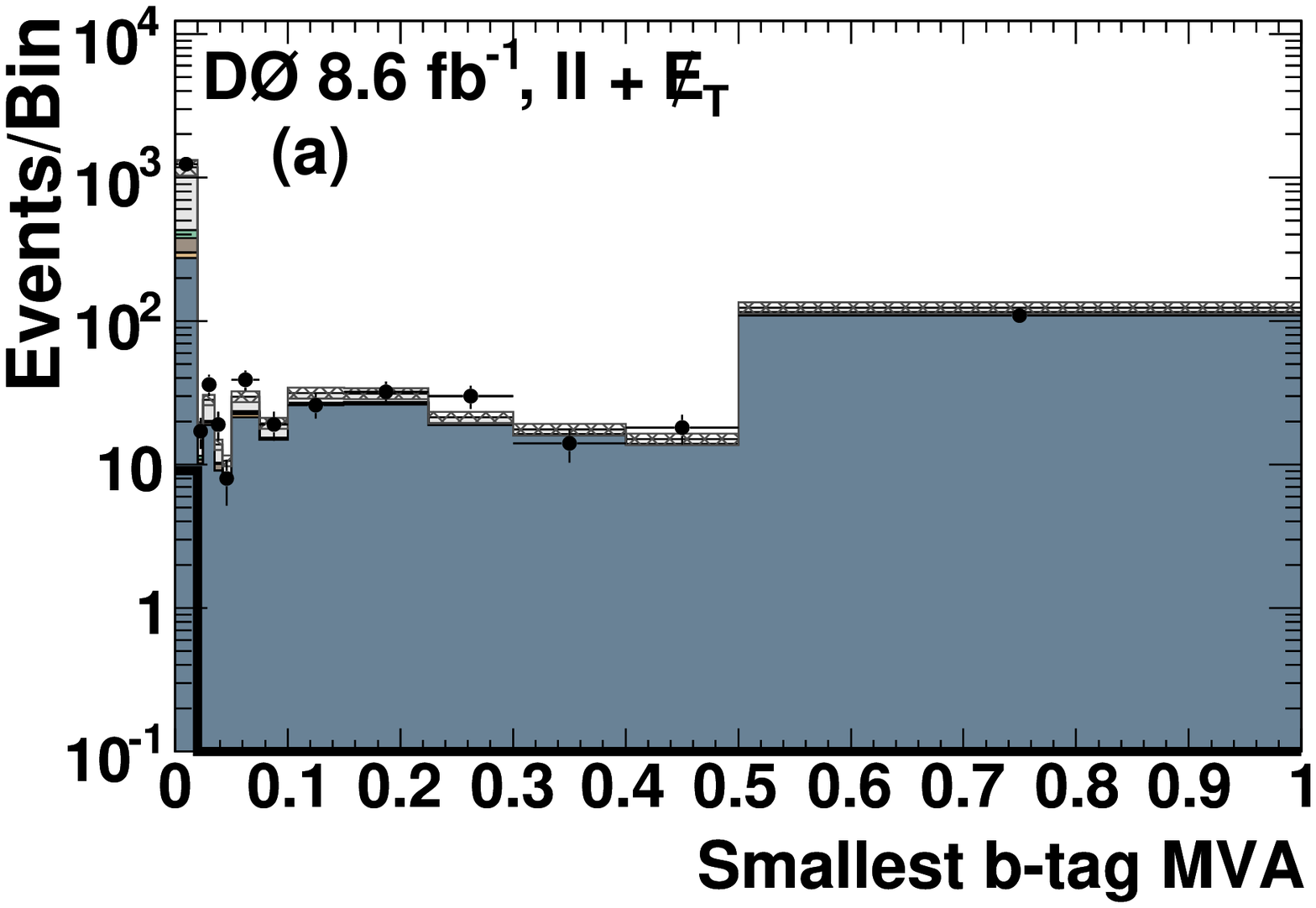}&
       \includegraphics[width=1.0\columnwidth]{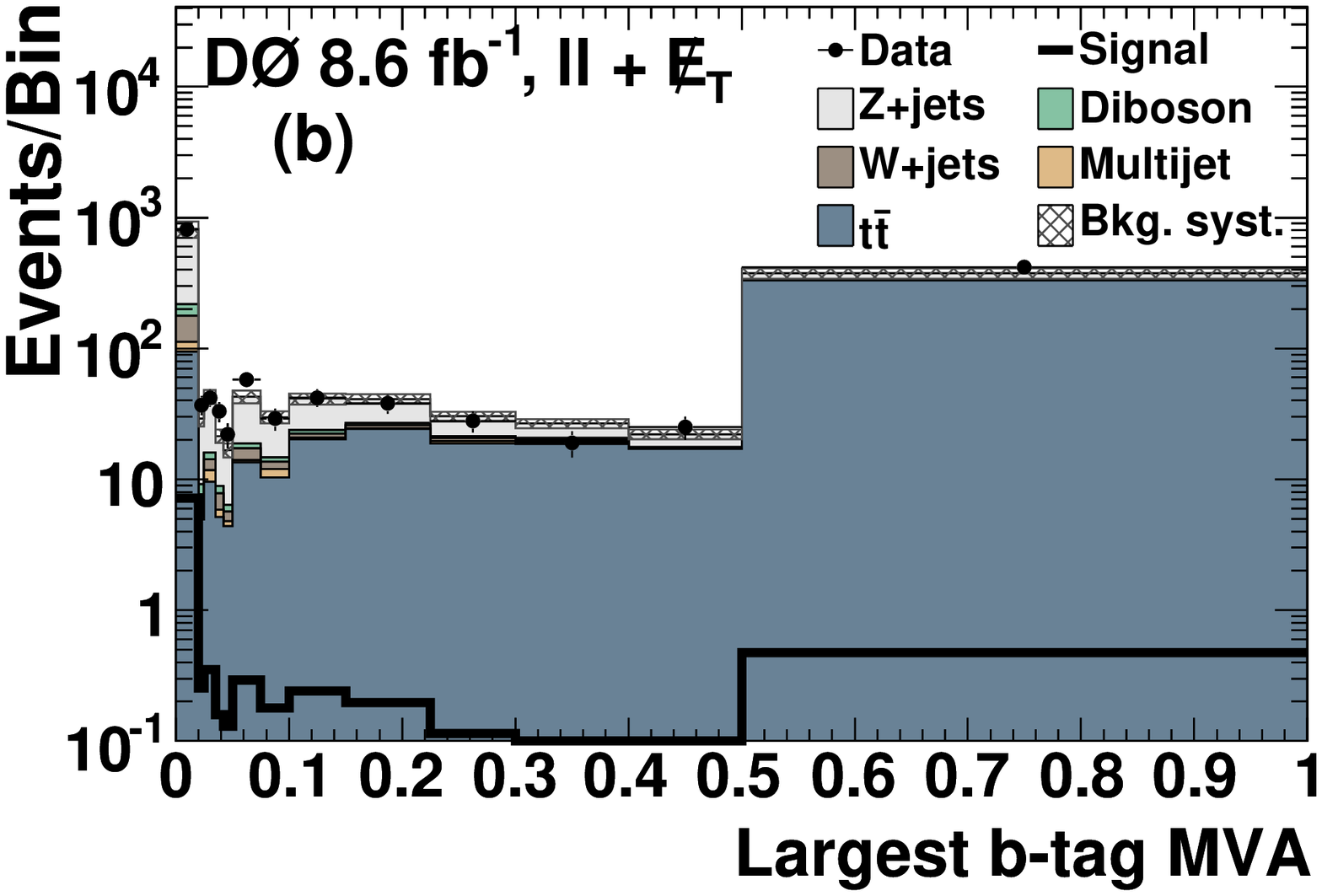} \\
    \end{tabular}
  \end{center}
  \caption{[color online] The (a) smallest $b$-tagging MVA output and
    (b) largest $b$-tagging MVA output for the $\ge 2$-jet
    multiplicity bin, for all channels summed up at the final
    selection stage. The signal distribution shown corresponds to a
    Higgs boson mass of 165\,GeV. The hatched bands show the total
    systematic uncertainty on the background prediction.}
  \label{fig:finalsel_bid_plots}
\end{figure*}

\subsection{Final Multivariate Discriminant}
In the final selection step, the signal is separated from the
remaining backgrounds using an additional random forest of BDTs. This
final random forest of BDTs referred to as FD-BDT, is trained for each
Higgs boson mass hypothesis and jet multiplicity bin, separately for
the three dilepton final states using signal and background samples,
which satisfy the final selection criteria, to differentiate between
all Higgs boson production processes and backgrounds. These decision
trees use as inputs all the variables from the DY-BDT listed above
with the addition of the following variables:

\begin{enumerate}[(i)]
\item electron quality likelihood output, ${\cal L}_8$; for the
  dielectron channel the lower value of the two electron quality
  likelihood outputs is used
\item a quality criterion based on the number of hits in the muon
  spectrometer characterized in four distinct categories; this
  parameter is referred to as ``muon quality'' and for the dimuon
  channel the lower quality of the two muons is used
\item number of Layer 0 hits in the SMT matched to each electron
\item track isolation variable of each muon
\item the product of charge and pseudorapidity, for both leptons
  $\ell_1$ and $\ell_2$
\item $b$-tag output: the output of a multivariate discriminant to
  separate jets originating from heavy flavor quarks ($b$ and $c$)
  from those originating from light partons; for the channels with
  $N_{\rm{jets}}\ge 2$, the smallest and largest $b$-tag outputs are
  used.
\end{enumerate}

Some representative input distributions to the FD-BDT at the final
selection stage with all jet multiplicity bins added in each
distribution are shown in Figs.\ \ref{fig:finalsel_plots_1} and
\ref{fig:finalsel_plots_2}.

Representative distributions of the electron and muon quality
variables, ${\cal L}_8$ and ``muon quality,'' are shown in Fig.\
\ref{fig:finalsel_minQual_plots}. These along with other variables
given in (iii) and (iv) gauge the quality of the reconstruction of the
lepton and are crucial to discriminate between true leptons and jets
misidentified as leptons originating from backgrounds like
$W+$jets. The distribution for the product of charge and
pseudorapidity, is symmetric in $\eta$ for the signal, however this is
not true for the background processes with misidentified leptons.

The output from $b$-tagging is used to separate the Higgs boson signal
from $t\bar{t}$ production, which is an important background in the 1
and 2 jet multiplicity bins.  An MVA-based $b$-tagging~\cite{bib:bid}
is employed in each of the dilepton final states to discriminate the
signal, which comprises primarily light flavor quarks, against the
heavy flavor jets arising from top quark decays. The distributions for
smallest and largest $b$-tagging output in the $\ge2$-jet multiplicity
bin are shown in Fig.\ \ref{fig:finalsel_bid_plots}.

The distributions of the final BDT discriminant for each channel and
Higgs boson masses of 125 GeV and 165 GeV are shown in
Figs.\ \ref{fig:final_bdt_em} -- \ref{fig:final_bdt_mm}.

\begin{figure*}[!]
  \begin{center}
    \begin{tabular}{cc}
      \includegraphics[width=1.0\columnwidth]{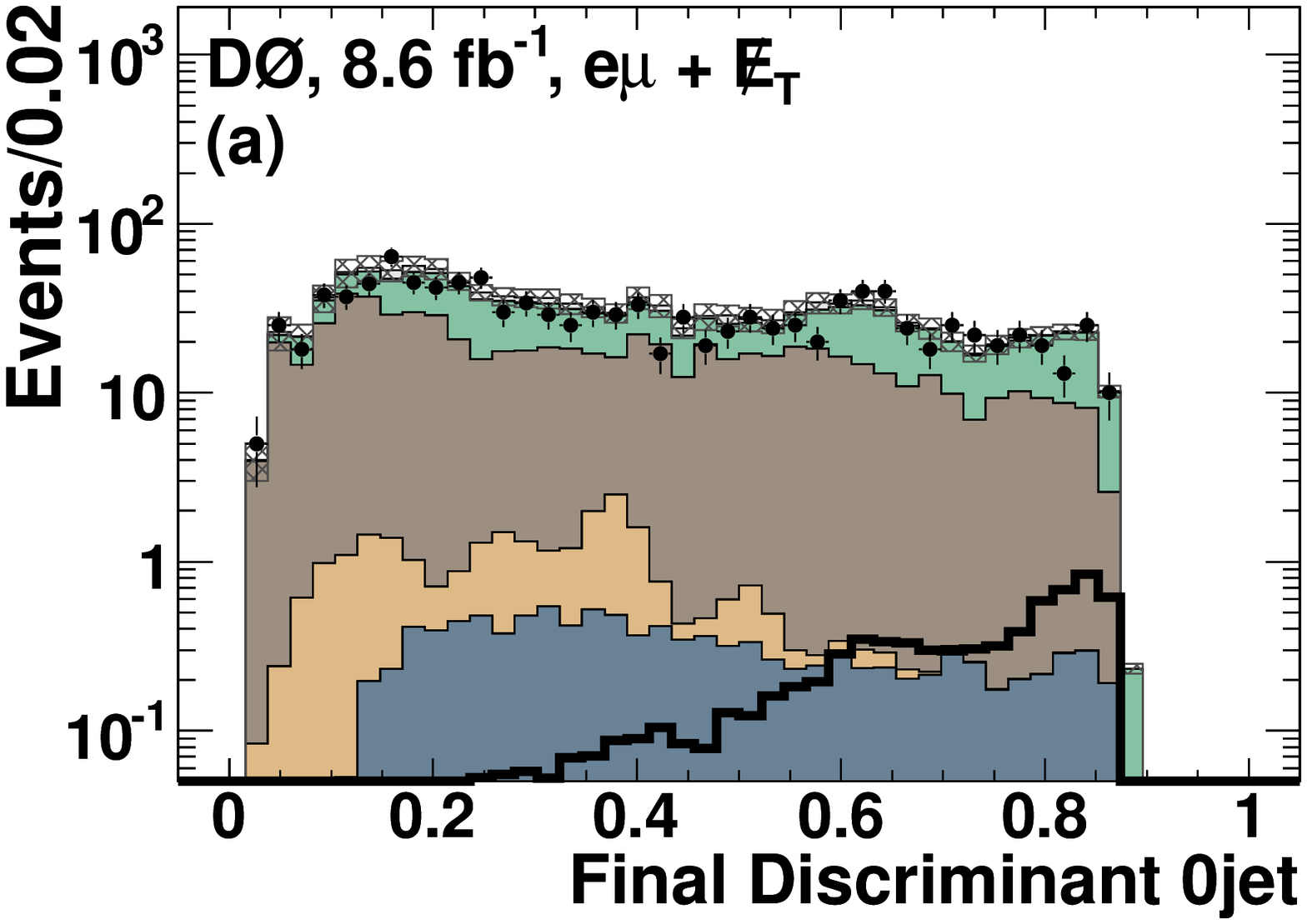} &
      \includegraphics[width=1.0\columnwidth]{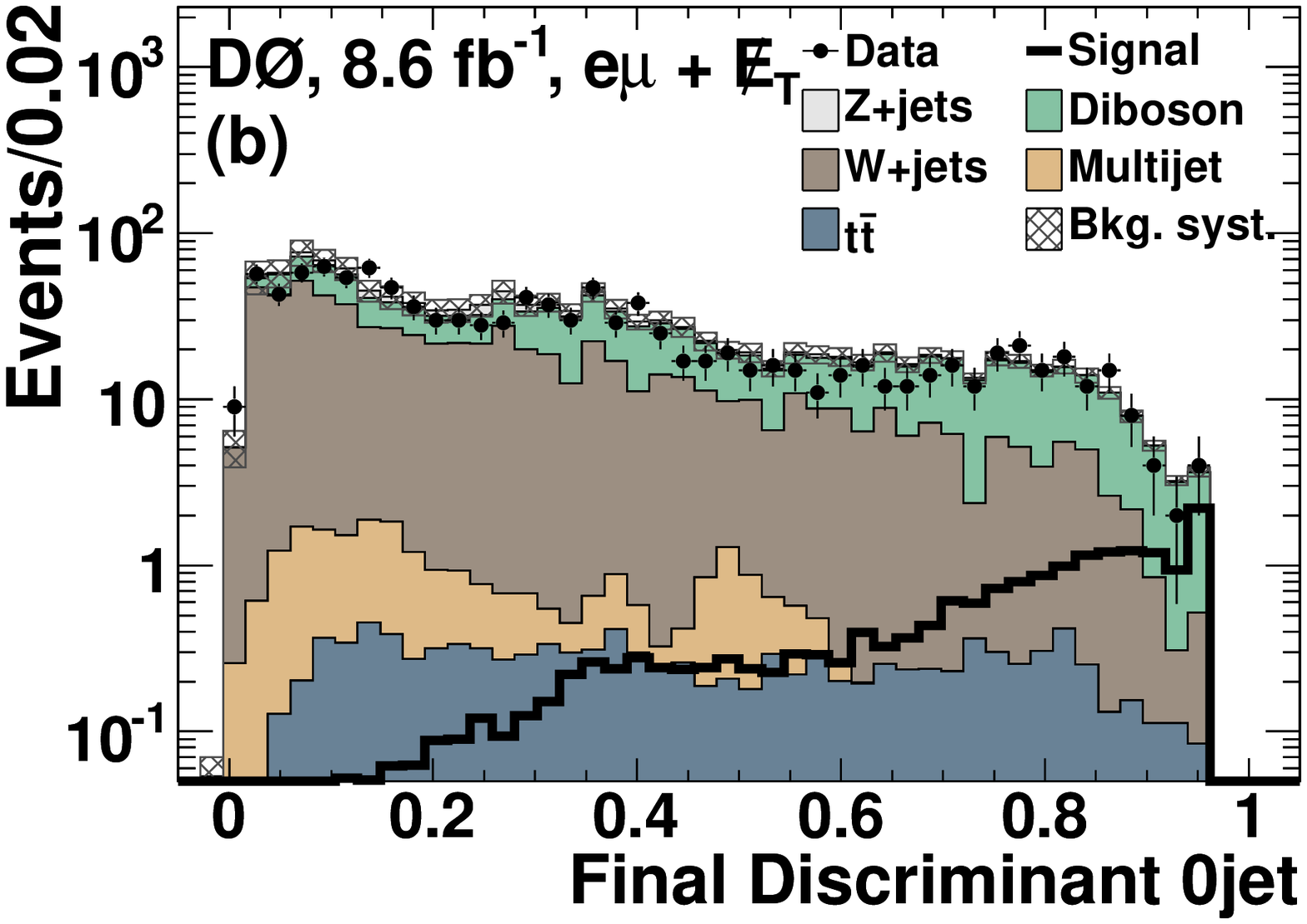} \\
      \includegraphics[width=1.0\columnwidth]{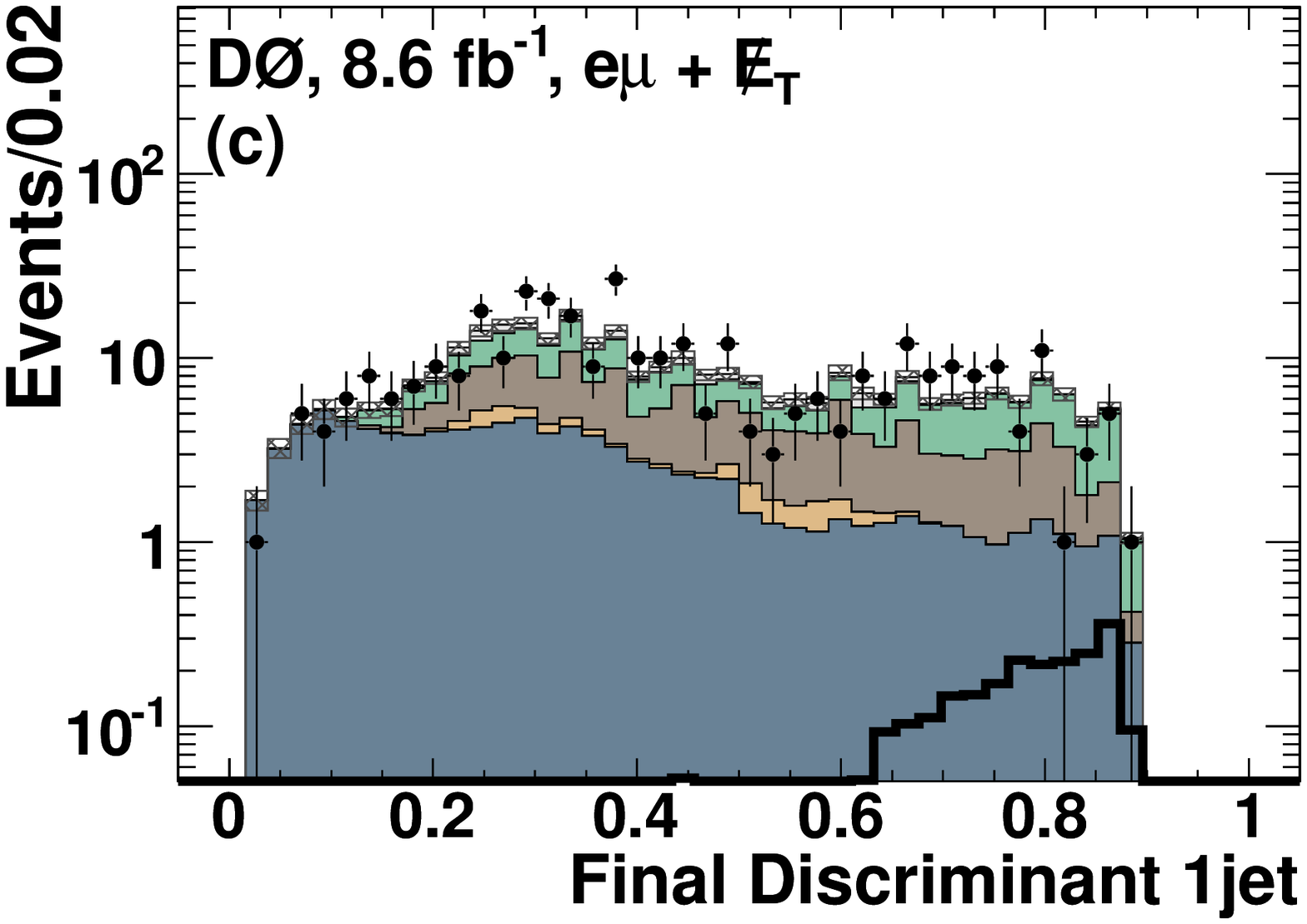} &
      \includegraphics[width=1.0\columnwidth]{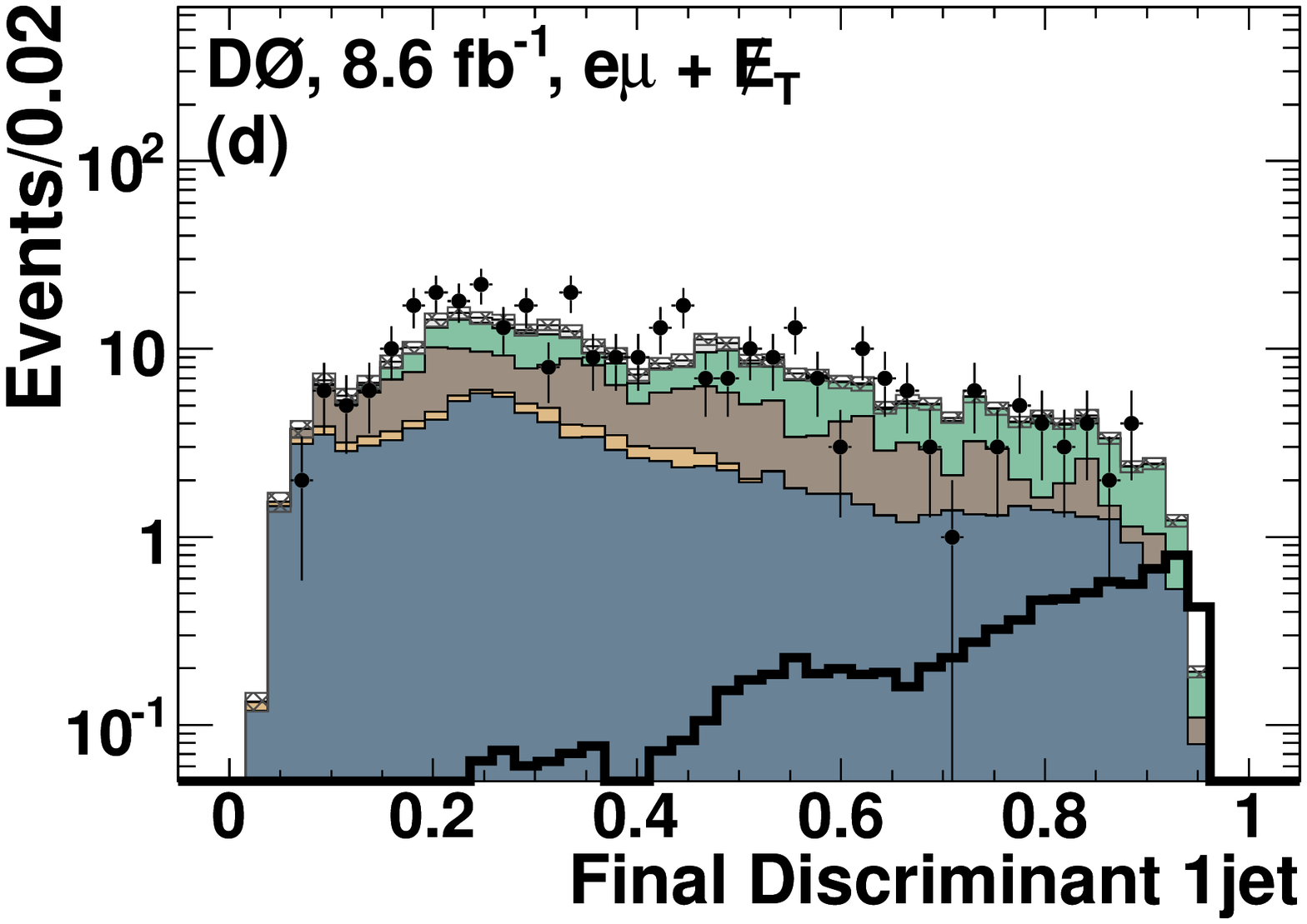} \\
      \includegraphics[width=1.0\columnwidth]{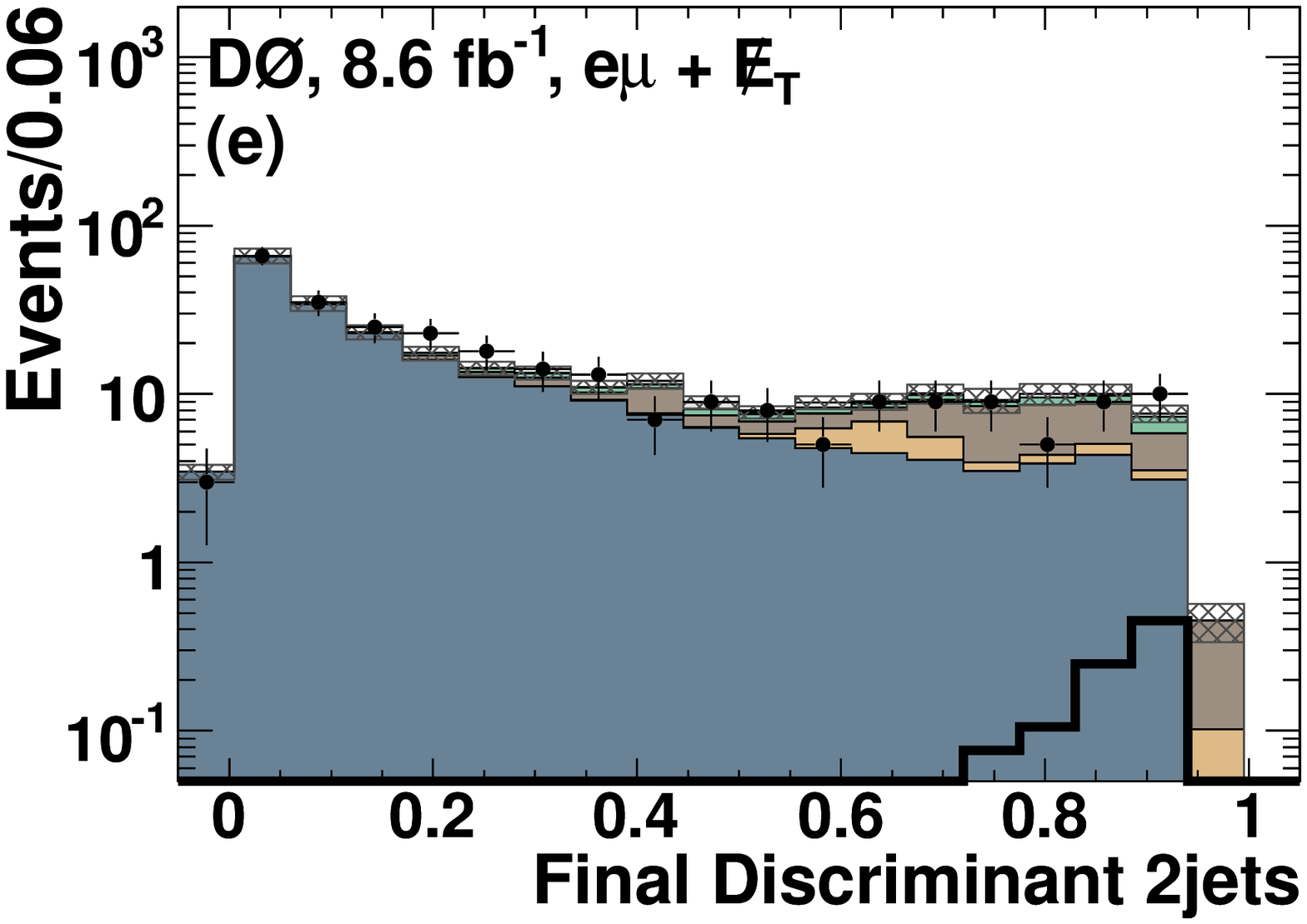} &
      \includegraphics[width=1.0\columnwidth]{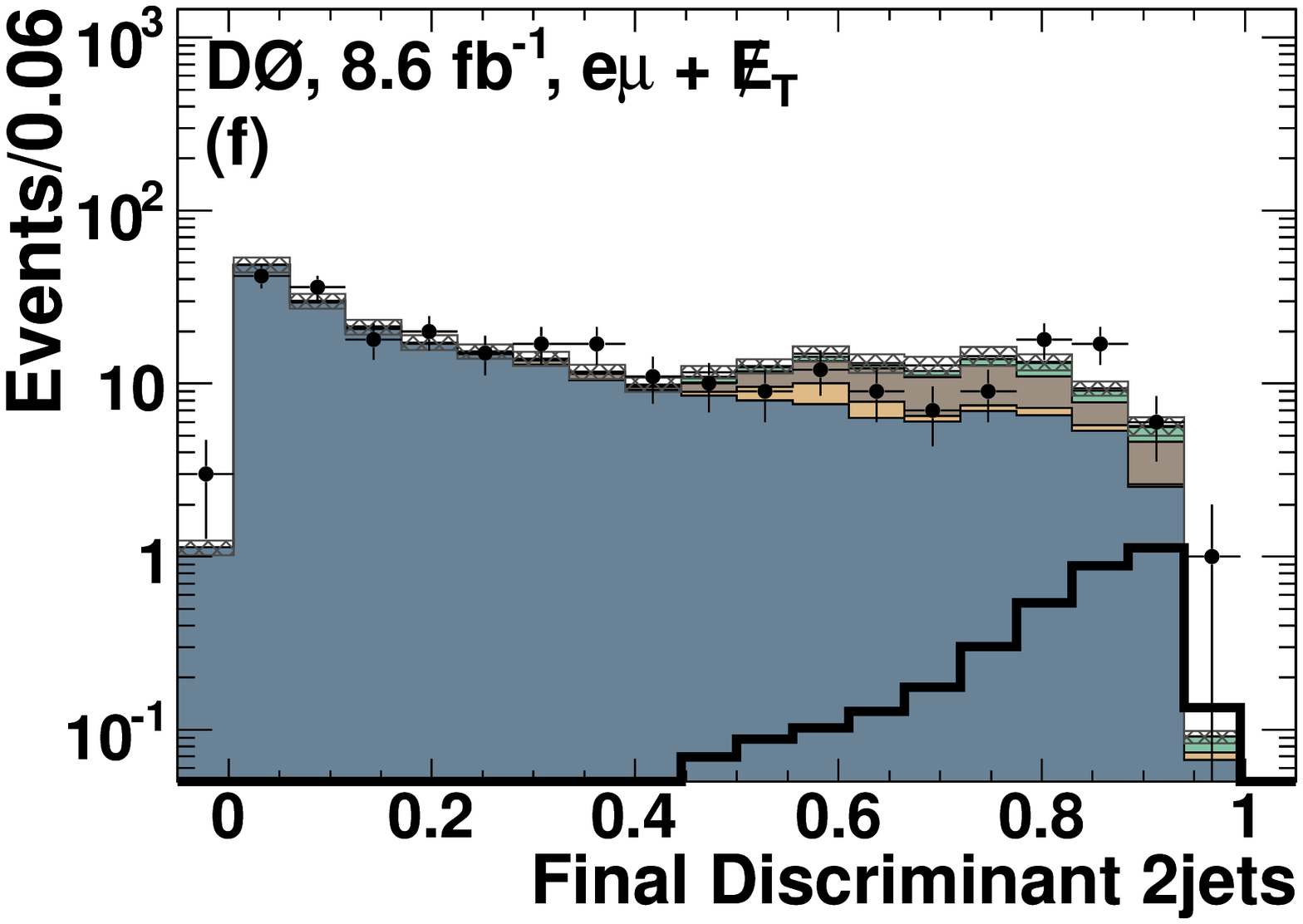} \\
    \end{tabular}
  \end{center}
  \caption{[color online] Final BDT discriminant for the (top-row)
    0-jet, (middle-row) 1-jet, and (bottom-row) $\ge 2$-jet bins for
    the \em\ final state for a Higgs boson masses of 125 GeV [left
    (a,c,e)] and 165 GeV [right (b,d,f)]. The hatched bands show the
    total systematic uncertainty on the background
    prediction.\label{fig:final_bdt_em}}
\end{figure*}

\begin{figure*}[!]
  \begin{center}
    \begin{tabular}{cc}
      \includegraphics[width=1.0\columnwidth]{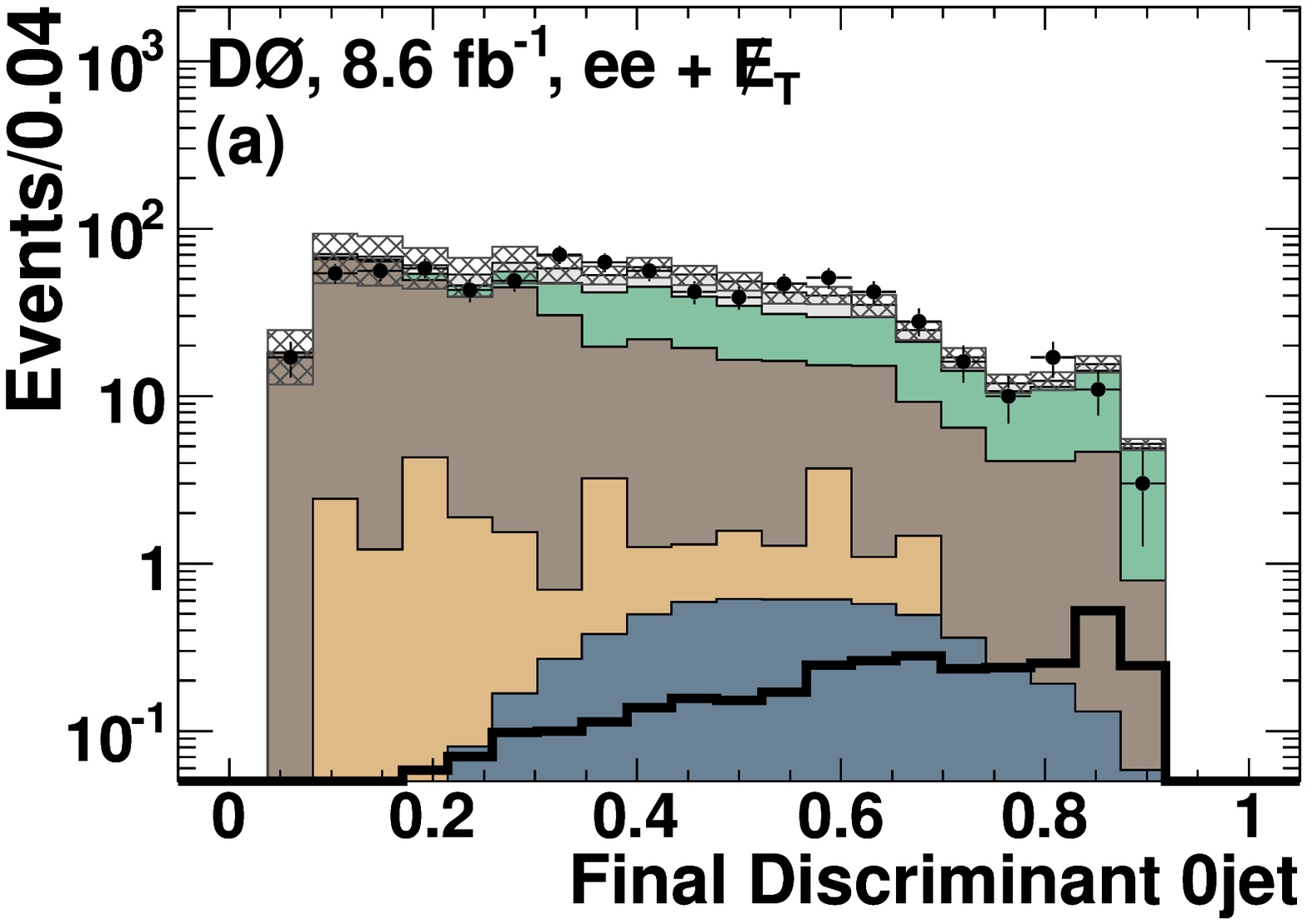} &
      \includegraphics[width=1.0\columnwidth]{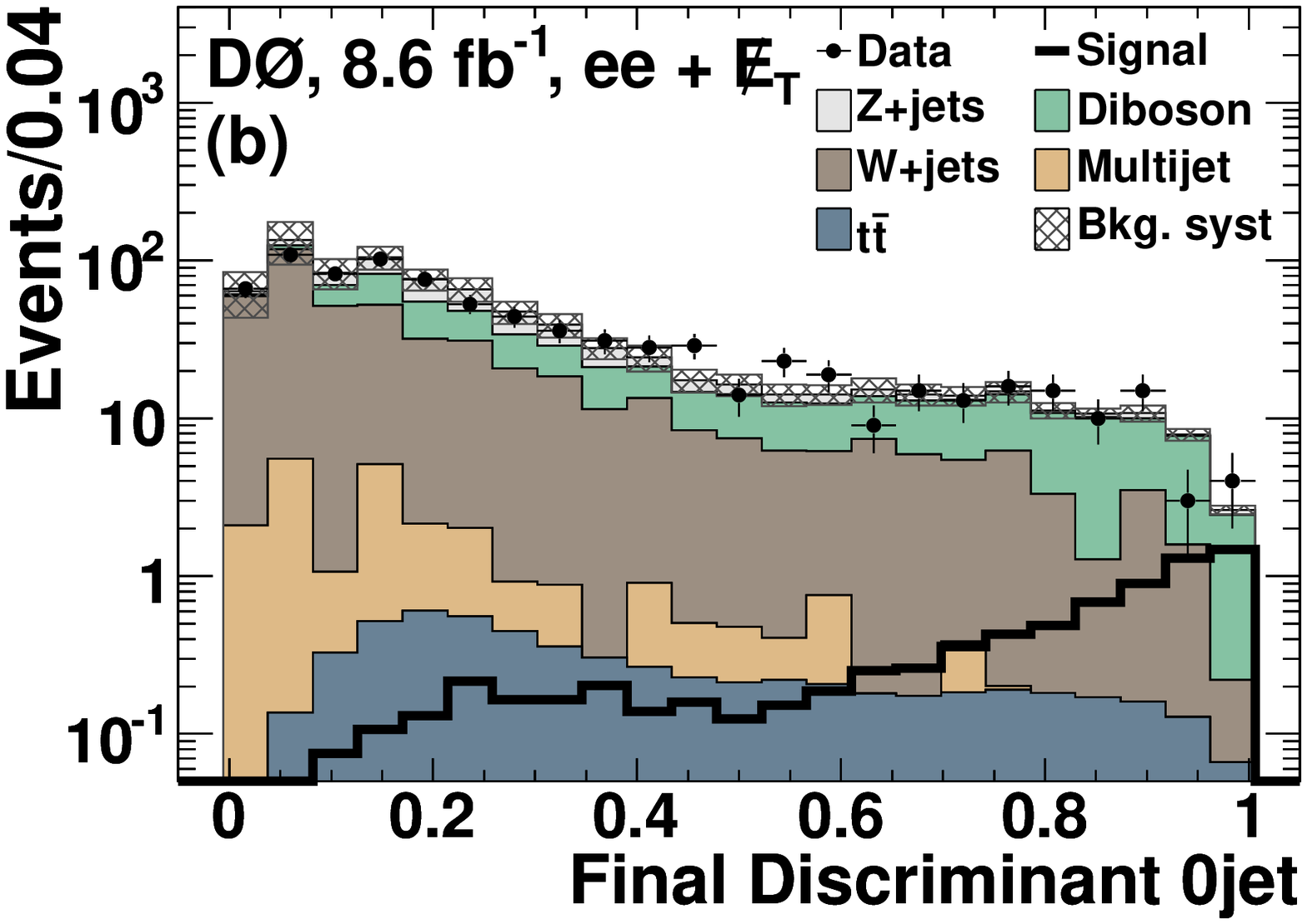} \\
      \includegraphics[width=1.0\columnwidth]{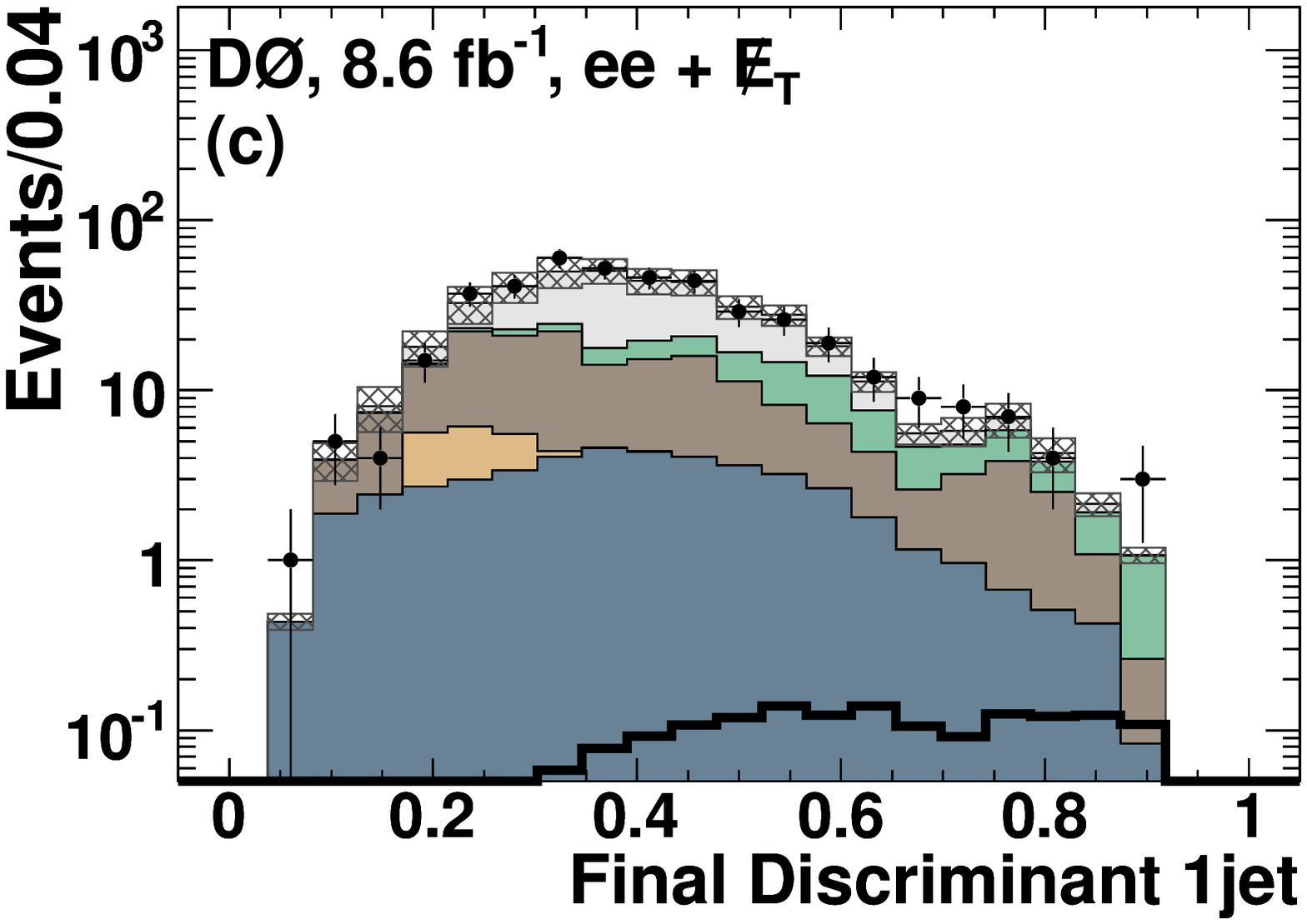} &
      \includegraphics[width=1.0\columnwidth]{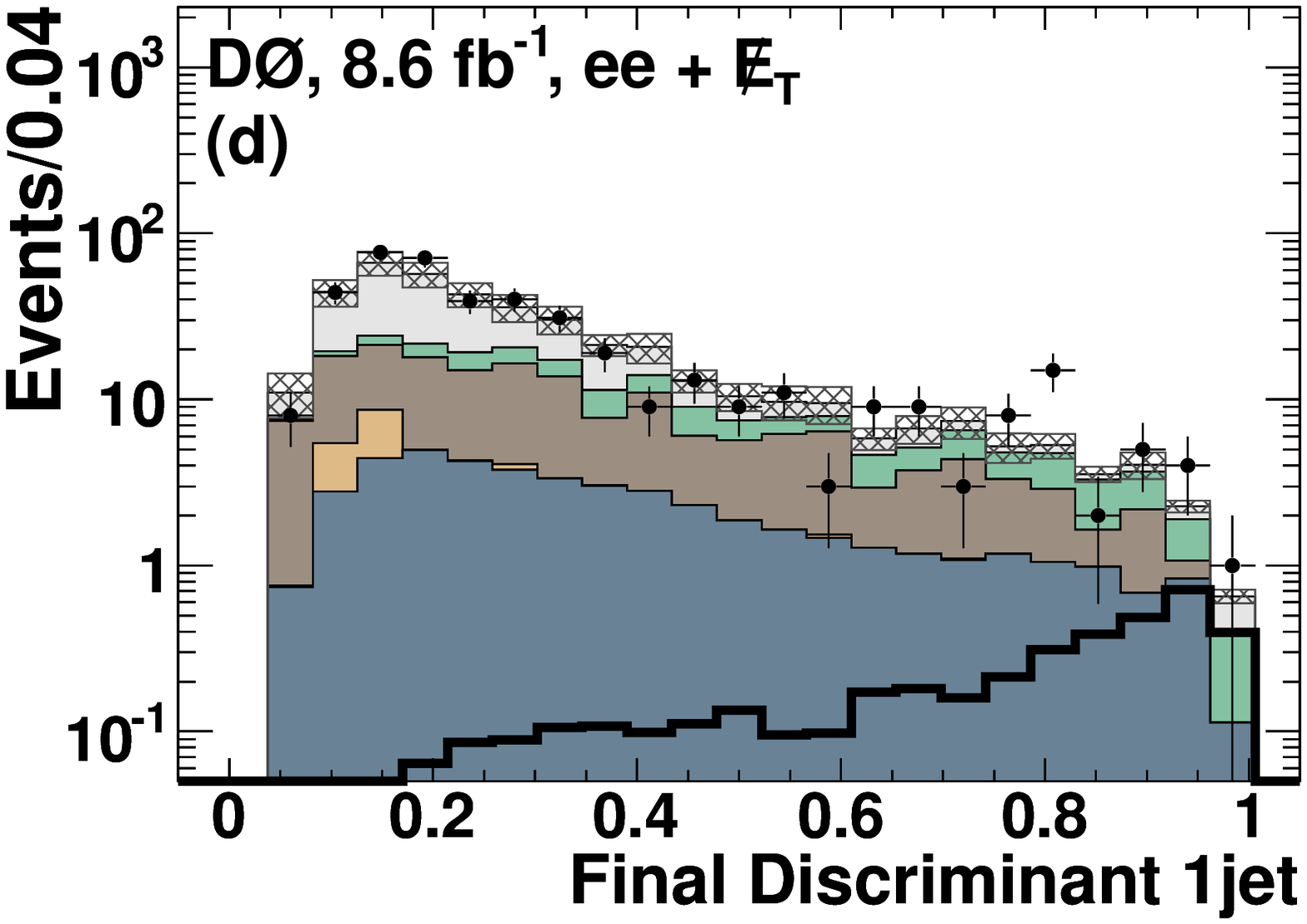} \\
      \includegraphics[width=1.0\columnwidth]{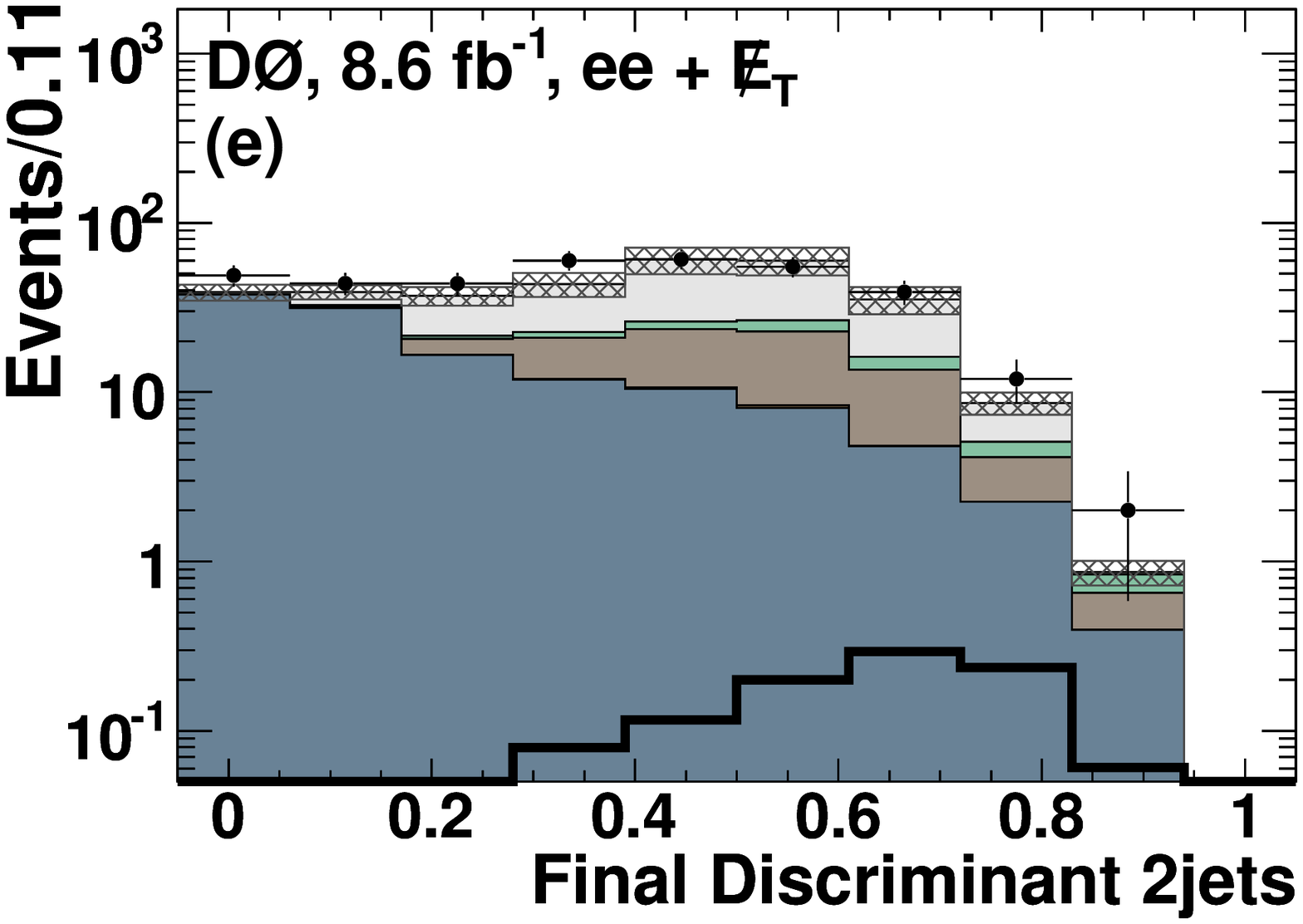} &
      \includegraphics[width=1.0\columnwidth]{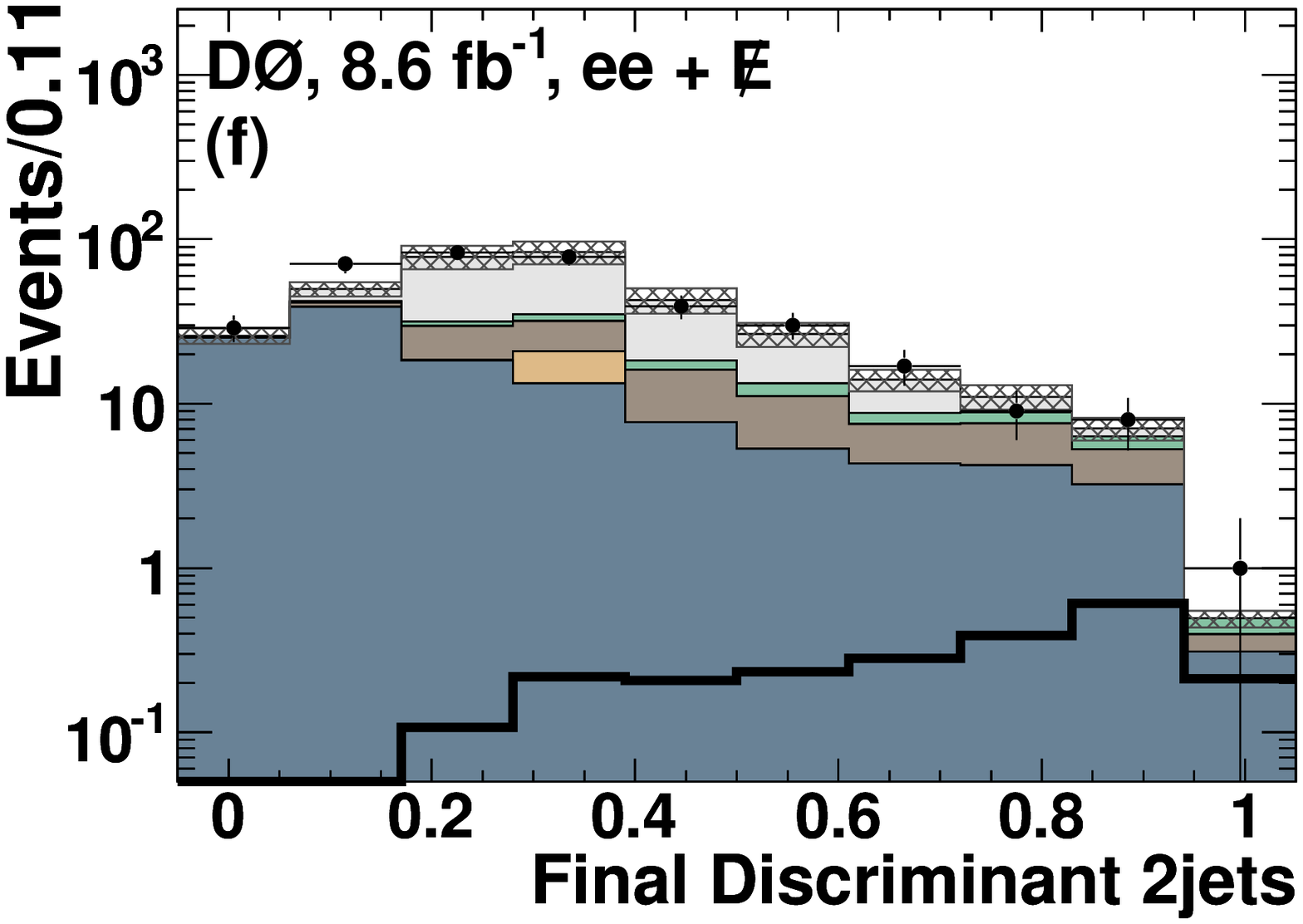} \\
      \end{tabular}
  \end{center}
 \caption{[color online] Final BDT discriminant for the (top-row)
    0-jet, (middle-row) 1-jet, and (bottom-row) $\ge 2$-jet bins for
    the \ee\ final state for a Higgs boson masses of 125 GeV [left
    (a,c,e)] and 165 GeV [right (b,d,f)]. The hatched bands show the
    total systematic uncertainty on the background
    prediction.\label{fig:final_bdt_ee}}
\end{figure*}

\begin{figure*}[!]
  \begin{center}
    \begin{tabular}{cc}
      \includegraphics[width=1.0\columnwidth]{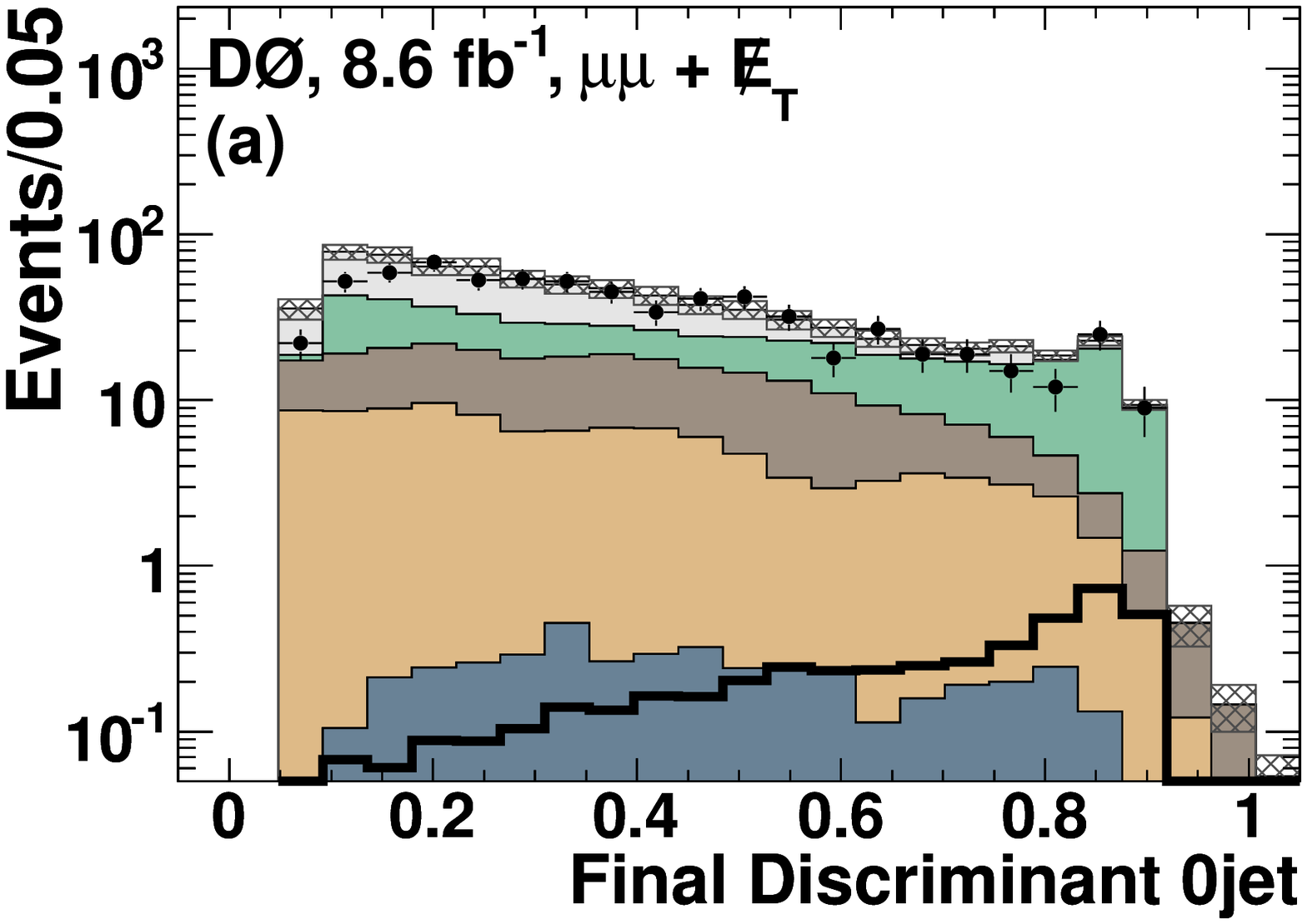} &
      \includegraphics[width=1.0\columnwidth]{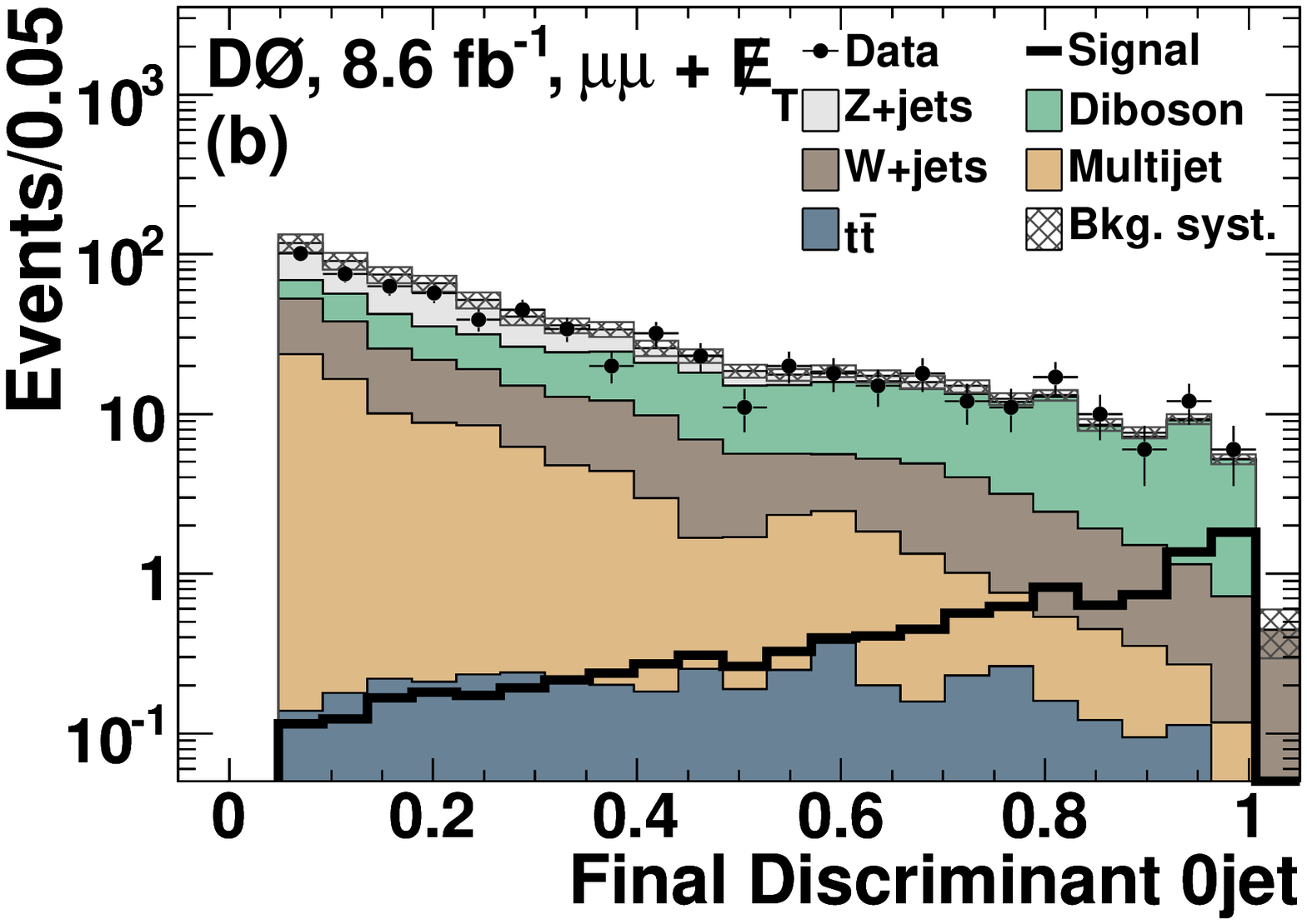}\\
       \includegraphics[width=1.0\columnwidth]{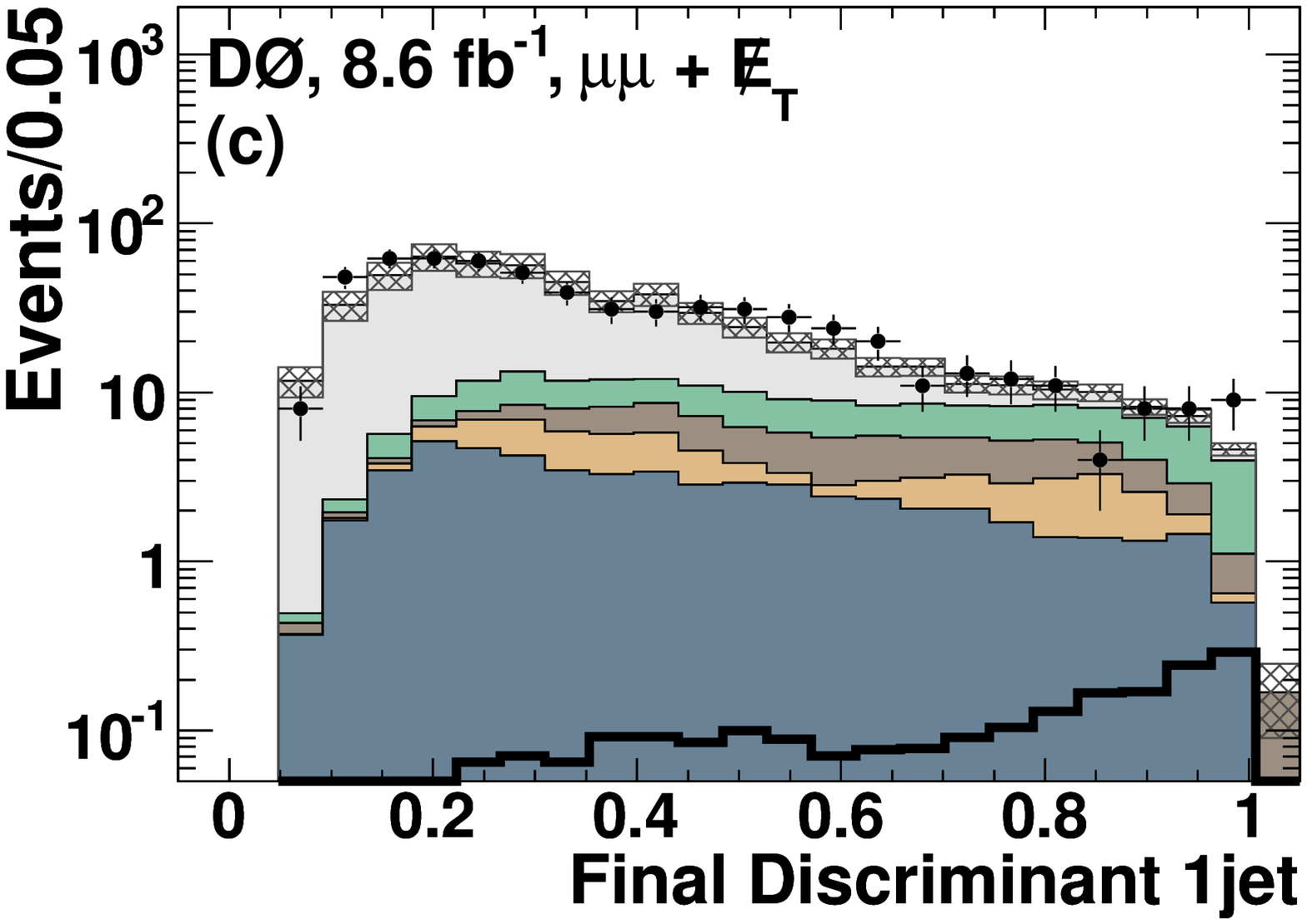}&
      \includegraphics[width=1.0\columnwidth]{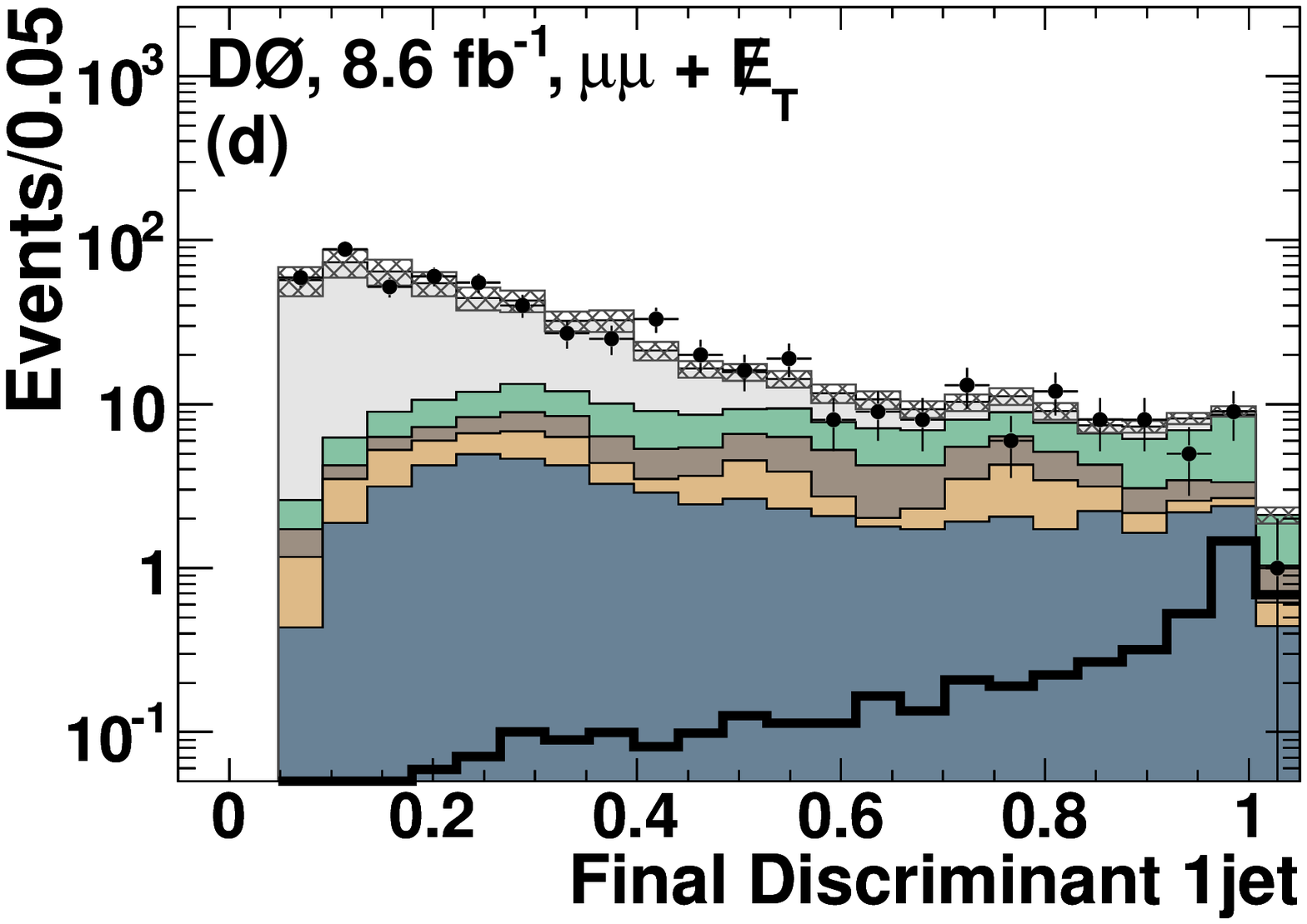} \\
       \includegraphics[width=1.0\columnwidth]{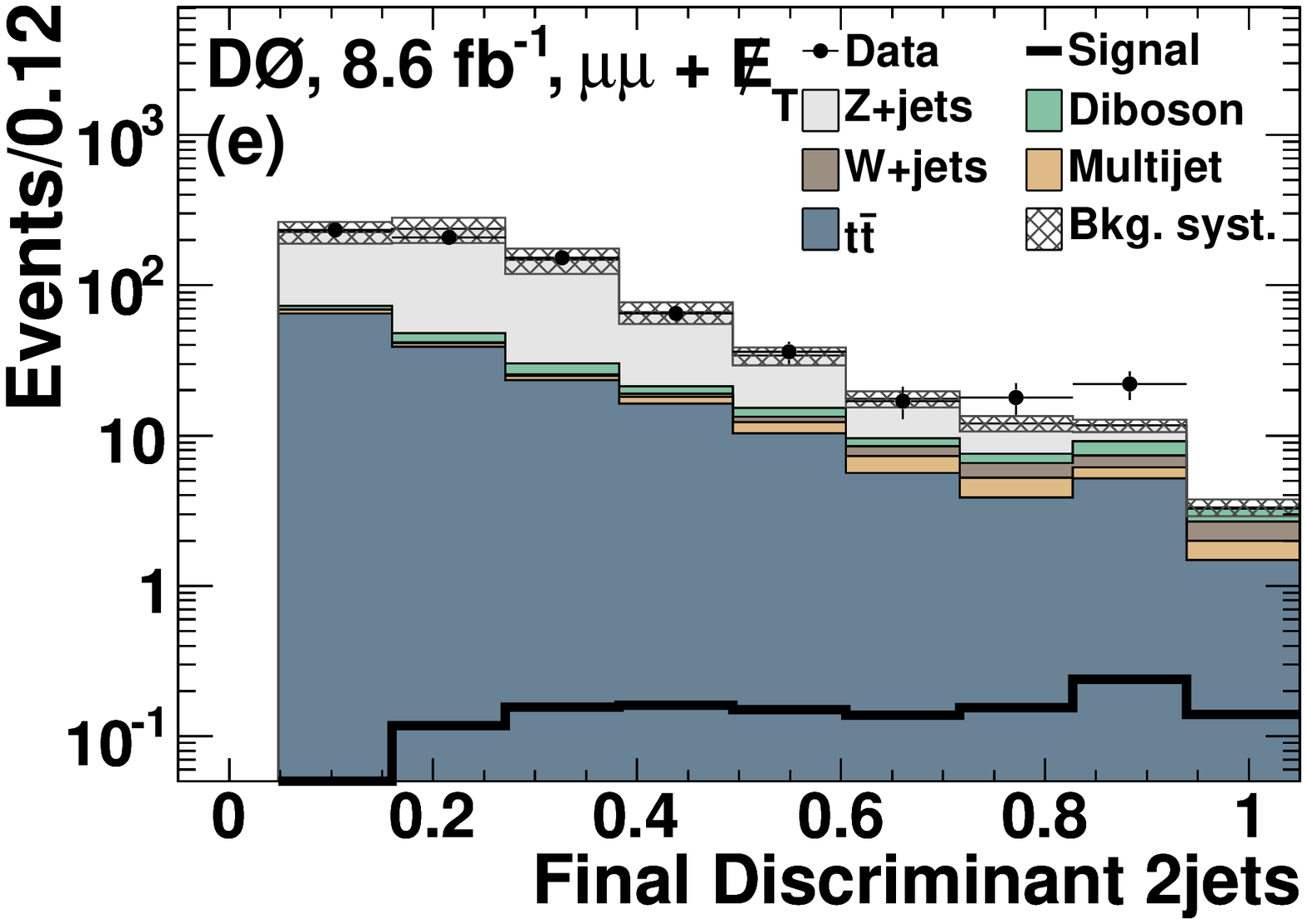}&
      \includegraphics[width=1.0\columnwidth]{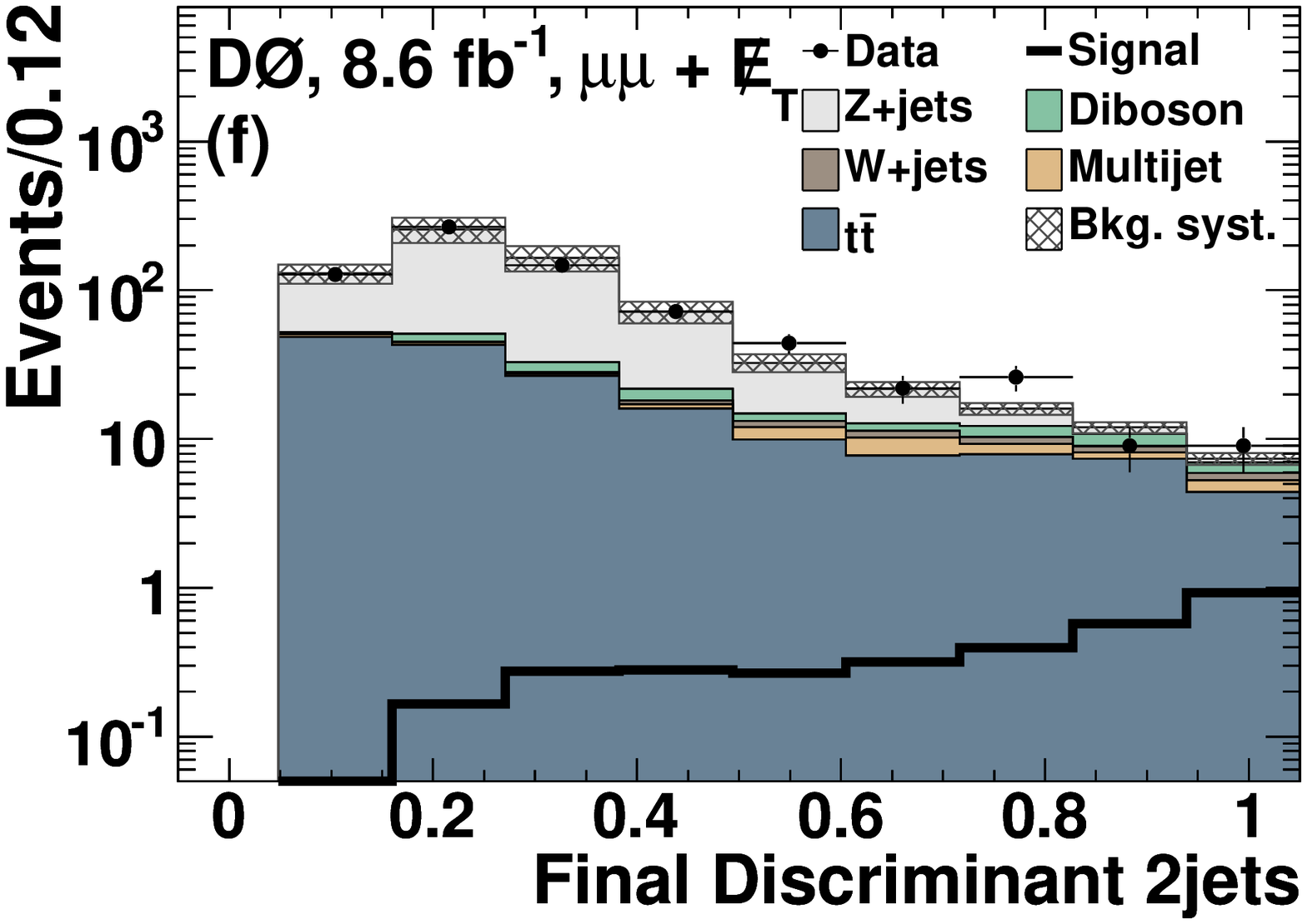} \\
    \end{tabular}
  \end{center}
   \caption{[color online] Final BDT discriminant for the (top-row)
    0-jet, (middle-row) 1-jet, and (bottom-row) $\ge 2$-jet bins for
    the \mm\ final state for a Higgs boson masses of 125 GeV [left
    (a,c,e)] and 165 GeV [right (b,d,f)]. The hatched bands show the
    total systematic uncertainty on the background
    prediction.\label{fig:final_bdt_mm}}
\end{figure*}

\begin{table*}[!]
 \caption{\label{tab:final_cutflow} Expected and observed numbers of
events after the final selection in the \em, \ee, and \mm\ final
states. The signal is for a Higgs boson mass of 165\,GeV.}
 \begin{ruledtabular}
    \begin{tabular}{cccllcccc}
      & Data & Total background & Signal & $Z/\gamma^*$ & $\ttbar$ & $W+\gamma/$jets & Dibosons & Multijet \\
      \hline
      \hline
      $e^\pm\mu^\mp$: & 1729 & 1806 & 30 & 94 & 335 & 766 & 584 & 29\\
      0 jets  &  1117 & 1222&  18 & 70  & 11 &  641 & 486 & 15 \\
      1 jet   & 335 &  307 &  8 &  19 & 98 & 94 & 87  & 10 \\
      $\ge 2$ jets &  277 & 277 & 4  & 5 & 226  & 31 & 11 &  4   \\
      \hline
      \ee:   & 1607 & 1644 & 14 & 466 & 200 & 658  & 288 & 33 \\
      0 jets      & 812    &881    &  8    &  135      &     6    & 499   & 222   & 20\\
      1 jet       & 430    &408    &  4    &  181      &    54    & 114   &  52   &  7\\
      $\ge 2$jets & 365    &355    &  2    &  150      &   140    & 45    &  14   &  6\\
      \hline
      \mm:    & 1950 & 1997 & 18 & 1101 & 231 & 198 & 328 & 140\\
      0 jets      & 645   &720    & 10    &  227      &     4     &  155    &  236   &  98\\
      1 jet       & 581   &564    &  5    &  376      &    56     &   35    &   68   &  30\\
      $\ge 2$jets & 724   &713    &  3    &  498      &   171     &    8    &   24   &  12\\
    \end{tabular}
    \end{ruledtabular}
 \end{table*}

\section{\label{sec:syst}Systematic Uncertainties}

Systematic uncertainties are characterized for each final state,
background, and signal processes. Uncertainties that modify only the
normalization and uncertainties that change the shape of the final
discriminant distribution are taken into account. Systematic
uncertainties that contribute only to the normalization are:
theoretical cross sections for diboson, 6\%, and $t\bar{t}$
production, 7\%; multijet normalization, 30\%; overall normalization,
4\%, which accounts for the uncertainty on the lepton
trigger/identiﬁcation efficiency and the integrated luminosity; and a
$Z$+jets jet-bin-dependent normalization (2--15)\%.

Since the analysis is split into categories depending on the number of
reconstructed jets, renormalization and factorization scale
uncertainties on $\sigma(gg\to H)$ are estimated following the
prescription described in Ref.~\cite{ref:errMatr}. By propagating the
uncorrelated uncertainties of the NNLL
inclusive~\cite{ref:Anastasiou,bib:gluon_fusion_xsec}, NLO $\ge 1$
jet\,\cite{bib:ggH01jetUncert}, and NLO $\ge 2$
jets\,\cite{ref:ggH2jetUncert} cross sections to the exclusive ${gg\to
H}+0$ jet, $\ge 1$ jet, and $\ge 2$ jets rates, the uncertainty matrix
shown in Table\,\ref{tab:ggHuncert} is built. The PDF uncertainties
for $\sigma({gg\to H})$, obtained using the prescription in
Refs.~\cite{bib:ggH01jetUncert,bib:gluon_fusion_xsec}, are also
summarized in Table\,\ref{tab:ggHuncert}. The uncertainties on the
inclusive $\sigma(VH)$ and $\sigma(qqH)$ are taken as 6\% and 5\%,
respectively.

Sources of systematic uncertainty that affect both the normalization
and the shape of the final discriminant distribution are: jet energy
scale (1--4)\% and jet energy resolution (1--3)\%, determined by
varying the parameters of the energy scale correction and the energy
resolution function within one standard deviation~(s.d.)~of the
uncertainty and repeating the analysis using the kinematics of the
modified jets; jet association to the $p\bar{p}$ interaction vertex
(1--2)\%, obtained by varying the correction factor within its
uncertainty; shape of the $b$-tagging discriminant associated with
heavy flavor jets (3--5)\%, determined by varying the correction
factor of the $b$-tagging neural network output within its
uncertainty; $W+$jets modeling (6--50)\%, depending on jet
multiplicity bin and final state, obtained by varying the correction
factors described in the ``Instrumental Backgrounds'' section within
their uncertainties. These uncertainties are presented in terms of the
average fractional change across bins of the final discriminant
distribution for all backgrounds and depend on the jet multiplicity.

Several systematic uncertainties are also included which have a small
$(<1\%)$ effect on the background model: modeling of diboson
production in terms of $p_T$($WW$), determined by taking the
fractional difference of the predicted final discriminant shape
between {\sc mc@nlo} and {\sc pythia} generators; modeling of diboson
production in terms of the impact of the gluon fusion production
process on the $\Delta\phi$ distribution between the leptons,
determined by taking the fractional difference of the predicted final
discriminant shape between {\sc mc@nlo} and {\sc
gg2ww}~\cite{bib:ggWW} generators; and the $p_T$ of the vector boson
from $W+$jets and $Z+$jets production. A summary of the dominant
systematic uncertainties is given in Table~\ref{tab:systematic}.

\begin{table*}[htpb]
  \caption{Elements of the uncertainty matrix of the scale ($\mu_{\rm
      R}$,$\mu_{\rm F}$) and PDF uncertainties on $\sigma(gg\to H)$
      for the three jet multiplicity categories considered, where
      $s_0$, $s_1$ and $s_2$ are the elements of the uncertainty
      matrix.}
  \label{tab:ggHuncert}
  \begin{ruledtabular}
    \begin{tabular}{ccccc}
      $\sigma$ $\mu_{R},\mu_{F}$ & $s_0$  & $s_1$ & $s_2$    &   PDF \\
      \hline						       
      $0$ jet      & $13.4\,\%$ & $-23.0\,\%$  & --          &  $7.6\,\%$\\
      $\ge 1$ jet  & --         & $35.0\,\%$   & $-12.7\,\%$ &  $13.8,\%$ \\
      $\ge 2$ jets & --         & --           & $33.0\,\%$  &   $29.7\,\%$\\
    \end{tabular}
    \end{ruledtabular}
\end{table*}

\begin{table*}[htpb]
  \caption{Summary of systematic uncertainties (in \%) for source
  categories. The jet, $b$-tagging and PDF related uncertainties are
  quoted for all the backgrounds combined.}
  \label{tab:systematic}
  \begin{ruledtabular}
    \begin{tabular}{cc}
      Source                          & Uncertainty (\%)     \\
      \hline
      \hline
      Overall normalization             & 4.0              \\
      $W+$jets  normalization           & 6.0--50.0                             \\
      Diboson cross section             & 6.0              \\
      $t\bar{t}$ cross section          & 7.0              \\
      Multijet normalization            & 30.0             \\
      $Z+$jets jet-bin normalization    & 2.0--15.0        \\
      $gg\rightarrow H$ cross section   & See Table\,\ref{tab:ggHuncert} \\
      $VH$ cross section                & 6.0              \\
      $qqH$ cross section               & 5.0              \\
      Jet energy scale                  & 1.0--4.0         \\
      Jet resolution                    & 1.0--3.0         \\
      Jet primary vertex association    & 1.0--2.0         \\
      $b$-tagging discriminant          & 1.0--2.0         \\
      PDF (background)                  & 2.5              \\
    \end{tabular}
    \end{ruledtabular}
\end{table*}

\section{\label{sec:result}Results}

The methodology of this search is validated by an independent
measurement of the $p\bar{p}\rightarrow W^+W^-$ cross section using
the analysis procedure described in the ``Event Selection'' section of
this Article, considering $WW$ events as the signal. This is motivated
by the fact that $WW$ production is the main contributor to the
diboson entry in both Tables~\ref{tab:presel_cutflow} and
\ref{tab:final_cutflow} compared to the expected yields from
$WZ$~and~$ZZ$ production backgrounds. Similarly to the Higgs boson
search, a dedicated BDT is constructed, but now it is trained to
separate $WW$ production signal from other SM processes.  For this
BDT, we use the identical input variables, the same separation method
in terms of jet multiplicity bins, and the same treatment of
systematic uncertainties as in the Higgs boson search.  The \ee~and
\mm~final states use only the $0$ and $1$ jet multiplicity bins while
the \em~final state uses all three jet multiplicity bins yielding a
total of seven analysis channels for the combination. The results
obtained for the $WW$ cross section in the individual final states and
their combination are summarized in Table~\ref{tab:wwXsec}. The
measured value of 11.1 $\pm$ 0.8 pb is in good agreement with the SM
prediction of 11.7 $\pm$ 0.8 pb~\cite{bib:dibo-xs}.  The presence of a
Higgs boson signal in the mass range $115<M_{H}<180$~GeV would bias
the cross section measurement result by 5\% at most.  This maximum
bias is reached for $M_{H}=165$~GeV, but at low masses
$(M_{H}<130$~GeV), the bias would be less than 2\%.

Figures~\ref{fig:bk_sub_WW} and \ref{fig:bk_sub_higgs} show the
expected $WW$ and Higgs boson signals, respectively, for the combined
decay channels in the analysis. In these distributions, the data is
shown, ordered in bins of increasing values of the $s/b$ ratio, after
the subtraction of the SM backgrounds. The background model is fit to
the data, and the uncertainties on the background are those after the
systematic uncertainties have been constrained by the fit.

\begin{figure*}[!]
  \begin{center}
    \includegraphics[width=1.0\columnwidth]{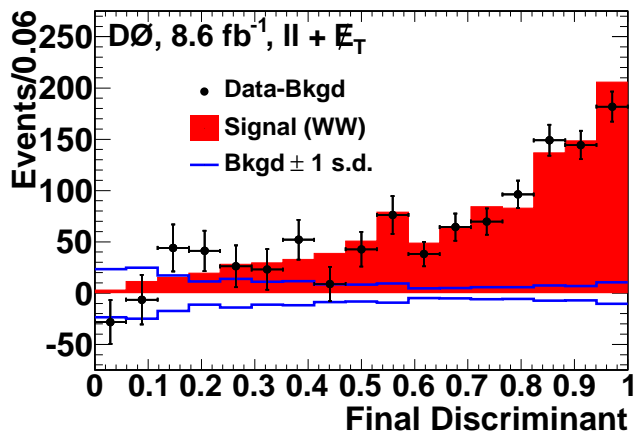}
  \end{center}
  \caption{[color online]\label{fig:bk_sub_WW} The
    background-subtracted data distribution for the final
    discriminant, summed in bins with similar signal to background
    ratio, for the $WW$ cross section measurement. The uncertainties
    shown on the background-subtracted data points are the square
    roots of the post-fit background predictions in each bin,
    representing the expected statistical uncertainty on the
    data. Also shown is the $\pm$1 standard deviation (s.d.) band on
    the total background after fitting.}
\end{figure*}

\begin{table*}[!]
\caption{Summary of the measurements of the $p\bar{p}\rightarrow
    W^+W^-$ cross section measurement (in pb) in all seven channels
    considered and their combination.}
 \label{tab:wwXsec} 
 \begin{ruledtabular}
   \begin{tabular}{cc}
    Channel & $\sigma(p\bar{p}\rightarrow W^+W^-)$ \\
    \hline
    \hline
    \em & 10.6 $\pm$ 0.6 (stat) $\pm$ 0.6 (syst)\\
    \ee & 12.4 $\pm$ 1.2 (stat) $\pm$ 0.9 (syst)\\
    \mm & 11.0 $\pm$ 0.9 (stat) $\pm$ 0.7 (syst) \\
    \hline
    Combined &11.1 $\pm$ 0.5 (stat) $\pm$ 0.6 (syst) \\
  \end{tabular}
   \end{ruledtabular}
 \end{table*}

The final multivariate discriminants of the SM Higgs boson search,
shown in Figs.~\ref{fig:final_bdt_em} -- \ref{fig:final_bdt_mm},
demonstrate that the data is well described by the sum of the
background predictions.  In the absence of an excess in the number of
observed events above the SM backgrounds, these BDT output
distributions are used to set upper limits on the Higgs boson
inclusive production cross section $\sigma(p\bar{p} \rightarrow H +
X)$ assuming SM values for the branching ratios and for the relative
cross sections of the various Higgs production mechanisms considered.
The limits are calculated using a modified frequentist method with a
log-likelihood ratio (LLR) test statistic~\cite{bib:modified_freq}.
The value of $CL_{s}$ is defined as $CL_{s} = CL_{s+b}/CL_{b}$, where
$CL_{s+b}$ and $CL_{b}$ are the $p$-values for the signal+background
and background-only hypotheses, respectively.  Expected limits are
calculated from the background-only LLR distribution whereas the
observed limits are quoted with respect to the LLR values measured in
data. They both are reported at the 95$\%$~C.L.

The multivariate discriminants corresponding to the nine individual
channels are all used to obtain upper limits on the Higgs boson
production cross section. Given the differences in the background
contributions to each of the channels, the nine BDT output
distributions are not combined in a single distribution for the limit
extraction, but treated separately. The degrading effects of
systematic uncertainties on the search sensitivity are minimized by
fitting individual background contributions to the data by maximizing
a profile likelihood function for the background-only and
signal+background hypotheses separately, taking into account
appropriately all correlations between the systematic
uncertainties~\cite{bib:collie}. Table~\ref{tab:alllimit} and Fig.\
\ref{fig:alllimit}~present expected and observed upper limits at the
95\% C.L. for $\sigma(p\bar{p} \rightarrow H + X)$ relative to SM
predictions for each Higgs boson mass considered.

The corresponding LLR distributions are shown in Fig.\
\ref{fig:allllr}.~Included in this plot are the median of the LLR
distributions for the background-only hypothesis $(LLR_b)$, the
signal-plus-background hypothesis $(LLR_{s+b})$, and the observed
value for the data $(LLR_{\mathrm{obs}})$. The shaded bands represent
one and two s.d.\,departures for $LLR_{b}$ centered on the median. The
separation between the $LLR_b$ and $LLR_{s+b}$ distributions provides
a measure of the discriminating power of the search. The current
result indicates that the signal+background model can be separated
from the background-only model by up to 1 s.d.\,over most Higgs boson
masses between 115 to 200 GeV while the level of separation increases
above 2 s.d.\,for Higgs boson masses between 160 to 170 GeV. The
sensitivity of the search reaches an expected exclusion of
$159<M_H<169$~GeV at 95\% C.L. However due to a slight excess in the
data, an observed exclusion is not obtained.

\begin{figure*}[!]
  \begin{center}
    \begin{tabular}{cc}
      \includegraphics[width=1.0\columnwidth]{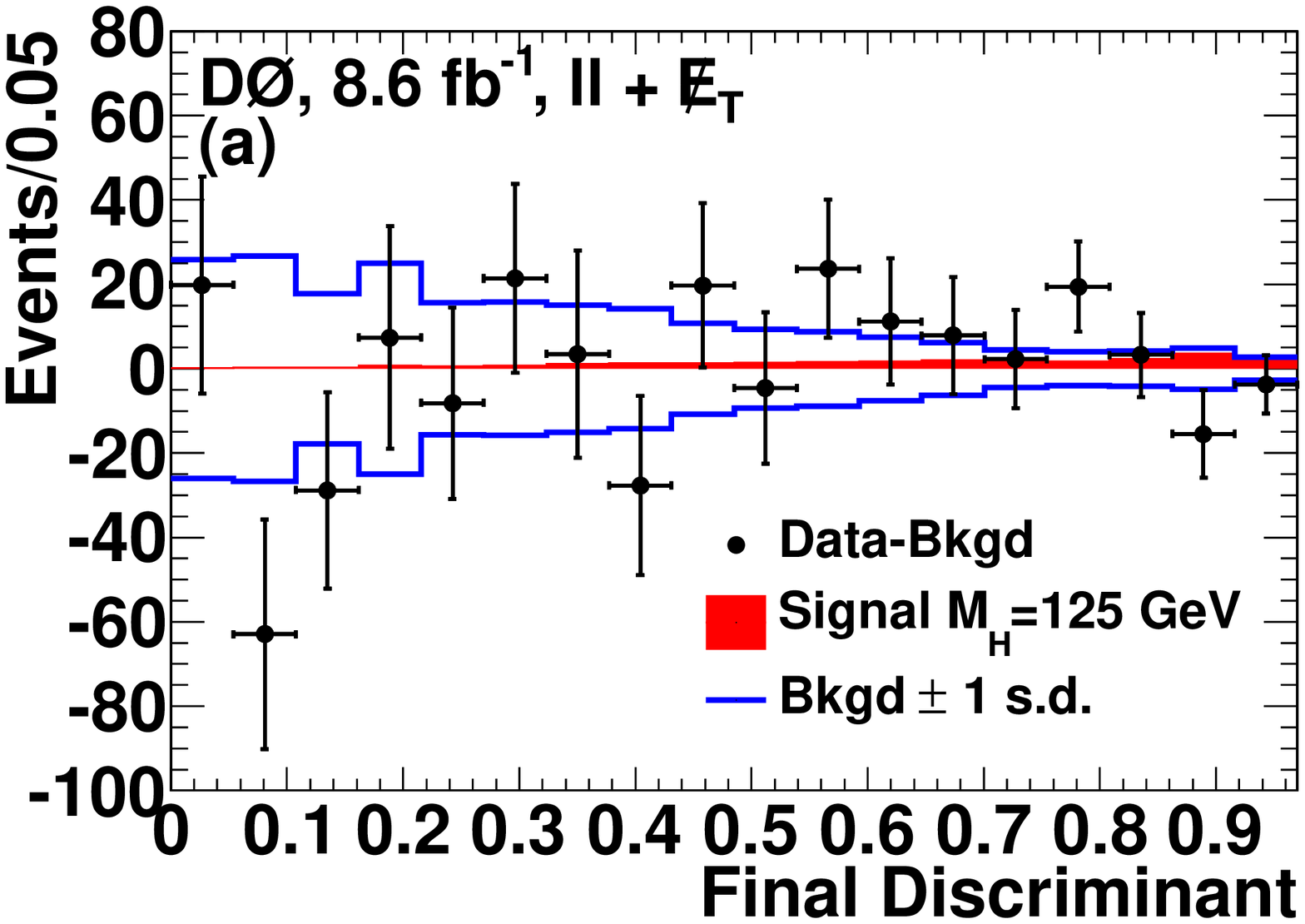}&
      \includegraphics[width=1.0\columnwidth]{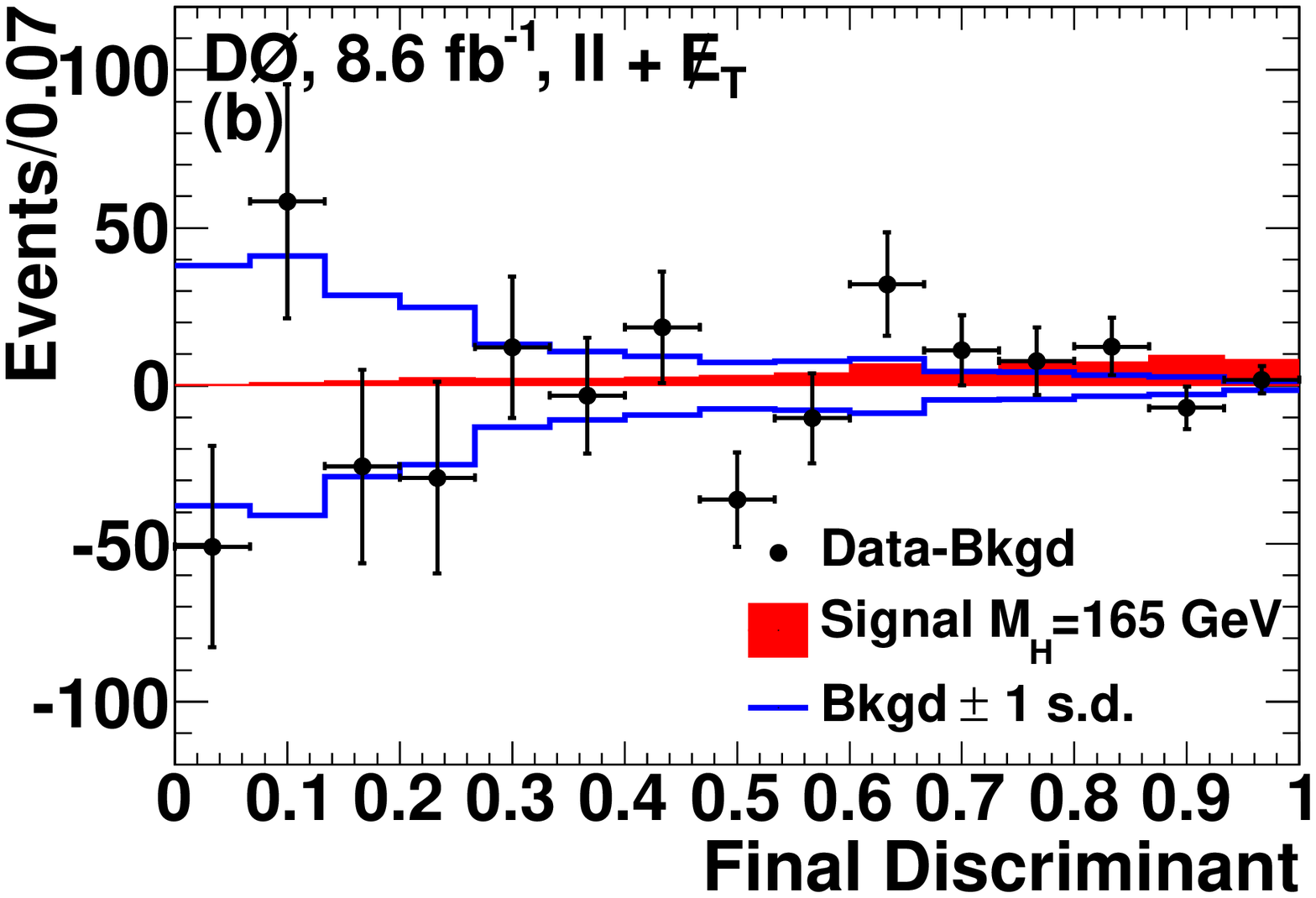}
      \\
    \end{tabular}
  \end{center}
  \caption{[color online]\label{fig:bk_sub_higgs} The
background-subtracted data distributions for the final discriminants,
summed in bins with similar signal to background ratio, for (a) $M_H
=125$~GeV and (b) $M_H =165$~GeV. The uncertainties shown on the
background-subtracted data points are the square roots of the post-fit
background predictions in each bin, representing the expected
statistical uncertainty on the data. Also shown is the $\pm$1 standard
deviation (s.d.) band on the total background after fitting.}
\end{figure*}

\begin{table*}[!]
\caption{\label{tab:alllimit} Expected and observed upper limits at the
95\% C.L. for $\sigma(p\overline{p}\rightarrow H+X)$ relative to the
SM for the total combination and separately for the \em, \ee\, and
\mm\ channels for different Higgs boson masses ($M_H$).}
\begin{ruledtabular}
\begin{tabular}{cccccccccccccccccccc}
$M_H$ (GeV)     &115   &120  &125  &130  &135  &140  &145  &150  &155  &160  &165  &170  &175  &180  &185  &190  &195  &200 \\
\hline
\hline
Exp. all:  &8.00 &5.37 &3.81 &3.02 &2.43 &2.09 &1.77 &1.53 &1.28 &0.92 &0.85 &1.05 &1.27 &1.49 &1.88 &2.48 &2.87 &3.32\\
Obs. all:  &13.27 &9.14 &5.00 &4.71 &3.93 &3.28 &2.13 &1.99 &1.75 &1.10 &1.17 &1.40 &1.40 &1.64 &1.91 &2.34 &2.87 &3.50\\
\hline
Exp. \em &11.25 &7.08 &5.07 &4.01 &3.18 &2.76 &2.29 &1.93 &1.60 &1.21 &1.13 &1.39 &1.64 &1.96 &2.48 &3.12 &3.66 &4.24 \\
Obs. \em &13.86 &8.50 &5.12 &4.62 &4.01 &2.61 &1.96 &1.68 &1.47 &1.10 &1.27 &1.38 &1.60 &1.68 &2.28 &2.52 &2.84 &3.39\\
\hline
Exp. \ee  &16.07 &11.53 &8.08 &6.30 &4.84 &4.05 &3.60 &3.12 &2.65 &1.92 &1.82 &2.11 &2.63 &3.07 &3.66 &4.76 &5.84 &6.52\\
Obs. \ee  &19.37 &13.93 &10.08 &9.12 &6.31 &6.65 &4.78 &4.95 &4.52 &2.61 &2.88 &3.35 &3.16 &4.82 &4.55 &7.12 &8.26 &9.24\\
\hline

Exp. \mm &  15.09 &9.97 &7.08 &5.44 &4.56 &3.92 &3.37 &2.93 &2.60 &1.99 &1.83 &2.29 &2.72 &3.24 &4.16 &5.08 &5.68 &6.89\\
Obs. \mm &  25.84 &18.83 &9.93 &8.34 &7.01 &7.11 &5.37 &4.45 &3.88 &2.99 &2.31 &3.22 &3.79 &4.19 &5.16 &5.78 &7.98 &8.42 \\

\end{tabular}
\end{ruledtabular}
\end{table*}

\begin{figure*}[!]
  \begin{center}
    \begin{tabular}{cc}
      \includegraphics[width=1.0\columnwidth]{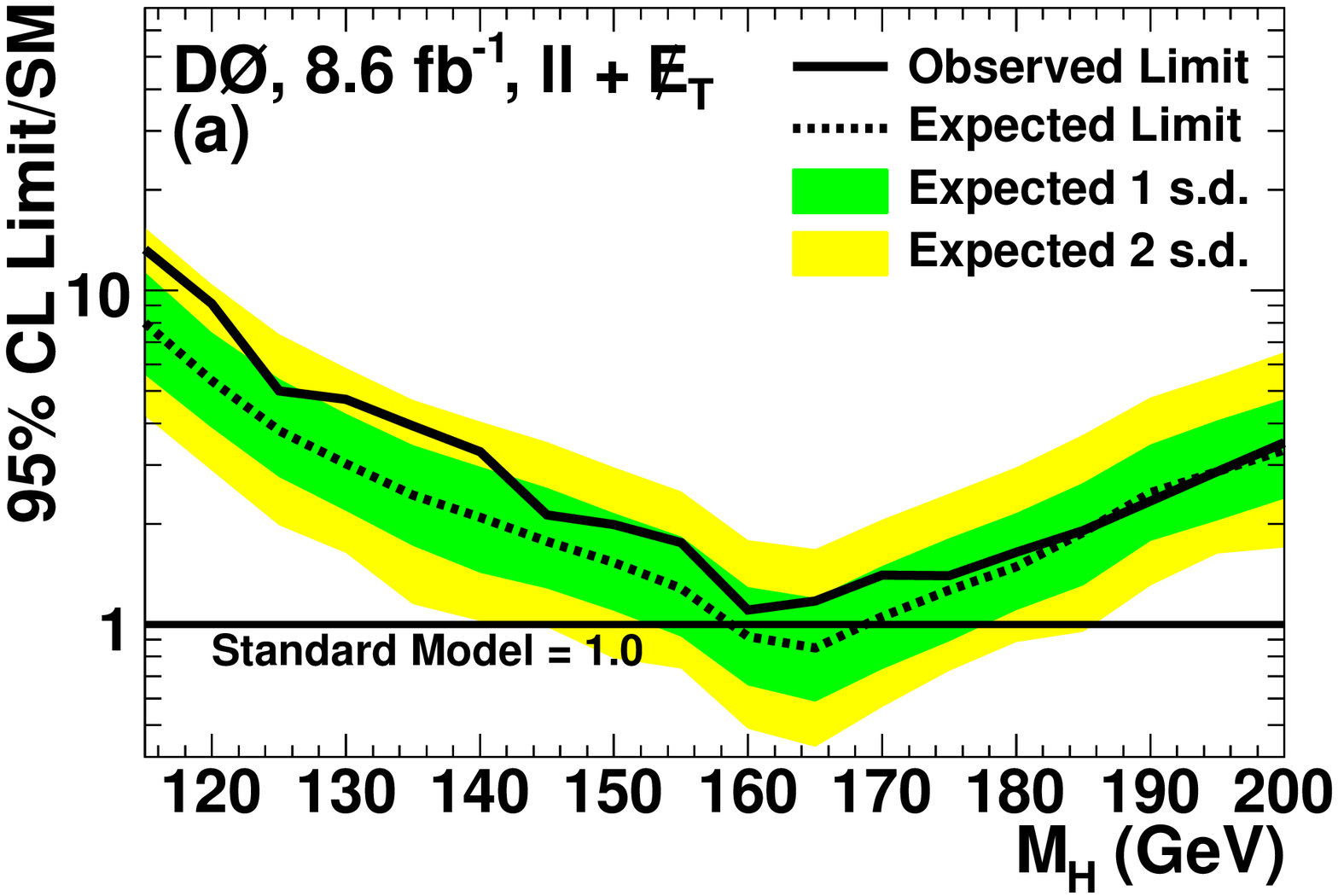} &
      \includegraphics[width=1.0\columnwidth]{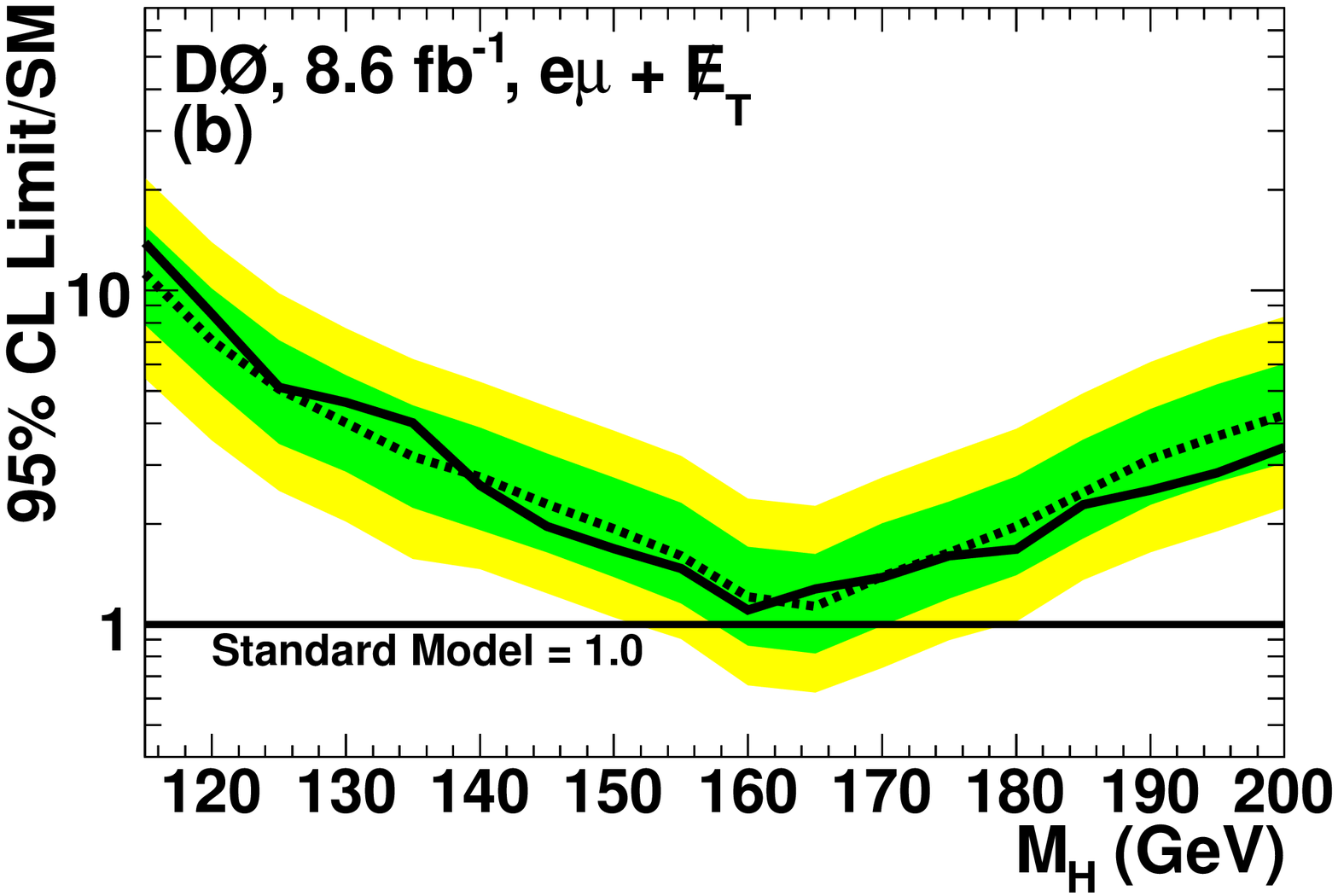} \\
       \includegraphics[width=1.0\columnwidth]{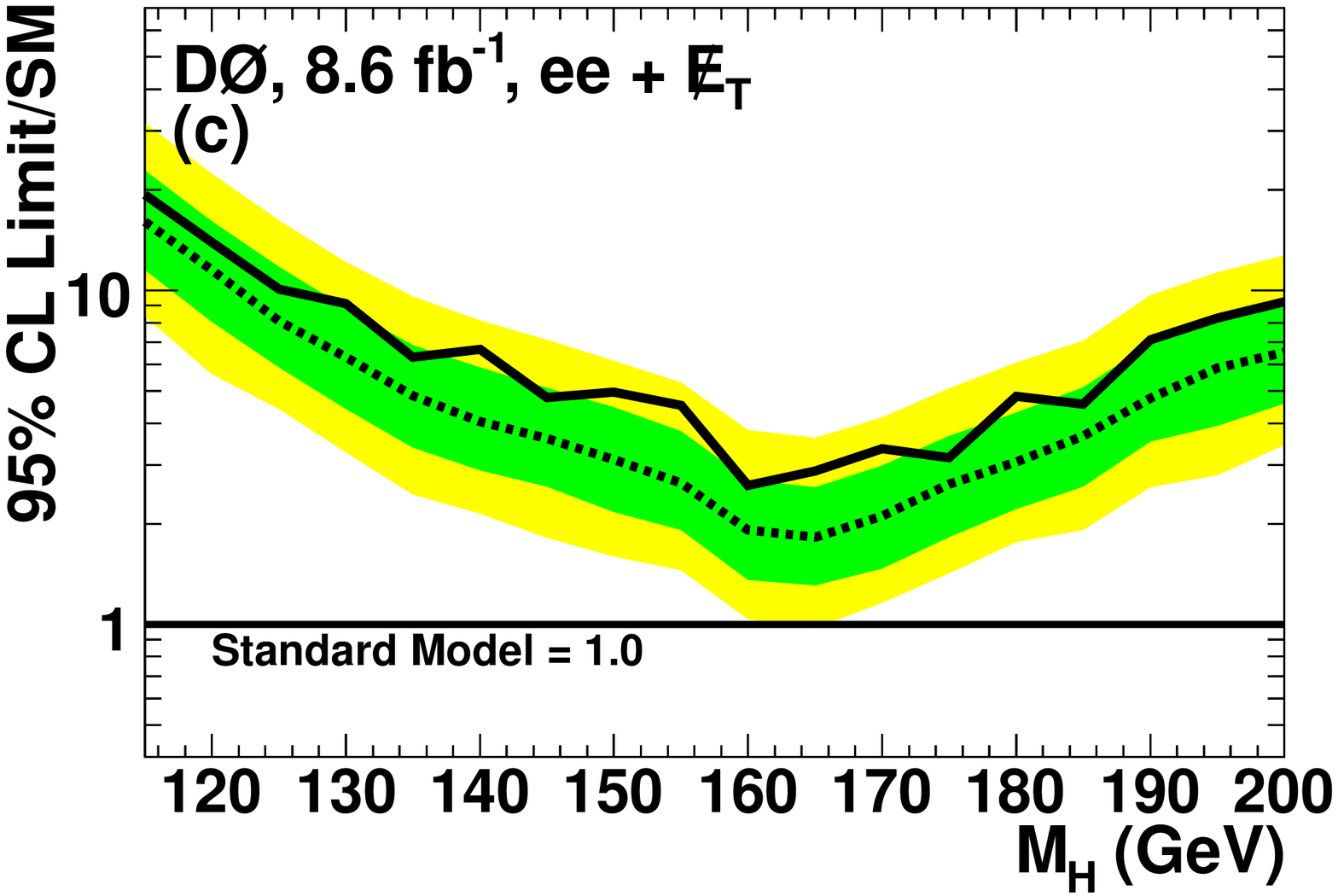} &
       \includegraphics[width=1.0\columnwidth]{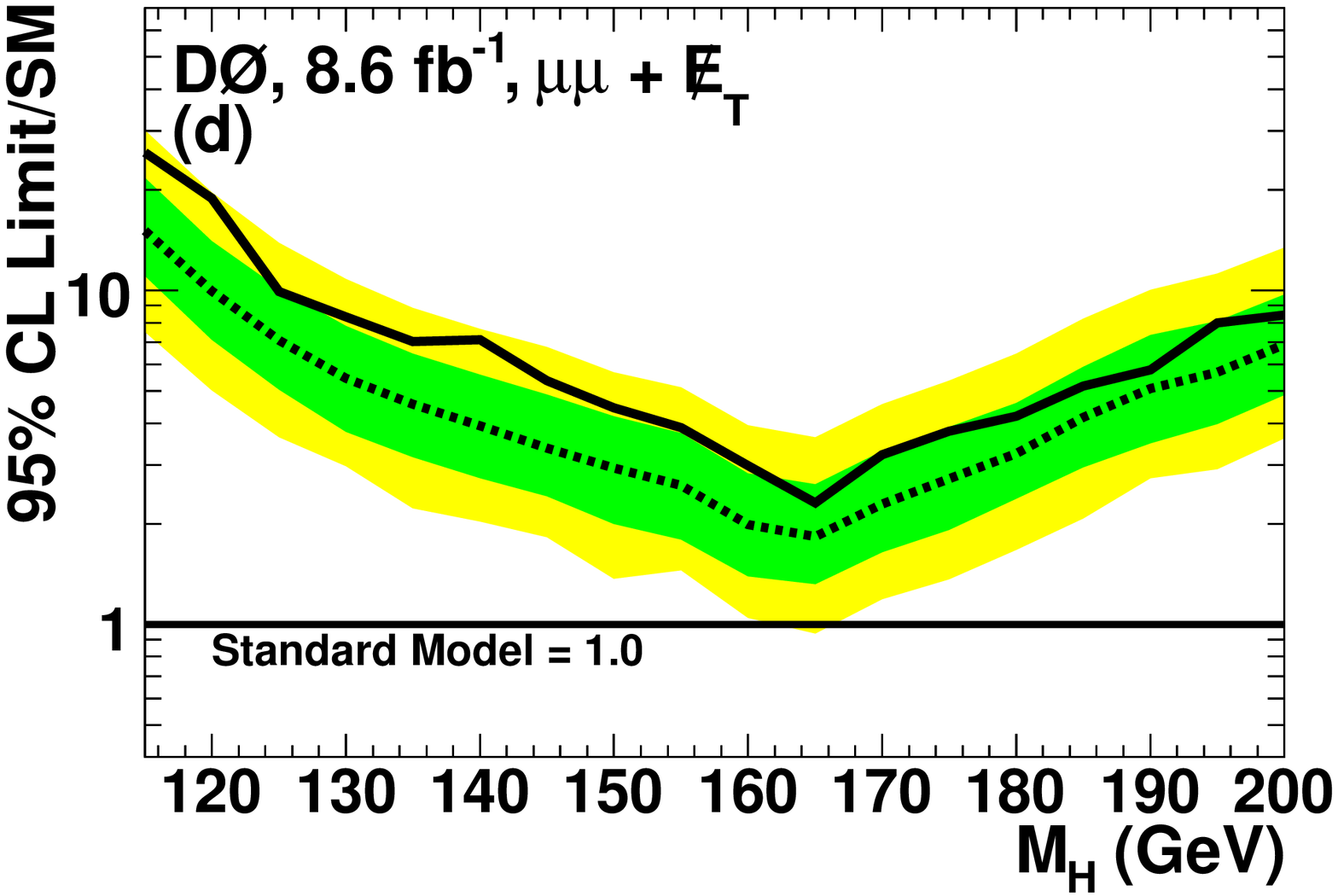} \\
    \end{tabular}
  \end{center}
  \caption{[color online]\label{fig:alllimit} Excluded cross section,
    $\sigma(p\overline{p}\rightarrow H+X)$, at the 95\% C.L. in units
    of the SM cross section as a function of $M_H$ using (a) all
    channels, (b) \em\ channel, (c) \ee\ channel, (d) \mm\ channel.}
\end{figure*}

\begin{figure*}[!]
  \begin{center}
    \begin{tabular}{cc}
      \includegraphics[width=1.0\columnwidth]{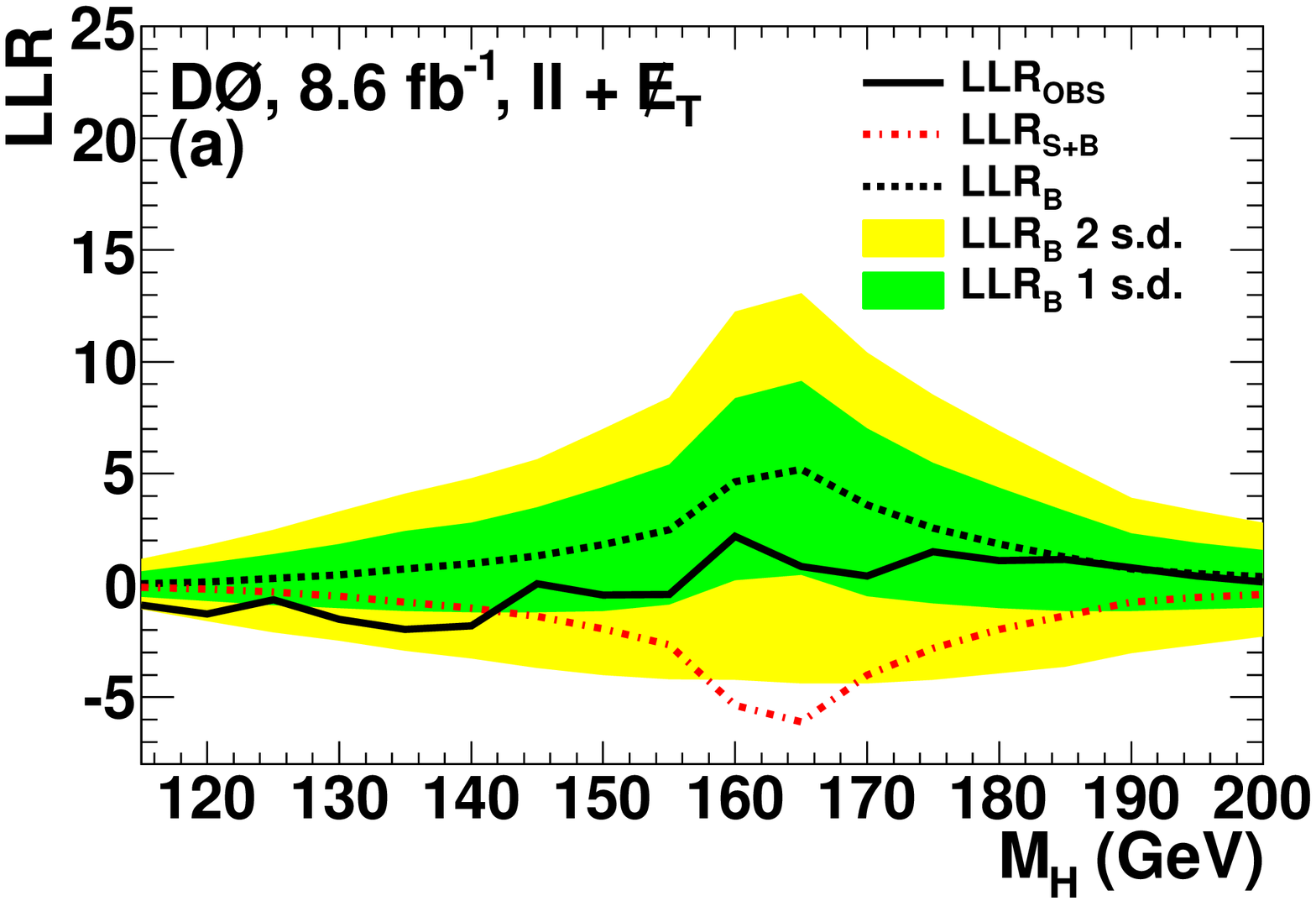} &
      \includegraphics[width=1.0\columnwidth]{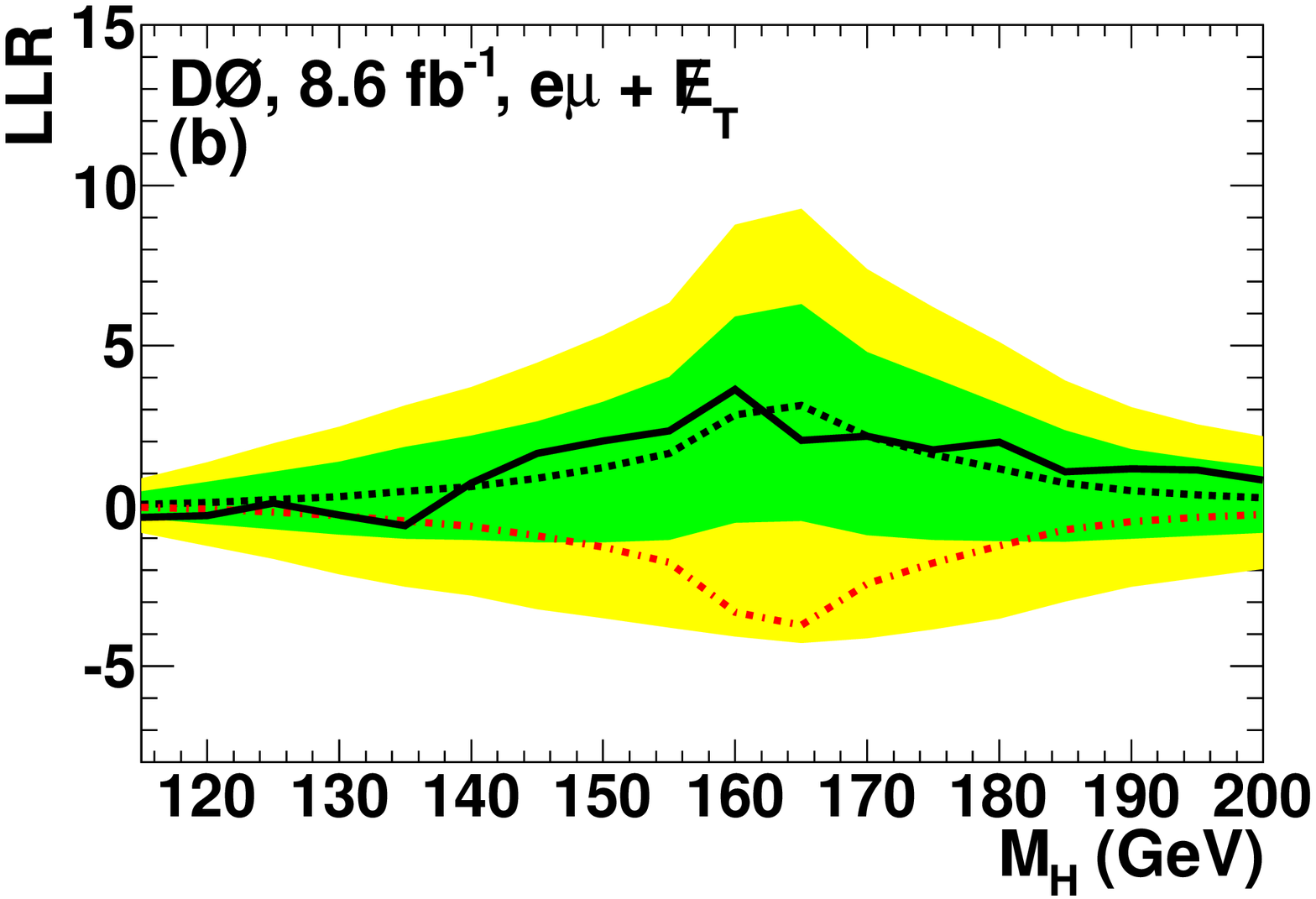} \\
		      \includegraphics[width=1.0\columnwidth]{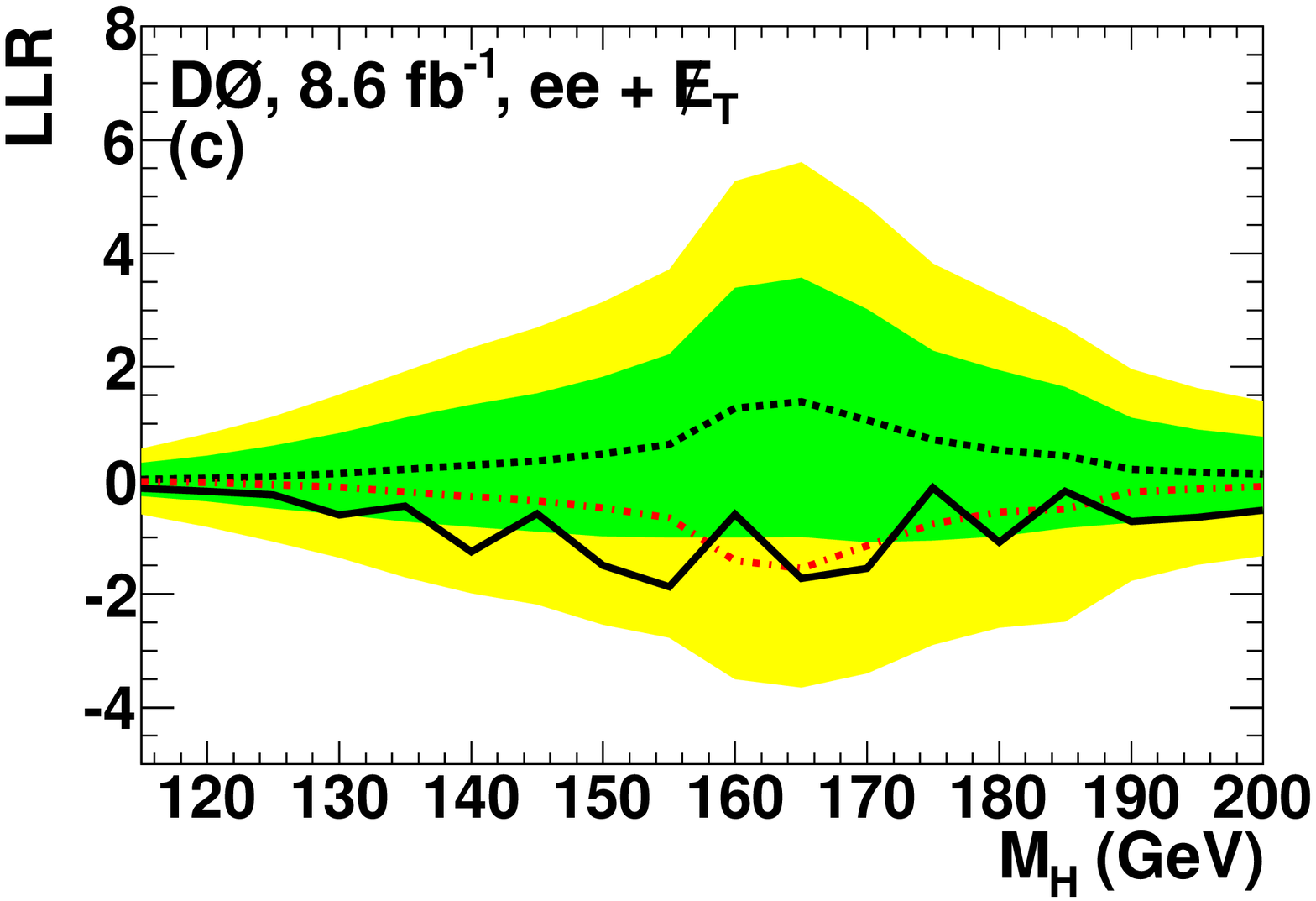} &
		      \includegraphics[width=1.0\columnwidth]{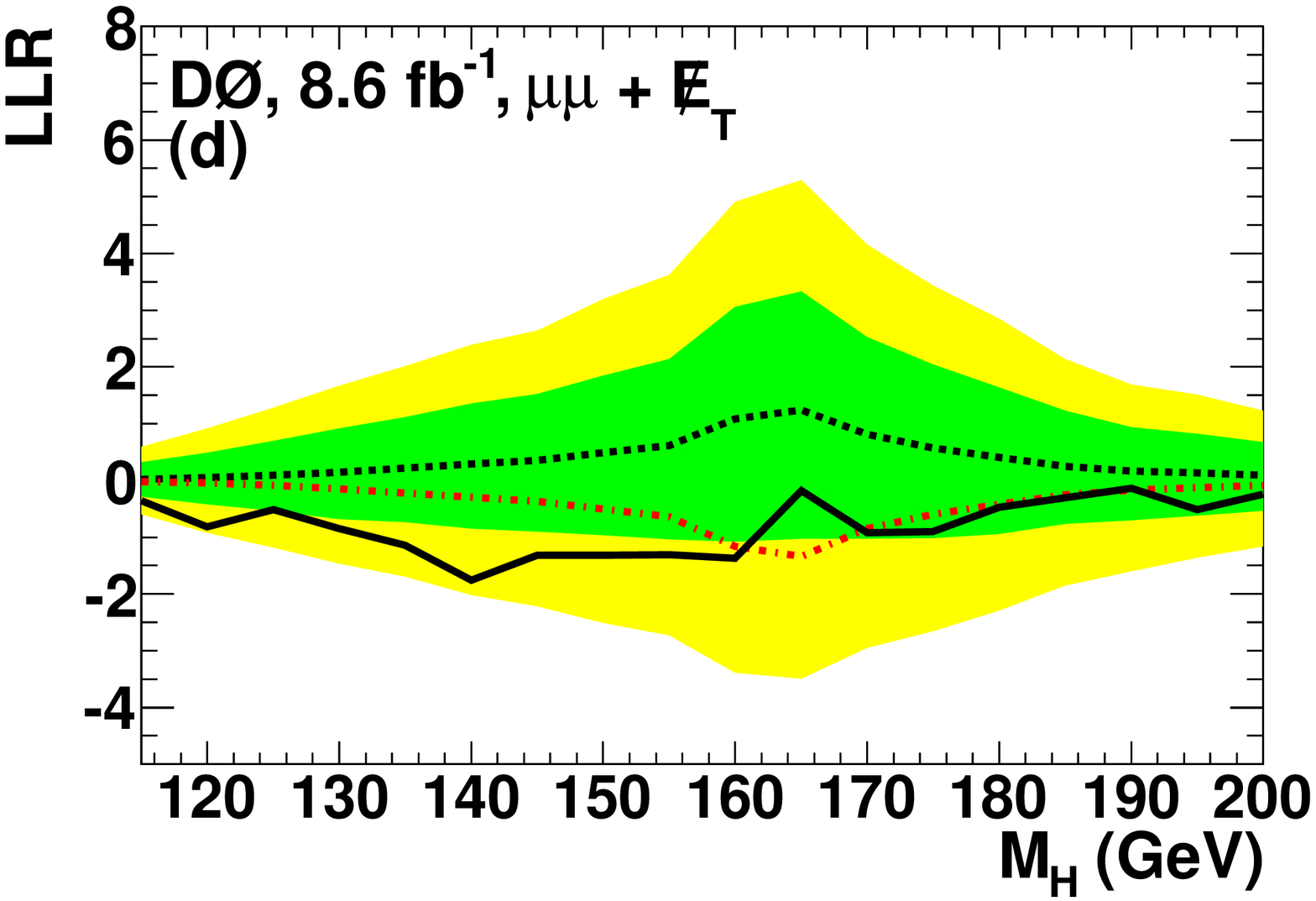} \\
    \end{tabular}
  \end{center}
  \caption{[color online]\label{fig:allllr} The observed LLR (solid
    line) as a function of $M_H$ using (a) all channels, (b) \em\
    channel, (c) \ee\ channel, (d) \mm\ channel.  Also shown are the
    expected LLR distributions for the background only hypothesis
    (dashed line) and for the signal+background (dash-dotted line)
    hypothesis, with the dark grey and light grey bands indicating
    $\pm1$ and $\pm2$ s.d. fluctuations of the expected LLR
    distributions for the background-only hypothesis, respectively.}
\end{figure*}

\section{\label{sec:conclusion}CONCLUSIONS}

We have performed a search for SM Higgs boson production using final
states with two oppositely charged leptons and large missing
transverse energy in the \em, \ee,~and \mm~channels.  After imposing
all selection criteria, no significant excess in data over expected SM
backgrounds is observed.  We set upper limits on Higgs boson
production at the 95\% C.L. The sensitivity of the search reaches an
expected exclusion of 159 $< M_{H} <$ 169~GeV. The best observed limit
is obtained at 160~GeV, where it reaches 1.1 times the SM
expectation. This channel is the single most sensitive channel when
the $H
\rightarrow WW$ branching ratio is dominant $(M_H > 135$~GeV$)$, and
for lower masses at $M_H=125$~GeV, this search still has a similar
sensitivity as a single major low mass channel ($WH$ or $ZH$) with an
expected limit of 3.8 times the SM expectation~\cite{bib:low_mass}.
The results and the analysis techniques are validated through an
independent measurement of the $WW$ production cross section, which
agrees with the NNLO calculation.

\section{\label{sec:ackn}ACKNOWLEDGMENTS}
%
We thank the staffs at Fermilab and collaborating institutions,
and acknowledge support from the
DOE and NSF (USA);
CEA and CNRS/IN2P3 (France);
MON, Rosatom and RFBR (Russia);
CNPq, FAPERJ, FAPESP and FUNDUNESP (Brazil);
DAE and DST (India);
Colciencias (Colombia);
CONACyT (Mexico);
NRF (Korea);
FOM (The Netherlands);
STFC and the Royal Society (United Kingdom);
MSMT and GACR (Czech Republic);
BMBF and DFG (Germany);
SFI (Ireland);
The Swedish Research Council (Sweden);
and
CAS and CNSF (China).
%


\begin{thebibliography}{99}
\bibitem{bib:eta}
The pseudorapidity is defined as $\eta = - \ln[\tan(\theta/2)]$, where
$\theta$ is the polar angle relative to the proton beam direction.
\end{thebibliography}
\end{document}
%